\documentclass[]{emulateapj}
\usepackage{natbib,amsmath}

\begin{document}

\def \nar  {NAR} 
\def \jcap  {JCAP}
\def \na  {Nature}
\newcommand\ncbin{5}
\newcommand\npair{427}
\newcommand\ncpair{137}
\newcommand\nsdss{78}
\newcommand\nciv{105}
\newcommand\vfct{0.73 \pm 0.10} 
\newcommand\dndz{2.1 \pm 0.2}
\newcommand\vro{7.5^{+2.8}_{-1.4} \, \mhMpc}
\newcommand\vgmm{1.7^{+0.1}_{-0.2}}
\def\N#1{{N({\rm #1})}}
\newcommand\mlstar{L^*}
\newcommand\lstar{$\mlstar$}
\def \mmccgm {M_{\rm CGM}^{\rm cool}}
\def \mccgm {$\mmccgm$}
\def \mnhalo {n_{\rm halo}}
\def \nhalo {$\mnhalo$}
\def \mmminq {M_{\rm min}^{\rm QPQ}}
\def \mminq {$\mmminq$}
\def \mncom {n_{\rm com}}
\def \ncom {$\mncom$}
\def \mrcom {R^{\rm com}_\perp}
\def \rcom {$\mrcom$}
\def \mzem {z_{\rm em}}
\def \zem {$\mzem$}
\def \zbg {$z_{\rm bg}$}
\def \zfg {$z_{\rm fg}$}
\def \mzfg {z_{\rm fg}}
\def \mzlya {z_{\rm Ly\alpha}}
\def \zlya {$\mzlya$}
\def \mavgz{\langle z \rangle}
\def \avgz{$\mavgz$}
\def \wlya {$W_{\rm Ly\alpha}$}
\def \wsubj {$W_{\rm Ly\alpha}^{\rm line}$}
\def \mwsubj {W_{\rm Ly\alpha}^{\rm line}}
\def \mwlya {W_{\rm Ly\alpha}}
\def \mrphys {R_\perp}
\def \rphys {$\mrphys$}
\def \lbol {$L_{\rm Bol}$}
\def \guv {$g_{\rm UV}$}
\def \nhi  {$N_{\rm HI}$}
\def \mnhi  {N_{\rm HI}}
\def \kms  {\, km~s$^{-1}$}
\def \mkms  {{\rm km~s^{-1}}}
\def \lya  {Ly$\alpha$}
\def \mlya  {{\rm Ly\alpha}}
\def \lyb  {Ly$\beta$}
\def \hMpc      {h^{-1}{\rm\ Mpc}}
\def \msol      {M_\odot}
\def \msun      {\msol}
\def\cm#1{\, {\rm cm^{#1}}}
\def \cgsflux   {{\rm erg\ s^{-1}\ cm^{-2}}}
\def \cgssflux   {{\rm erg\ s^{-1}\ Hz^{-1} cm^{-2}}}
\def\sci#1{{\; \times \; 10^{#1}}}
\def \msnlya {{\rm S/N}_{\rm Ly\alpha}}
\def \snlya {$\msnlya$}
\newcommand{\civt}{\ion{C}{4}~1548}
\newcommand{\ciit}{\ion{C}{2}~1334}
\newcommand{\mgiit}{\ion{Mg}{2}~2796}
\newcommand{\nvt}{\ion{N}{5}~1238}
\newcommand\mwciv{W_{1548}}
\newcommand\wciv{$\mwciv$}
\newcommand\mwcii{W_{1334}}
\newcommand\wcii{$\mwcii$}
\newcommand\avgcii{$\left < \mwcii \right >$}
\newcommand\mwmgii{W_{2796}}
\newcommand\wmgii{$\mwmgii$}
\newcommand\mavmgs{\left < \mwmgii \right >^{\rm S}}
\newcommand\avmgs{$\mavmgs$}
\newcommand\mavmgw{\left < \mwmgii \right >^{\rm w}}
\newcommand\avmgw{$\mavmgw$}
\newcommand\mtavmg{\left < \mwmgii \right >}
\newcommand\tavmg{$\mtavmg$}
\newcommand\mwnv{W_{1238}}
\newcommand\wmwnv{$\mwnv$}
\def \mfc {f_C}
\def \fc {$\mfc$}
\def \mfcs {f_C}
\def \fcs {$\mfcs$}
\newcommand\mmetal{$M_{\rm metal}^{\rm cool}$}
\newcommand\mwlim{W^{\rm lim}}
\newcommand\wlim{$\mwlim$} 
\newcommand\mxiq{\xi_{\rm Q-Q}}
\newcommand\xiq{$\mxiq$}
\newcommand\mxic{\xi_{\rm CIV-Q}}
\newcommand\xic{$\mxic$}
\newcommand\mxia{\xi_{\rm CIV-CIV}}
\newcommand\xia{$\mxia$}
\newcommand\mxixi{(\bar\mxiq \bar\mxia)^{1/2}} 
\newcommand\xixi{$\mxixi$}
\newcommand\mlxmg{\ell(X)_{\rm MgII}}
\newcommand\lxmg{$\mlxmg$}
\newcommand\mlxciv{\ell(X)_{\rm CIV}}
\newcommand\mlciv{\ell(z)_{\rm CIV}}
\newcommand\lciv{$\mlciv$}
\def \mhMpc{h^{-1}{\rm\ Mpc}}
\def \hMpc{$h^{-1}{\rm\ Mpc}$}
\def \mrvir {r_{\rm vir}}
\def \rvir {$\mrvir$}
\def\ltk{\left [ \,}
\def\ltp{\left ( \,}
\def\ltb{\left \{ \,}
\def\rtk{\, \right  ] }
\def\rtp{\, \right  ) }
\def\rtb{\, \right \} }

\title{Quasars Probing Quasars VII. The Pinnacle of the Cool
  Circumgalactic Medium Surrounds 
  Massive $z \sim 2$ Galaxies}

\author{
J. Xavier Prochaska\altaffilmark{1,2},
Marie Wingyee Lau\altaffilmark{1},
Joseph F. Hennawi\altaffilmark{2}
}
\altaffiltext{1}{Department of Astronomy and Astrophysics, UCO/Lick
  Observatory, University of California, 1156 High Street, Santa Cruz,
  CA 95064}
\altaffiltext{2}{Max-Planck-Institut f\"ur Astronomie, K\"onigstuhl 17,
  D-69115 Heidelberg, Germany} 

\begin{abstract}
We survey the incidence and absorption strength of the metal-line
transitions \ciit\ and \civt\ from the circumgalactic medium (CGM)
surrounding $z \sim 2$ quasars, which act as signposts for massive
dark matter halos $M_{\rm halo} \approx 10^{12.5} \msun$.  On scales
of the virial radius ($\mrvir \approx 160$\,kpc), we measure a high
covering fraction $f_C = \vfct$ to strong \ciit\ absorption (rest
equivalent width $\mwcii \ge 0.2$\AA\ ), implying a massive reservoir
of cool ($T \sim 10^4$\,K) metal enriched gas.  We conservatively
estimate a metal mass exceeding $10^8 \msol$. 
We propose these metals trace enrichment of
the incipient intragroup/intracluster medium that these halos
eventually inhabit.  
This cool CGM around quasars is the pinnacle amongst
galaxies observed at all epochs, as regards covering fraction and
average equivalent width of \ion{H}{1} \lya\ and low-ion metal
absorption.  
We argue that the properties of this cool CGM primarily 
reflect the halo mass, and that other factors such as feedback,
star-formation rate, and accretion from the intergalactic medium are
secondary.  
We further estimate, that the CGM of massive, $z \sim 2$ galaxies
accounts for the majority of strong \ion{Mg}{2} absorption along
random quasar sightlines.  Lastly, we detect an excess of strong
\civt\ absorption ($\mwciv \ge 0.3$\AA) over random incidence to
1\,Mpc physical impact parameter and measure the quasar-\ion{C}{4}
cross-correlation function: $\mxic(r) = (r/r_0)^{-\gamma}$ with $r_0 =
\vro$ and $\gamma = \vgmm$.  Consistent with previous work on larger
scales, we infer that this highly ionized \ion{C}{4} gas traces
massive ($10^{12} \msun$) halos.
\end{abstract}


\section{Introduction}


The detection of a strong emission feature at 7\,keV from the hot plasma
surrounding the Perseus cluster revealed that its intracluster medium
(ICM) is enriched in heavy elements \citep{mitchell76}.
This marked the discovery that gas within dark matter halos was
enriched over the past 13\,Gyr. Modern X-ray experiments have 
extended the measurements to clusters at $z \gtrsim 1$ and have
recorded metallicities of $\approx 1/3$ solar across this $\approx
10$\,Gyr time period \citep[e.g.][]{baldi+12,andreon12}.
If one extrapolates these data to the virial radius,
the total metal mass is extraordinary -- $4\sci{10} \msun$ in Fe alone
\citep[e.g.][]{siv09}
-- and offers a challenging constraint to models of chemical enrichment
\citep[e.g.][]{zaritsky04,bonaparte+13}.
The tension is sufficiently strong that researchers continue to
debate/discuss
whether one must invoke non-standard initial mass functions
(IMFs) and/or stellar yields for galaxies in cluster environments
\citep[e.g.][]{portinari04,arrigoni10b,loewenstein13,yates14}.
Recently,
the latest generation of X-ray telescopes have extended such
studies to lower mass systems ($M \approx 10^{13} \msun; kT \sim
1$\,keV).  These `groups' also exhibit a highly enriched medium 
to at least \rvir/2, with a mass
in Fe {\it alone} that exceeds $10^9 \msun$ \citep{sms14}.

Despite the ongoing arguments over the IMF and yields, these 
enrichment models all agree that
the gas that forms the ICM must have been polluted at early times, when
massive galaxies form the majority of their stars.
In the models of \cite{portinari04}, for example, over $50\%$ of the Fe and O
is injected into the ICM after only a few Gyr.
With early enrichment of the ICM a uniform
prediction, one is well motivated to search for direct evidence 
of this activity.  The cleanest signature might be the
observation of a large mass in metals in the proximity of massive
high-$z$ galaxies, far beyond the stars that drive
the enrichment.  
Gas in this environment is commonly referred to as halo gas or
the circumgalactic medium (CGM).
Unfortunately, such an experiment
cannot be carried out through X-ray emission 
because of surface brightness dimming, 
the likelihood that the gas is too cool, and the redshifting of
the putative emission to energies $\lesssim 2$\,keV.
And while deep \ion{H}{1} \lya\ imaging 
has proven successful in discovering
diffuse and extended emission around galaxies \citep[e.g.][]{matsuda04,cantalupo14}, 
metals from such nebulae are rarely detected. Furthermore, estimating the gas
metallicity and metal mass would be very challenging \citep{fab14}.

For these reasons, one is compelled to perform the search in
absorption using background sources whose sightlines pass close to
massive, high-$z$ galaxies.  This experiment is also difficult to
perform, in particular to 
discover and characterize high-$z$ galaxies that lie foreground
to sufficiently bright, background sources. 
Steidel and collaborators have spent 100+ nights on the 10m
Keck telescopes to perform their Keck Baryonic Structure Survey
\citep[KBSS;][]{adel05,steidel+10,rakic12} which focuses on the
star-forming, Lyman break galaxies (LBGs) at $z \sim 2$.   
Analysis of the observed clustering of LBGs, however, implies dark
matter halo masses $M_{\rm halo} \lesssim 10^{12} \msol$
\citep{adel05,bielby+13}.  While relatively massive, the majority of
LBGs are not predicted to evolve into the massive galaxies
representative of clusters and massive groups \citep{conroy+08}.
Furthermore, despite the tremendous observational
investment, the sample of quasar sightlines probing the galactic halos of
LBGs remains modest \citep[$\sim 10$;][]{rudie12,turner14} 
and other groups have
provided only a few additional cases \citep{cwg+06,simcoe06,crighton+11}.
Nevertheless, the results to date do indicate a halo of enriched, cool gas
traced extending to an impact parameter $\mrphys \approx 100$\,kpc.  
\cite{crighton13} has studied one sightline in great
detail, finding a low metallicity for the less ionized material
and a near solar metallicity for a more highly ionized phase. 
Ongoing analysis will examine similar properties for the
full sample to characterize (with large sample variance) the
enrichment of this halo gas. 

For studying ICM enrichment, 
one would preferably perform a similar experiment with
a large sample of more massive galaxies, 
i.e.\ $M_{\rm halo} > 10^{12} \msol$ at $z=2$. 
This is a primary goal of our Quasar Probing 
Quasars\footnote{http://www.qpqsurvey.org} (QPQ) survey.
Over the past decade, we have discovered a large sample of close quasar
pairs through a dedicated, follow-up effort on candidates drawn from
the SDSS, BOSS, and 2dF surveys and confirmed with spectroscopy
from 4m-class telescopes including: the 3.5m telescope
at Apache Point Observatory (APO), the Mayall 4m telescope at Kitt
Peak National Observatory (KPNO), the Multiple Mirror 6.5m Telescope,
and the Calar Alto Observatory (CAHA) 3.5m telescope 
\citep{thesis,hso+06,hennawi10,bovy11,bovy12}. 
A large fraction of the confirmed pairs are physically unassociated, i.e.\ the quasars have
distinct redshifts indicating that they lie at cosmologically large
separations.  With this sample of projected pairs, 
we recognized that one could explore the CGM of massive galaxies at $z \sim 2$ using
the f/g quasars as ``signposts''. 
Assuming that the scaling
relations between massive black holes and galaxies holds at this
epoch, these active nuclei tag massive galaxies in the young universe.
Indeed, the galaxy masses inferred from the measured clustering of quasars
yields estimates for the dark matter halos $M_{\rm halo} \approx
10^{12.5} \msun$ \citep{white12,vikas+13}. 
In this manner, we established the 
QPQ survey, a dedicated analysis of the properties of gas surrounding
quasars on scales of a few tens kpc to $\sim 1$\,Mpc.  
This probes gas on scales within the virial radius of massive
galaxies ($\mrvir^{\rm QPQ} \approx 160$\,kpc) and beyond.
One predicts that the galaxies hosting quasars
evolve into the massive dark matter halos associated to groups and
clusters in the modern universe \citep[$M > 10^{14}
\msun$;][]{fanidakis13}. 
In these respects, the QPQ experiment addresses the enrichment and
heating of the incipient ICM and intragroup medium (IGrM).

Thus far, the QPQ survey
has focused primarily on \ion{H}{1} \lya\ absorption.
We recognized in our first analyses \citep[QPQ1,QPQ2:][]{qpq1,QPQ2}
that the incidence of \ion{H}{1} optically thick absorbers (aka Lyman limit
systems or LLS) about quasars was significantly
higher than random, and measured a cross-correlation 
clustering amplitude of $r_0 = 9.2 \mhMpc$ \citep{QPQ2}.  We then conducted
a detailed study of the physical properties of one of these LLS
\citep[QPQ3:][]{qpq3}, finding
a near solar metallicity, a significantly ionized medium, and extreme
kinematics.  In QPQ4 \citep{qpq4}, we searched sensitively for faint
\lya\ emission associated with these LLS and found no positive
detections, 
implying that the low-ionization material frequently detected in our b/g sightlines 
is very likely shadowed from the quasar radiation by the same
obscuring medium invoked in unified models of AGN
\citep[e.g.][]{Anton93}. 
The most recent publications 
studied further the CGM via \ion{H}{1} \lya\ absorption
\citep[QPQ5,QPQ6:][]{qpq5,qpq6}.
We confirmed the high incidence of optically thick gas, with an
excess above the random incidence
to at least 1\,Mpc (physical).  Similarly, the data exhibit
excess \ion{H}{1} \lya\ absorption on all scales within 1\,Mpc in the
transverse direction, with a stronger signal than 
complimentary measurements from any other galactic population.
This follows from the fact that $z \sim 2$ quasars are hosted by
massive galaxies and, apparently, their radiative emission does not
suppress cool gas absorption in the surrounding medium (transverse to
the sightline).
This large reservoir of \ion{H}{1} runs contrary, however, to the
predictions from zoom-in simulations of galaxy formation in massive 
dark matter halos even if one ignores quasar radiation \citep{fhp+14}.  
This suggests that the models lack key
aspects of the astrophysics that determine the properties of halo gas
\cite[see also QPQ6;][]{cantalupo14}.

It is possible that
the presence of a quasar within these massive galaxies may
complicate interpretation of the results.  Wherever such luminous
sources shine, they will over-ionize the medium to distances of $\gg
100$\,kpc \citep[e.g., QPQ2;][]{cmb+08}.  In \ion{H}{1} \lya\ absorption, one detects a
line-of-sight proximity effect to distances of approximately 15\,Mpc
\citep{bdo88,sbd+00}.  In the transverse direction, however, we and
others have found little evidence for over-ionization; studies
report the absence of a transverse proximity effect
\citep[TPE;][QPQ6]{Crotts89,moller92,Croft04}.
We have concluded that quasars emit their ionizing radiation
anisotropically (or on timescales of $t \ll 10^5$\,yr).
In this manuscript, we proceed under the expectation that the quasar
has little radiative impact on the majority of gas surrounding it.  A
major motivation of the QPQ survey, however, is to test this
hypothesis and we search for evidence here of a significant transverse
proximity effect.

We acknowledge that 
the presence of a luminous quasar undoubtedly represents a
special epoch (likely repeated) in the life-cycle of massive
galaxies.  
Even allowing for anisotropic emission, the number density of luminous
quasars is at least an order of magnitude smaller than the number
density of massive halos ($> 10^{12} \msun$).
Nevertheless,
studies of the galaxies hosting quasars
indicate that these systems have rather standard characteristics.  
Estimations of their star formation rate (SFRs) from sub-mm
observations, for example, 
show that the galaxies in a quasar phase lie along the so-called `main
sequence' of star formation \citep{mainieri11,rtl+13}.  
Searches for signatures that the quasar phase is
triggered by an ongoing or recent galaxy-galaxy merger generally yield
null results \citep[e.g.][but see also Canalizo et al.\ 2007; Bennert
et al.\ 2008]{dunlop03,veilleux09}.
Lastly, searches for CO emission from the galaxies hosting quasars do
not find extreme gas masses  nor unusual kinematics \citep{coppin08,simpson12}. 
The growing consensus is that quasar activity does not imply special
conditions on galactic scales\footnote{Note, however, that AGN
  discovered in galaxies selected for extreme properties \citep[e.g.\ bright
  sub-mm emission;][]{wang08} are obvious exceptions.}.

This manuscript expands greatly on our study of metal-line absorption in the
environments of $z \sim 2$ massive galaxies, as first presented in QPQ3 and QPQ5. 
We focus primarily on the transitions of \ciit\ and the 
\ion{C}{4}~$\lambda\lambda 1548,1550$ doublet, which are frequently
observed in the ISM and IGM and whose larger rest-wavelengths often place them 
outside the \lya\ forest.
These ions also span a wide range of ionization conditions, probing
gas that is predominantly neutral to highly
ionized material.  Our analysis presents an assessment of the gas enrichment to
1\,Mpc, i.e.\ scales characteristics of the ICM/IGrM.  These data also 
provide an assessment of the ionization state, albeit limited by the
low-resolution of the spectra which preclude precise estimates of the
column densities.
An analysis of the smaller set of high-resolution spectra that provide
ionic column density measurements will be the focus
of a future work (QPQ8: Lau et al., in prep.).

Another motivation for studying the CGM of massive, $z \sim 2$
galaxies is to explore the origin of halo gas in individual galaxies.  
Given the positive detections of enriched halo gas in the ICM and IGrM, 
it is reasonable to expect that the dark matter
halos hosting individual galaxies like the Milky Way also contain a hot and
enriched plasma.  
Unfortunately, this medium is either too cool or has
too low mass to permit regular detection with X-rays, even using modern technology
and techniques \citep[e.g.][]{anderson13}. 
Instead, 
scientists have demonstrated that these lower mass halos 
do harbor a substantial reservoir of heavy elements by linking
metal absorption systems discovered in quasar spectra to galaxy halos
\citep{bergeron86}. 
Initially traced by the \ion{Mg}{2} doublet and now studied with other
lines, this halo gas has a much lower
temperature and ionization state than the hot ICM discovered in
X-rays.  This `cool CGM'
was first predicted by \cite{bs69} and decades of absorption-line research have
since statistically mapped the incidence and average surface density of
ions ranging from Mg$^+$ to O$^{+5}$, in galaxies large and small 
\citep[e.g.][]{chg+10,pwc+11,ttw+11,stocke13,werk+13}.
The growing consensus from these programs is that halo gas has a metal
mass of at least $10^7 \msun$ \citep{werk+14}, 
exceeding the mass in metals of the ISM and possibly even that in stars
\citep{peeples+14}.   

With the basic observables of the $z \sim 0$, metal-enriched CGM nearly
well-defined, researchers are now striving to understand its origin and 
to resolve the physical phenomena that shape
it across cosmic time.  First and foremost, one requires a source of
metals, i.e.\ stars, and we presume that these form and die primarily
within galaxies \citep[but see][]{zaritsky04}.  
Under this expectation, one must identify a mechanism to
transport the metals to large ($> 100$\,kpc) separations from
the stellar systems.  Theorists have proposed a variety
of such processes, e.g., galactic scale winds, AGN feedback, cosmic ray
pressure, tidal stripping. 
But while most of these have been `caught in the act' driving
gas/metals from galaxies \citep[e.g.][]{veilleux13,ebeling14,rubin+14}, their
relative importance and impact over the lifetime of a galaxy 
are poorly constrained.  In part, this
results from the great difficulty in modeling such processes.
The time-scales and large dynamic range required to capture the
astrophysics of feedback
remain an intractable computational problem, even in idealized
simulations \citep[e.g.][]{creasey13}.  As such, numerical simulations
inevitably resort to sub-grid prescriptions to model such
processes. 
In this manuscript, we draw comparisons between the cool CGM of
massive, $z \sim 2$ galaxies and that for galactic populations with a
diverse range of masses, ages, and SF histories. 
These comparisons offer insight into the origin and evolution of cool gas around
galaxies. 

This manuscript is summarized as follows.  We present a summary of the
data sample in Section~\ref{sec:data} and the equivalent width
measurements in Section~\ref{sec:ew}.
The primary observational results are given in
Section~\ref{sec:results} and these are discussed in
Section~\ref{sec:discuss}.   A summary and concluding remarks
completes the main body of the manuscript (Section~\ref{sec:summary}).
An Appendix provides our own survey for strong \civt\ absorption in
random sightlines and details on semi-empirical models for the cool,
enriched gas surrounding LRGs.
Throughout this manuscript, we adopt a $\Lambda$CDM cosmology with
$\Omega_M = 0.26, \Omega_\Lambda = 0.74$, and $H_0 = 70 {\rm km \,
  s^{-1} \, Mpc^{-1}}$.  
When referring to comoving distances we include 
explicitly an $h_{70}^{-1}$ term and follow modern convention of scaling
to a Hubble constant of 70\kms\,Mpc$^{-1}$.  
All equivalent width measurements are
presented in the rest-frame, unless otherwise specified.


\begin{deluxetable*}{lcccccccccc}
\tablewidth{0pc}
\tablecaption{QPQ7 Spectra\label{tab:qpq7_spectra}}
\tabletypesize{\footnotesize}
\tablehead{\colhead{b/g Quasar} &
\colhead{f/g Quasar} & \colhead{$\mzfg$} & 
\colhead{\rphys} &
\colhead{Instr.} & \colhead{$\lambda_r^a$} 
\\
&&&(kpc)&&(\AA)
}
\startdata
J000216.66$-05$3007.6&J000211.76$-05$2908.4&2.8190&768&SDSS&1093-1738&\\
J000432.76$+00$5612.5&J000426.43$+00$5703.5&2.8123&882&BOSS&1049-1638&\\
J000531.32$+00$0838.9&J000536.29$+00$0922.7&2.5224&725&BOSS&1099-1747&\\
J000633.35$-00$1453.3&J000629.92$-00$1559.1&2.3327&711&BOSS&1146-1638&\\
J000838.30$-00$5156.7&J000839.31$-00$5336.7&2.6271&841&BOSS&1023-1715&\\
J001025.73$-00$5155.3&J001028.78$-00$5155.7&2.4268&387&BOSS&1097-1728&\\
J001634.35$-01$0622.7&J001631.71$-01$0426.9&3.5359&932&BOSS&834-1633&\\
J001641.17$+01$0045.2&J001637.10$+00$5936.7&2.6690&760&BOSS&1044-1763&\\
J002123.80$-02$5210.9&J002126.10$-02$5222.0&2.6912&299&BOSS&1059-1806&\\
J002610.92$-01$0213.0&J002613.63$-01$0132.2&2.3260&491&BOSS&1178-2798&\\
J002802.60$-10$4936.0&J002801.18$-10$4933.9&2.6591&175&Keck/ESI&1200-2756&\\
J003138.17$+00$3333.2&J003135.57$+00$3421.2&2.2328&529&BOSS&1135-2826&\\
J003423.40$-10$4956.3&J003423.06$-10$5002.0&1.8364& 67&Keck/LRISb-1200/3400&1124-1365&\\
&&&&Keck/LRISr-1200/7500&1694-2084&\\
&&&&Keck/LRISr-600/10000&2331-3188&\\
J003922.69$+00$2642.9&J003924.54$+00$2501.5&3.2980&819&BOSS&1068-1767&\\
J004603.29$+00$0729.4&J004600.47$+00$0543.6&2.4526&960&BOSS&1066-1643&\\
J004732.73$+00$2111.3&J004738.19$+00$1955.3&2.4826&940&BOSS&1031-2805&\\
J010130.26$+03$1822.0&J010134.93$+03$1701.4&2.2718&912&BOSS&1097-2820&\\
J010900.86$+00$0137.1&J010857.21$+00$0154.5&2.4169&486&BOSS&1147-2832&\\
J011138.20$+14$0414.9&J011144.06$+14$0401.6&2.0019&749&Keck/LRISb-1200/3400&1038-1345&\\
&&&&Keck/LRISr-1200/7500&1564-1569&\\
&&&&Keck/LRISr-600/10000&2771-2848&\\
J011149.39$+14$0215.7&J011150.06$+14$0141.3&2.4732&301&Keck/LRISb-1200/3400&950-1244&\\
&&&&Keck/LRISr-1200/7500&1548-1561&\\
&&&&Keck/LRISr-600/10000&2777-2834&\\
\enddata
\tablenotetext{a}{Rest-frame wavelengths at the f/g quasar redshift where the spectra of the b/g quasar exceeds $S/N > 5$ per pixel.}
\tablecomments{[The complete version of this table is in the electronic edition of the Journal.  The printed edition contains only a sample.]}
\end{deluxetable*}

\section{Data Sample}
\label{sec:data}

The sample of quasar pairs analyzed here is a subset of the sample
studied in QPQ6 for \ion{H}{1} \lya\ absorption.  Specifically, we
have restricted the current study to those pairs where the
signal-to-noise (S/N) ratio at \ion{H}{1} \lya\ exceeds 9.5 per
rest-frame \AA.  This facilitates a more precise
evaluation\footnote{In fact, this was the cutoff adopted in QPQ6 for
  the analysis of individual systems.} 
of \ion{H}{1}~\lya\ and generally insures sufficient S/N redward of \lya\
for the metal-line analysis.
Quasar emission redshifts are taken directly from QPQ6, following the
methodology described in that manuscript.
Briefly, we adopt a custom line-centering algorithm to centroid one or
more far-UV emission lines and adopt the analysis of \cite{shen07} to
combine these measurements and assess systematic uncertainty in the
final value.
The median emission redshift of the \npair\ pairs is $\mzem^{\rm
  median} = 2.35$ and the median uncertainty in the redshift
measurements is $\approx 520 \mkms$.
The impact parameters range from $\mrphys = 39$\,kpc to 1\,Mpc, with
52~pairs having $\mrphys < 200$\,kpc.  

Generally, we have analyzed the same spectrum of the b/g quasar used
to measure \ion{H}{1} \lya, as listed in Table~4 of QPQ6.  The 
exceptions are primarily for data obtained with Keck/LRIS and Magellan/MIKE
which employ dual cameras and a beam-splitting dichroic.  For those
pairs, key metal-line transitions (e.g.\ the \ion{C}{4} doublet) tend
to occur in spectra taken with the red camera.  For MIKE, the spectral
resolution and S/N of the data from the red camera
are comparable to the data from the blue camera.
For LRIS, we employed a number of gratings with the red camera.  
In approximately half of the pairs observed with Keck/LRIS 
we employed only the 300/7500 grating, which we consider too low of
resolution to analyze metal-line transitions, or the 
wavelength coverage is uninformative with the chosen grating.  
We supplement the analysis
of these pairs, when possible, with BOSS or SDSS spectra
\citep{sdssdr7,boss_dr9}.  

For completeness, Table~\ref{tab:qpq7_spectra} lists all of the spectra
analyzed in this study.
In cases where multiple spectra covered the same transition, we gave
preference to the higher spectral resolution unless the S/N was very
poor.
Details related to data reduction and calibration of the 1D spectra
are provided in QPQ6.
As a starting point, we employed the continua generated in QPQ6 for
the \ion{H}{1} analysis following standard principal component
analysis (PCA) techniques \citep{lee+12,lee+13}.

\begin{deluxetable*}{lccccccccccccccc}
\tablewidth{0pc}
\tablecaption{QPQ7 EW Measurements\label{tab:qpq7_EW}}
\tabletypesize{\scriptsize}
\tablehead{\colhead{f/g Quasar} &
\colhead{b/g Quasar} & \colhead{\rphys} &
\colhead{\zfg} &
\colhead{$\delta v^a$}
& \colhead{$W_{1334}$}
& \colhead{$\sigma(W_{1334})$}
& \colhead{$W_{1548}$}
& \colhead{$\sigma(W_{1548})$}
\\
&& (kpc) &&& (\AA) & (\AA) & (\AA) & (\AA)}
\startdata
J000211.76$-05$2908.4&J000216.66$-05$3007.6&768&2.819& -232& 0.001&0.035& 0.044&0.066\\
J000426.43$+00$5703.5&J000432.76$+00$5612.5&882&2.812& -429& 0.051&0.132&&\\
J000536.29$+00$0922.7&J000531.32$+00$0838.9&725&2.522&  317& 0.160&0.084& 0.113&0.262\\
J000629.92$-00$1559.1&J000633.35$-00$1453.3&711&2.333&   40&&& 0.084&0.268\\
J000839.31$-00$5336.7&J000838.30$-00$5156.7&841&2.627&  551& 0.069&0.054&&\\
J001028.78$-00$5155.7&J001025.73$-00$5155.3&387&2.427& -797&&& 0.049&0.136\\
J001631.71$-01$0426.9&J001634.35$-01$0622.7&932&3.536&  578&-0.048&0.144& 0.089&0.094\\
J001637.10$+00$5936.7&J001641.17$+01$0045.2&760&2.669& -925&&& 0.368&0.113\\
J002126.10$-02$5222.0&J002123.80$-02$5210.9&299&2.691&  841&&& 0.351&0.282\\
J002613.63$-01$0132.2&J002610.92$-01$0213.0&491&2.326& -239&&& 1.956&0.140\\
\enddata
\tablenotetext{a}{Velocity offset from \zlya\ to \zfg.  Note that absorption lines were assessed only for data with $\msnlya>9.5$.}
\tablecomments{[The complete version of this table is in the electronic edition of the Journal.  The printed edition contains only a sample.]}
\end{deluxetable*}

\section{Equivalent Width Measurements}
\label{sec:ew}

The heart of this analysis is the search for and measurement of
metal-line transitions at $z \approx \mzfg$ in each quasar pair.
Owing to the substantial uncertainty in quasar emission redshifts
($\approx 500 \mkms$), we
performed the search in 3000\kms\ velocity windows centered on \zfg.
The strongest transitions commonly detected in quasar spectra (e.g.\
\ion{Si}{2}~1260, \ciit, and the \ion{Si}{4}, \ion{C}{4} doublets)
were explored first, restricting to spectral regions outside the \lya\
forest.  This effort was first completed in support of the QPQ6
analysis, in particular to assist the assessment of whether the
associated \ion{H}{1} gas was optically thick.  
For this study, we have re-inspected all of the
data and have performed new analysis as warranted. 

In cases where one or more metal-line transitions were 
identified, we defined a narrower velocity region encompassing the
absorption (typically $100-200 \mkms$)  
for subsequent equivalent width measurements.
As the default, one
window was adopted for all metal-line transitions of a given quasar
pair.
In a small fraction of the cases, absorption from unrelated systems
(typically \ion{Mg}{2} or \ion{C}{4} doublets) lay coincident to the
key transitions for the pair analysis.  If the blending was modest,
then these spectral regions were simply excluded in the equivalent
width integrations.  If the blending was severe, we excluded the
transition from further analysis.
If no strong absorption was evident in the search window
for any of the metal-line transitions (corresponding to an $\approx
3\sigma$ detection threshold), we adopted an analysis window
roughly matching that used in the \ion{H}{1} \lya\ analysis of QPQ6
(see their Table~4).
This generally corresponds to the absorption profile of the strongest
\lya\ line within the 3000\kms\ window centered on \zfg.
During the search, the local continuum was frequently adjusted to
more accurately normalize the b/g quasar flux.

The equivalent width measurements $W_\lambda$
were derived from simple boxcar
integrations within the analysis regions and uncertainties were
estimated from standard error propagation using the $1\sigma$
uncertainties in the normalized flux.  These do not include
uncertainties related to continuum placement, which contribute $\approx
0.05$\AA\ of systematic error.  Because the majority of our spectra
have modest resolution ($R \sim 2000$), the metal-line transitions are
typically unresolved and a simple Gaussian fit to the observed
profiles would yield similar values for the equivalent widths.

The rest-frame equivalent width measurements and errors for \ciit\ and \civt\ (\wcii,\wciv)
are presented in Table~\ref{tab:qpq7_EW}. 
Positive detections are considered those with $W > 3\sigma$ and
non-detections are reported as $2\sigma$ limits or the measured value
if $2\sigma < W < 3\sigma$.
Our analysis focuses on these two transitions because they span a wide
range of ionization states (the ionization potentials to produce
these ions are 11.3\,eV and 47.9\,eV respectively)
and are two of the strongest lines commonly
available in our sample.
Note that for the strongest \ion{C}{4} systems ($\mwciv > 1$\AA), the
\civt\ transition is frequently blended with the \ion{C}{4}~1550
transition and one may adopt an additional 15\%\ error in the \wciv\
measurement (not included explicitly in our analysis).  
Similarly, there may be weak and unresolved 
\ion{C}{2}$^*$~1335 absorption present (e.g.\ QPQ3) that contributes 
to the \wcii\ measurements. 

\begin{figure*}
\includegraphics[scale=0.65,angle=90]{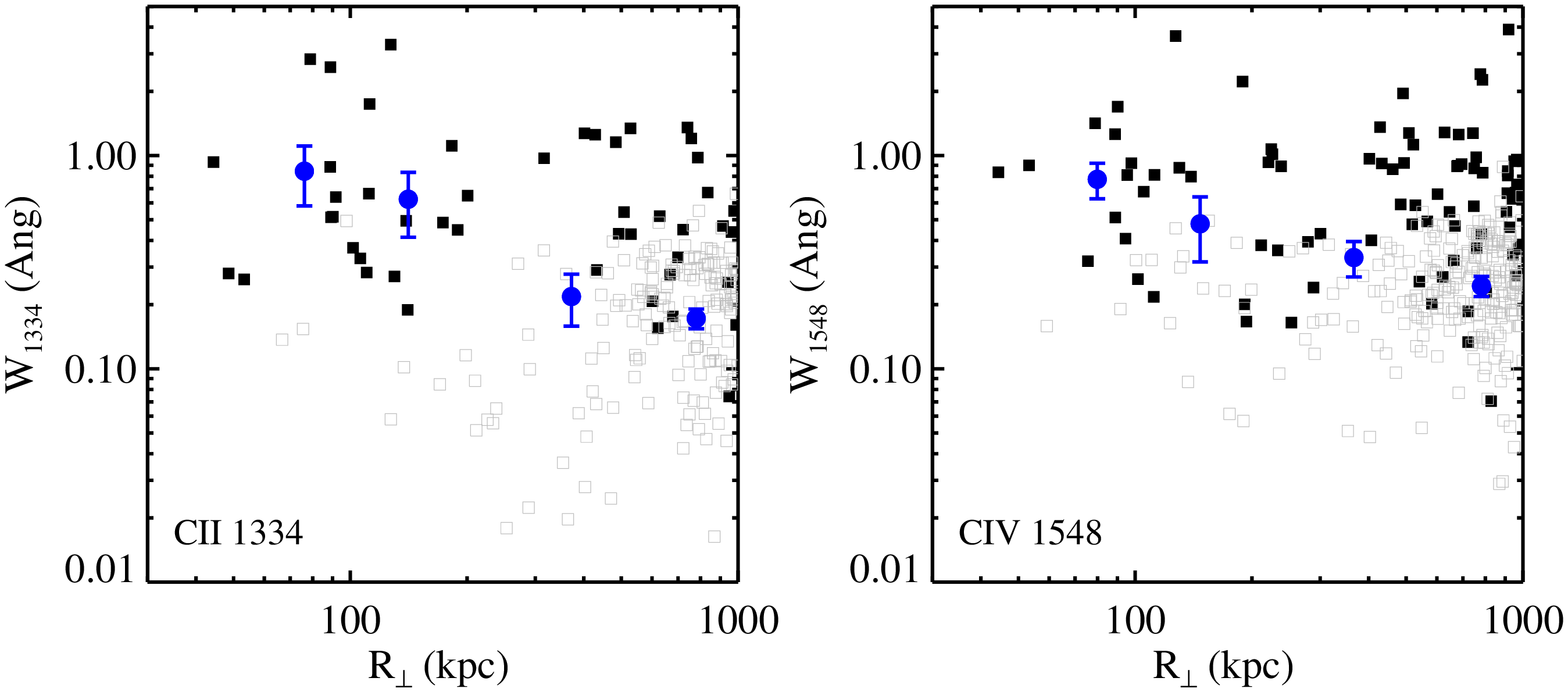}
\caption{Rest-frame equivalent widths for \ciit\
  (left) and \civt\ (right) measured at $z \approx \mzfg$ of the f/g
  quasar in our quasar pair sample, restricted to the analysis of
  spectra outside the \lya\ forest of the b/g quasar.
  These are plotted against the projected pair separations \rphys,
  measured at \zfg.
  Detections ($3\sigma$) are the filled black symbols and
  non-detections are open gray squares, plotted at their
  $2\sigma$ values.
  The blue symbols express the average $W_\lambda$ values and RMS
  scatter, taking non-detections at their measured values (i.e.\
  column 7 in Table~\ref{tab:qpq7_EW}) in bins of 
  $\mrphys = [0,100], [100,200], [200,500], [500,1000]$\,kpc.  
  Both transitions exhibit large equivalent widths ($W_\lambda
  \approx 1$\AA) at the smallest impact parameters. Both
  distributions also exhibit an anti-correlation with 
  \rphys\ (at $>99.99\%$ confidence).
  We conclude that $z \sim 2$ quasars are enveloped in a highly
  enriched, circumgalactic medium. 
}
\label{fig:ewscatt}
\end{figure*}

Figure~\ref{fig:ewscatt} presents scatter plots of the \wcii\ and
\wciv\ measurements
against projected quasar pair separation \rphys,
calculated at \zfg. 
Filled symbols indicate positive, $3 \sigma$ detections.  The number
of detections per logarithmic \rphys\ interval is roughly constant to 1\,Mpc.  The number of
pairs, however, increases rapidly with increasing \rphys\ indicating a declining
detection rate.  Overplotted on the figure are the unweighted
mean\footnote{ 
We also considered the mean weighted by the square inverse of the
error in the equivalent width measurements.  This gives qualitatively
similar results but smaller values and uncertainties that do not
accurately reflect the sample variance.}
and RMS of
the equivalent widths in bins of \rphys\ (see also
Table~\ref{tab:avgew}).  
These were calculated from the actual measurements, not the limits
(e.g.\ negative EWs were permitted and included).
Both transitions
exhibit a monotonic decline in the average EW with increasing
\rphys, although the averages still exceed 0.1\AA\ even
at $\mrphys \sim 1$\,Mpc.
This is partly due to the few positive detections with $W >
0.5$\AA\ but even after eliminating those 
measurements the average
EWs exceed\footnote{ 
We were concerned that these averages were biased high by errors in
continuum placement, but see the analysis of stacked profiles below.}
0.1\AA\ at all \rphys. 
Of course, a non-zero mean is expected if not required, especially
given the large analysis windows adopted from our
analysis.  
This occurs because of the high incidence of intervening systems
(especially for \ion{C}{4}) and the coincidence of absorption-lines
from unrelated systems (an effect we minimized by excluding
strongly-blended systems).
That is, even `random' spectral regions should exhibit a
non-zero mean equivalent width.  In later sections, we examine whether
the quasar pair measurements exceed such random expectation.

\begin{figure*}
\includegraphics[scale=0.65,angle=90]{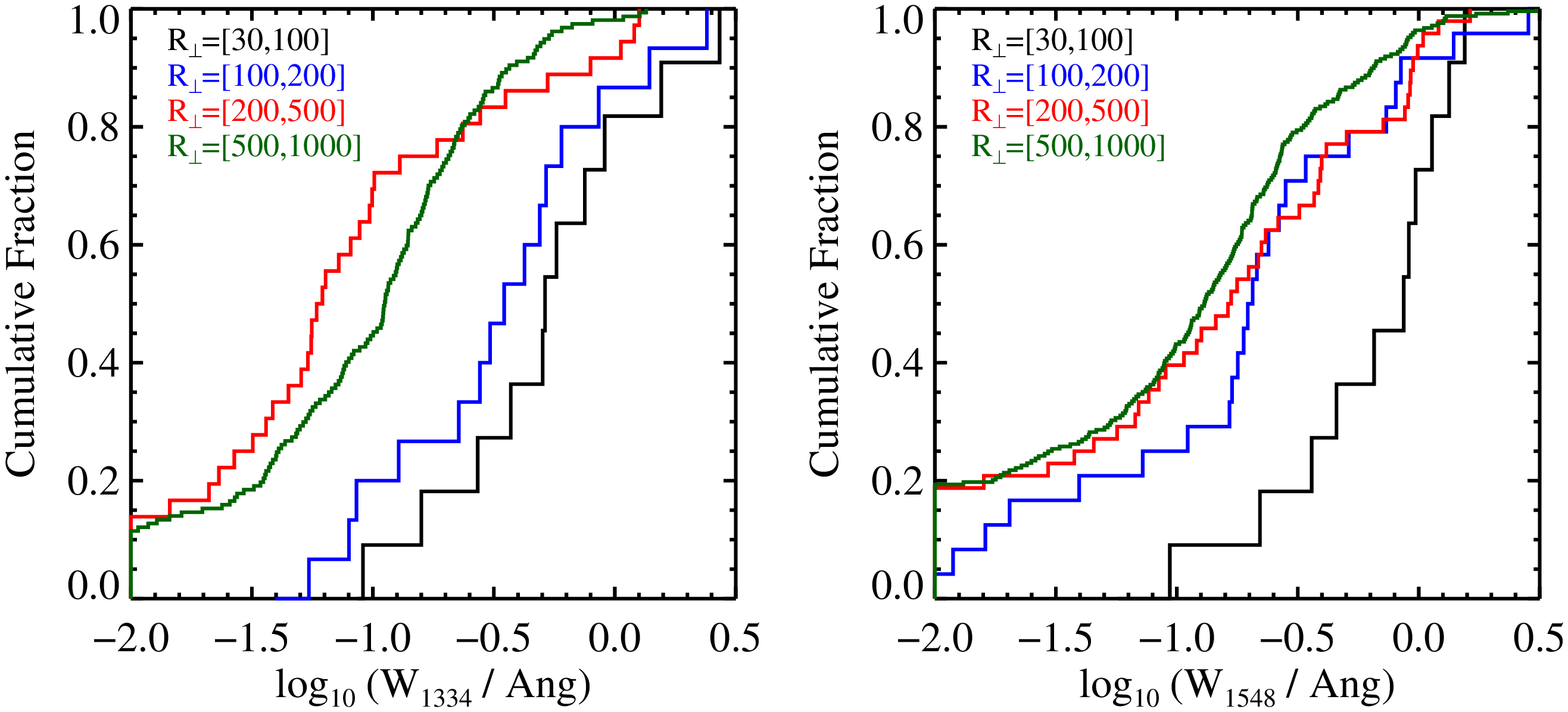}
\caption{Cumulative distributions of the rest-frame
  equivalent width measurements
  for \ciit\ (left) and \civt\ (right) in bins of \rphys.  
  For \ciit, we find much higher \wcii\ values for $\mrphys <
  200$\,kpc with little dependence below this impact parameter.
  This implies a cool and enriched CGM on scales characteristic of the
  estimated virial radius $\mrvir^{\rm QPQ} \approx 160$\,kpc.  In
  contrast, the \civt\ equivalent
  widths are lower at $\mrphys > 100$\,kpc but then decline less rapidly to
  larger \rphys.  This implies that beyond \rvir, the gas remains
  enriched on average but becomes more highly ionized.
  Note that the distributions begin at non-zero values owing to the
  sensitivity limit of the spectra.
}
\label{fig:ew_cumul}
\end{figure*}

Statistically, a generalized Kendall's tau test, which accounts for
upper limits in the EW measurements, gives a $>99.99\%$ probability 
that the null hypothesis of no correlation 
between \wcii\ (or \wciv) and \rphys\ is ruled out.
We conclude that the observed absorption is at least statistically (if
not physically) associated to the f/g quasar environment.
We further explore evolution in the metal-line absorption by comparing
the combined distribution of measurements and upper limits in bins of
\rphys\ (Figure~\ref{fig:ew_cumul}). 
Regarding \ciit, the \wcii\ distributions are nearly identical for the two
bins spanning $\mrphys = 0 - 200$\,kpc revealing strong, low-ion
absorption even beyond the estimated virial radii of the dark matter
halos hosting $z \sim 2$ quasars ($\mrvir^{\rm QPQ} = 160$\,kpc; see the
Introduction).
Beyond 200\,kpc, the incidence of strong \ciit\ absorption drops
sharply, even suddenly.  Furthermore, the two outer bins show very
similar distributions (at $\mwcii \approx 0.1$\AA)
with the difference 
driven primarily by the difference in data quality and therefore the
upper limit values for the non-detections (i.e., a comparison is fair
only for EW values above the detection limit).
This dominates the behavior in the \wcii\ distribution at low values.
We conclude that the CGM of galaxies hosting $z \sim 2$ quasars is
significantly enriched and also sufficiently self-shielded from the
EUVB to show substantial \ion{C}{2} absorption \citep[e.g.][]{werk+13}. 
We further conclude that this cool
and enriched medium extends to $\mrphys \approx 200$\,kpc and beyond
that separation there is a significant
decrease in enrichment, a sharp decline in the total gas surface
density, and/or an increase in the ionization
level of the gas.

At the smallest \rphys\ (less than 100\,kpc), the results are similar
for \civt, i.e.\ there is a preponderance of strong absorption
($\mwciv \gtrsim 1$\AA).  
The incidence modestly
decreases for $\mrphys = [100,200]$\,kpc and also as one extends 
to $\mrphys = [200,500]$\,kpc (again, the
differences at $\mwciv < 0.1$\AA\ are driven by differences in the
spectral S/N). 
We stress, however, that
there is no statistically significant difference in the distribution for
these two intervals.  Comparing to the \ciit\ measurements, the results
indicate that the gas in the quasar environments 
extending to $\mrphys \approx 500$\,kpc
remains substantially enriched in carbon but that the average
ionization state is significantly higher in the outer regions.  Of
course, this material is
unlikely to be physically bound to the quasar host, or to have originated
from within it, but may be associated to neighboring galaxies in the
extended environment.  We
explore this result further in the following sections. 

\begin{figure}
\includegraphics[width=3.7in]{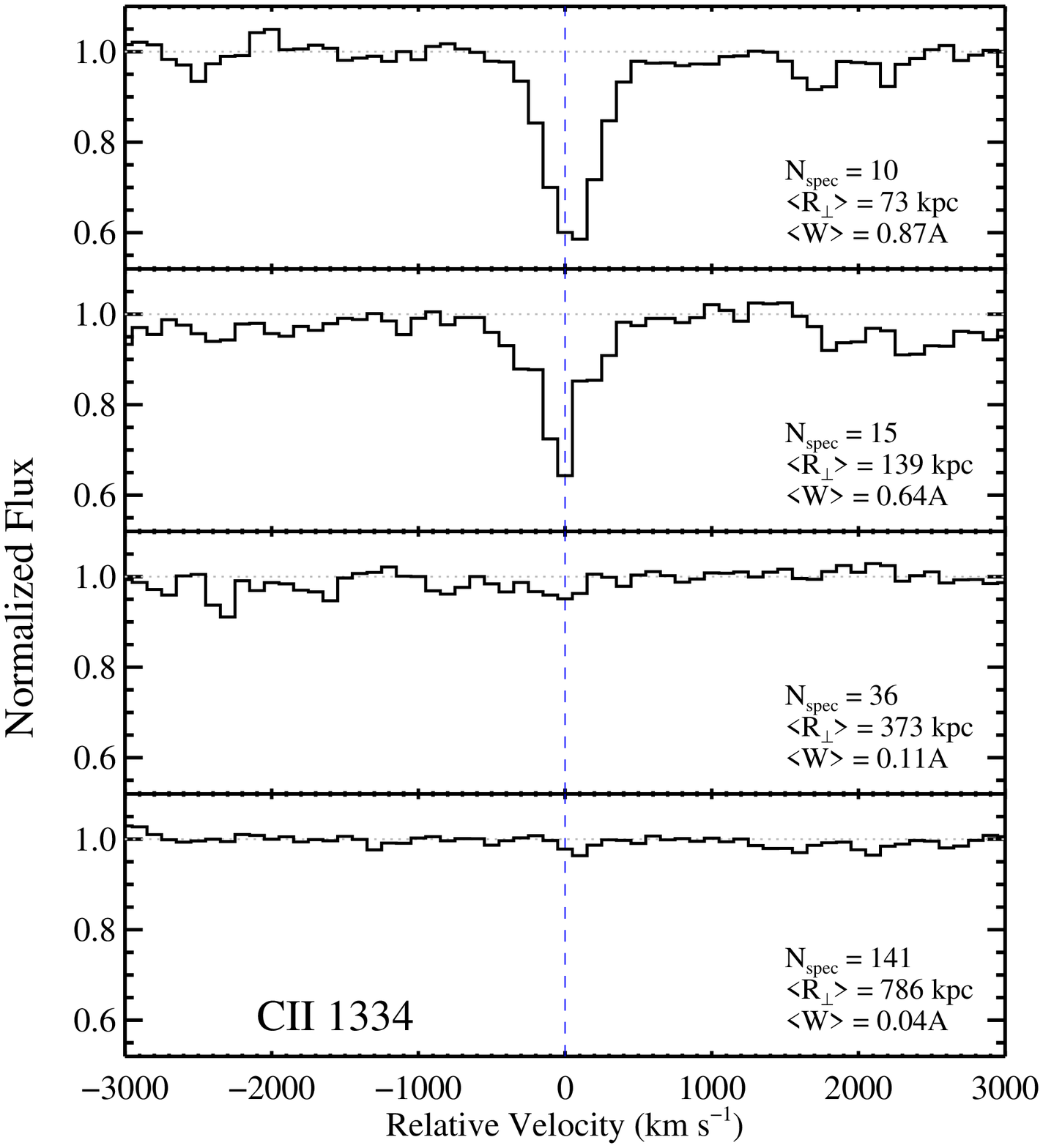}
\caption{Average absorption-line spectra for \ciit\ in bins of 
  $\mrphys = [0,100], [100,200], [200,500], [500,1000]$\,kpc, stacked
  at $v=0\,\mkms$ corresponding to the strongest \lya\ absorber within
  1500\kms\ of \zfg\ (\zlya; defined in QPQ6).  
  One observes large, average equivalent widths at $\mrphys <
  200$\,kpc and the disappearance of \ion{C}{2} 
  absorption at larger separations. Note that the strong
  absorption ($\mwcii \gtrsim 0.5$\AA)
  is not driven by a small set of events;  a median stack
  yields qualitatively similar results.  The equivalent widths
  reported were measured in $\delta v = \pm 800\,\mkms$ windows.  The
  uncertainty is dominated by sample variance in the first two bins (we
  estimate a 20\%\ error) and by a combination of statistical and
  systematic error (continuum fitting) for the larger \rphys\
  intervals.  We estimate 50m\AA\ and 30m\AA\ for the
  highest \rphys\ intervals.
}
\label{fig:CII_stack}
\end{figure}

\begin{figure}
\includegraphics[width=3.7in]{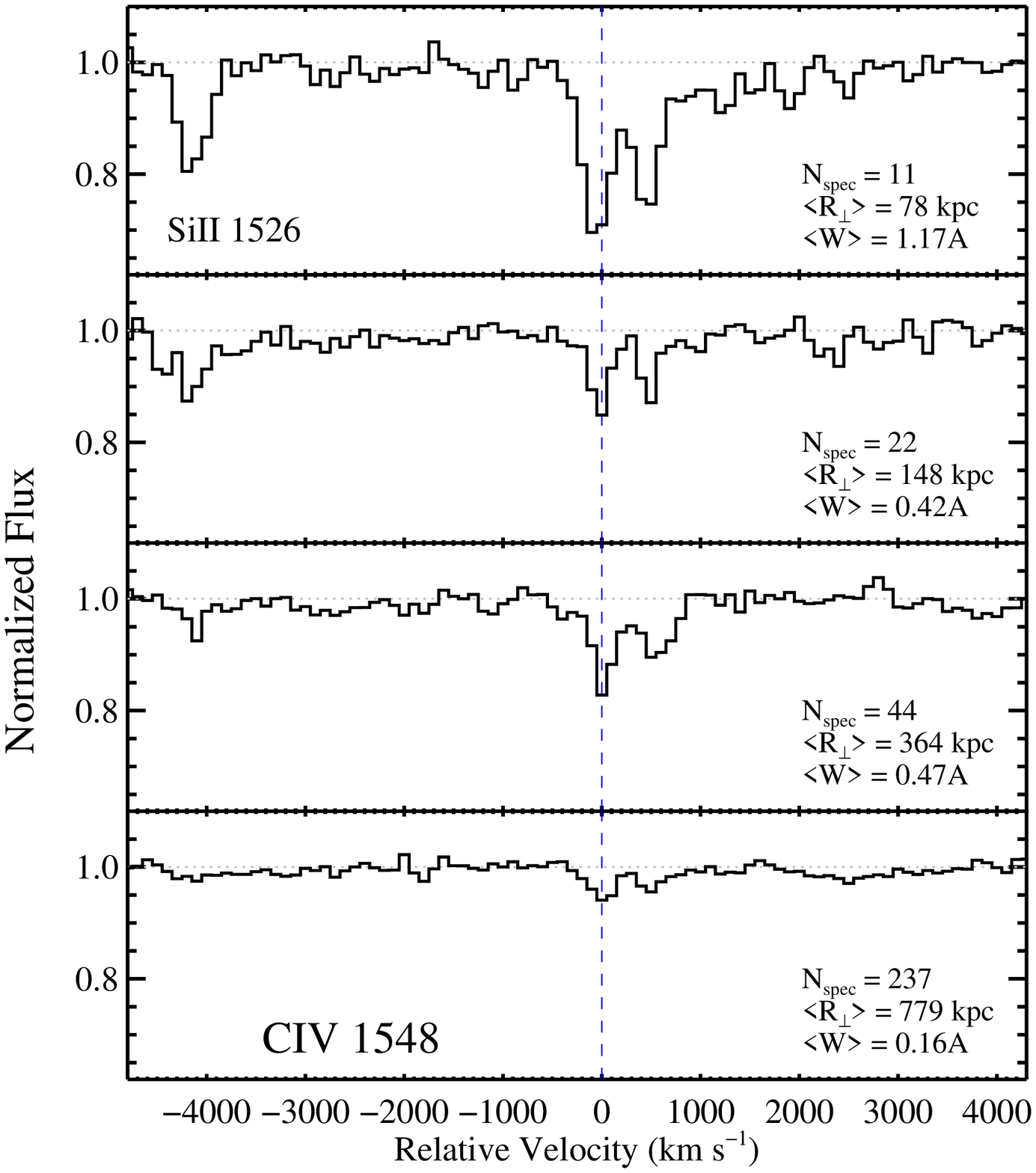}
\caption{Same as for Figure~\ref{fig:CII_stack} but for \civt.  
  In this case one also observes absorption from \ion{C}{4}~1550 at
  $v \approx +500 \mkms$ and \ion{Si}{2}~1526 at $v \approx
  -4200\mkms$.  In contrast to \ciit, the average \civt\ absorption
  remains substantial to $\mrphys \approx 500$\,kpc.
}
\label{fig:CIV_stack}
\end{figure}

The trends described above may be further illustrated by constructing
average (i.e.\ stacked) profiles in bins of \rphys.
Figures~\ref{fig:CII_stack} and \ref{fig:CIV_stack} show the stacked
spectra for \ciit\ and \civt\ with $v = 0 \, \mkms$ corresponding to $z =
\mzlya$, the approximate redshift centroid of the strongest \ion{H}{1}
\lya\ absorption within a 3000\kms\ window centered on \zfg\ (QPQ6).  
To generate the stacks, we rebinned each original spectrum onto a grid
centered on $v = 0\, \mkms$ with a dispersion of $\delta v = 100\, \mkms$.
The average was taken with uniform weighting of the pairs.
For this analysis, we did not include transitions flagged as having
blends with other, coincident absorption lines.
Lastly, a low-amplitude (i.e.\ near unit value), 
low-order polynomial was used to normalize the resultant stacks which
we associate to small errors in the continuum normalization.

\begin{deluxetable*}{lccccccccccc}
\tablewidth{0pc}
\tablecaption{QPQ7 Average Equivalent Width Values\label{tab:avgew}}
\tabletypesize{\scriptsize}
\tablehead{\colhead{$\mrphys^{\rm min}$} & 
\colhead{$\mrphys^{\rm max}$} 
&
\colhead{$m_{\rm pair}^a$} & 
\colhead{$\langle \mrphys \rangle$} & 
\colhead{$W_{1334}$} & 
\colhead{$\sigma(W)^b$} 
&
\colhead{$m_{\rm pair}^a$} & 
\colhead{$\langle \mrphys \rangle$} & 
\colhead{$W_{1548}$} & 
\colhead{$\sigma(W)^b$} 
\\
(kpc) & (kpc) & & (kpc) & (\AA) & (\AA) && (kpc) & (\AA) & (\AA)}
\startdata
  30& 100&  12& 76&0.84&0.26&  12& 80&0.79&0.15\\
 100& 200&  16&141&0.63&0.21&  25&147&0.48&0.16\\
 200& 500&  37&372&0.21&0.06&  49&367&0.33&0.06\\
 500&1000& 158&780&0.17&0.02& 249&781&0.24&0.03\\
\enddata
\tablecomments{Equivalent width values are rest-frame measurements.}
\tablenotetext{a}{Number of pairs analyzed.}
\tablenotetext{b}{Calculated from the dispersion of the individual measurements.  We also estimate an additional systematic uncertainty of 20\%\ from continuum placement.}
\end{deluxetable*}
 
\begin{figure*}
\includegraphics[scale=0.65,angle=90]{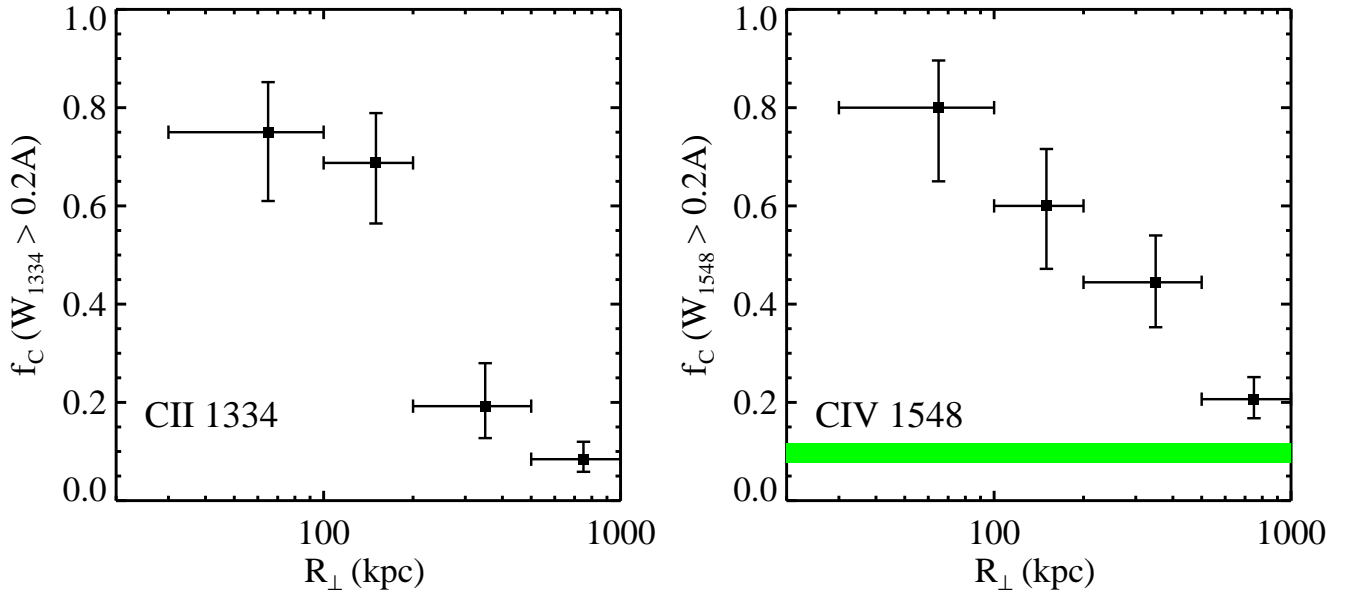}
\caption{Covering fractions \fc, estimated from the fraction of pairs
  exhibiting $W_\lambda > 0.2$\AA, 
  in bins of impact parameter \rphys 
for (a) \ciit\ and (b) \civt.
  These provide another
  perspective on the chief results of this work: 
  (i) strong and nearly ubiquitous \ciit\ absorption to $\mrphys
  \approx 200$\,kpc followed by a rapid drop-off;
  (ii) a more gradual decline in the incidence and strength of \civt\
  absorption.   In panel (b), the green band shows an estimate of the
  covering fraction for random 3000\kms\ intervals in the IGM based on
  our $\ell(z)$ measurement (see the Appendix).
}
\label{fig:fC}
\end{figure*}

The averaged profiles exhibit strong \ciit\ absorption for $\mrphys <
200$\,kpc which essentially vanishes at larger impact parameters (there
is a weak `line' with $\mwcii \approx 0.2$\AA\ present at $\mrphys >
200$\,kpc).  
These trends, of course, follow the results observed for the
individual \wcii\ measurements.
Turning to \civt, we observe strong absorption from both members of the
\ion{C}{4} doublet to $\mrphys = 500$\,kpc, and then weaker but
non-negligible absorption for $\mrphys = [500,1000]$\,kpc.  Again, the
striking difference with the \ciit\ results is the sustained, strong
absorption to well beyond 200\,kpc.  
Statistically, the extended environments of quasars is substantially
enriched to distances greatly exceeding the virial radius of the host dark
matter halo.  We further emphasize that this cannot be simple ISM gas
from neighboring galaxies, the \ion{H}{1} column densities are far too
low.  Such galaxies may, however, be the source of this enrichment.
In the largest \rphys\ interval gas unassociated with the f/g quasar
but within our $\Delta v = \pm 1500 \mkms$ window contributes to the
signal.  As discussed in the Appendix, the random incidence of
\ion{C}{4} systems with $\mwciv > 0.3$\AA\ is 2.1 absorbers per
$\Delta z = 1$ interval at $z_{\rm abs} \approx 2.1$.
Within our 3000\kms\ window, we expect 0.065~systems on average giving
$\mwciv \approx 0.05$\AA.  This represent approximately one-third of
the signal in our largest bin.
The \ion{C}{4} stacks also exhibit significant absorption at $v
\approx -4200\mkms$, corresponding to the low-ion \ion{Si}{2}~1526
transition.  Its behavior qualitatively tracks the trends observed for \ciit,
as expected.

\begin{deluxetable*}{lccccccccccc}
\tablewidth{0pc}
\tablecaption{QPQ7 \fc\ Values\label{tab:qpq7_fc}}
\tabletypesize{\scriptsize}
\tablehead{\colhead{$\mrphys^{\rm min}$} & 
\colhead{$\mrphys^{\rm max}$} 
&
\colhead{$m_{\rm pair}^a$} & 
\colhead{$\mfc^{1334}$} & 
\colhead{$+1\sigma^b$} & 
\colhead{$-1\sigma^b$} 
&
\colhead{$m_{\rm pair}^a$} & 
\colhead{$\mfc^{1548}$} & 
\colhead{$+1\sigma^b$} & 
\colhead{$-1\sigma^b$} 
\\
(kpc) & (kpc) &}
\startdata
\cutinhead{$W_{\rm lim} = 0.2$\AA}
  30& 100&  12&0.75&0.14&0.10&  10&0.80&0.15&0.10\\
 100& 200&  16&0.69&0.12&0.10&  15&0.60&0.13&0.12\\
 200& 500&  27&0.19&0.06&0.08&  27&0.44&0.09&0.10\\
 500&1000&  83&0.08&0.03&0.04&  92&0.21&0.04&0.04\\
\cutinhead{$W_{\rm lim} = 0.3$\AA}
  30& 100&  12&0.58&0.14&0.13&  12&0.83&0.13&0.08\\
 100& 200&  16&0.56&0.12&0.12&  19&0.32&0.09&0.11\\
 200& 500&  33&0.15&0.05&0.07&  38&0.34&0.07&0.08\\
 500&1000& 138&0.08&0.02&0.03& 170&0.15&0.03&0.03\\
\enddata
\tablenotetext{a}{Number of pairs analyzed.}
\tablenotetext{b}{Confidence limits from Binomial statistics (Wilson score) for a 68\%\ interval.}
\end{deluxetable*}

\section{Results}
\label{sec:results}

\subsection{Covering Fractions (\fc)}
\label{sec:fC}

To further characterize the metal-absorption associated with the
measurements of galaxies hosting $z \sim 2$ quasars and to facilitate
comparisons to CGM research for other galaxy samples and epochs, we
have made estimates for the covering fractions \fc\ of the gas as a
function of \rphys.  
These are calculated simply by taking the ratio of the pairs where the
EW measurements exceed a given threshold to the total number.
Uncertainties are estimated from a standard Wilson score.
Before presenting the results, we emphasize that
the \fc\ values recovered are sensitive to both the assumed \rphys\ bins and the
(rather arbitrary) equivalent width limit \wlim.  For the former, we
maintain the \rphys\ intervals presented in the previous section which
are a compromise between bin-width and sample size.  
For \wlim,
we are driven primarily by data quality.  In the following we present
results for $\mwlim = 0.2$\AA\ and restrict the analysis to the spectra
satisfying $\sigma(W) \le 0.1$\AA\ (134 pairs for \ciit\ and 135 pairs for
\civt), i.e. we adopt a $2\sigma$ detection threshold.  

Figure~\ref{fig:fC}a present the \fc\ measurements for \ciit\ (see
also Table~\ref{tab:qpq7_fc}).  The
covering fraction is near unity for $\mrphys < 200$\,kpc, consistent
with previous results that the CGM of quasar hosts
exhibits a high covering fraction to 
gas that is optically thick at the \ion{H}{1} Lyman limit
(QPQ1, QPQ2, QPQ5, QPQ6).  
Indeed, these strong \ciit\ detections contributed in part to those
prior conclusions because we adopted strong \ciit\ as an indicator
that the gas is optically thick.  
Independently, we conclude that the gas is
significantly enriched in heavy elements.  Furthermore, the steep drop
in $\mfc^{1334}$ at $\mrphys > 200$\,kpc requires that this C$^+$ gas lies
predominantly within the host halo.
The extent, however, well exceeds any reasonable estimations for the
ISM of the host galaxy and very few sightlines exhibit \ion{H}{1}
columns reflective of ISM gas (QPQ6).  Instead, this material must
be halo gas\footnote{Including material associated to satellite
  galaxies.}.  
We also emphasize that 0.2\AA\ well exceeds the average equivalent
width for \ciit\ in `random' LLS ($\approx 0.06$\AA), 
at $z \approx 3$ with $\mnhi =
10^{17} - 10^{19} \cm{-2}$ \cite{fop+13}.  The gas probed here 
represents a more highly enriched and/or dynamic medium than
`typical', optically thick gas.

Regarding \fc\ for \civt, the results follow the conclusions drawn in
the previous section, i.e.\ that the \civt\ absorption extends to
$\approx 500$\,kpc with a declining incidence that is shallower than
that observed fo \ciit.  
The results are well-described by a single power-law, 
$f_C^{\rm 1548} (W>0.2{\rm \AA}) \sim 0.4 (\mrphys/300\,{\rm
  kpc})^{1/2}$
In the following sub-section, we measure the cross-correlation
function between \ion{C}{4} and quasars, finding enhanced absorption
to at least 1\,Mpc.

\subsection{\ion{C}{4}-Quasar Two-Point Cross-correlation}
\label{sec:cross}

The results presented above demonstrate clearly that the environments
surrounding the host galaxies of luminous $z \sim 2$ quasars exhibit
an excess of \ion{C}{4} absorption to scales of at least 500\,kpc.  We
may quantify the excess by estimating the two-point cross-correlation
function between \ion{C}{4} absorbers and quasars, \xic.  Our approach
follows the maximum likelihood methodology presented in QPQ2 and QPQ6
used to assess the cross-correlation of \ion{H}{1} absorption to
quasars.  To briefly summarize, we parmaterize $\mxic(r)$ as a
power-law $(r/r_0)^{-\gamma}$, project the 3D correlation function
along the quasar sightlines to determine the transverse correlation
function $\chi_\perp(\mrphys)$, and find the values of $r_0$ and
$\gamma$ which maximize the likelihood of recovering the observed
incidence of strong \ion{C}{4} absorbers with \rphys.  Following
standard convention for clustering analysis, the calculations are
performed and reported in comoving coordinates in units of
$h^{-1}\,{\rm Mpc}$ with $H_0 = 100 h^{-1} {\rm km \, s^{-1} \,
  Mpc^{-1}}$.

Central to the evaluation of \xic\ is a precise and accurate estimate for the 
incidence of `random' \ion{C}{4} absorbers along quasar sightlines.
And, in our analysis this background may contribute significantly
because of the large uncertainty in quasar redshifts which requires
that we evaluate $\chi_\perp$
over relatively large velocity windows around each f/g quasar ($\Delta
v = \pm 1500 \mkms$).  
While a number of \ion{C}{4} surveys have been previously performed
\citep{boks03,dcc+10,cooksey+13}, none of these is
entirely satisfactory for the \ion{C}{4} equivalent widths
characterizing our study (i.e.\ $\mwciv \approx 0.5$\AA).  
Therefore, we
carried out our own survey for strong \ion{C}{4} systems, 
as described in the Appendix.
The primary result is a `random' incidence for strong \ion{C}{4}
absorbers (per unit redshift; $\mwciv > 0.3$\AA) of $\mlciv = 2.1$
at $\mavgz \approx 2.1$ with a 10\%\
statistical uncertainty and a comparable, estimated systematic error.
We assume no redshift evolution in \lciv;  \cite{cooksey+13} find a
small ($\approx 30\%$) decrease in \lciv\ from $z =2$ to 3.

\begin{figure}
\includegraphics[width=3.7in,angle=90]{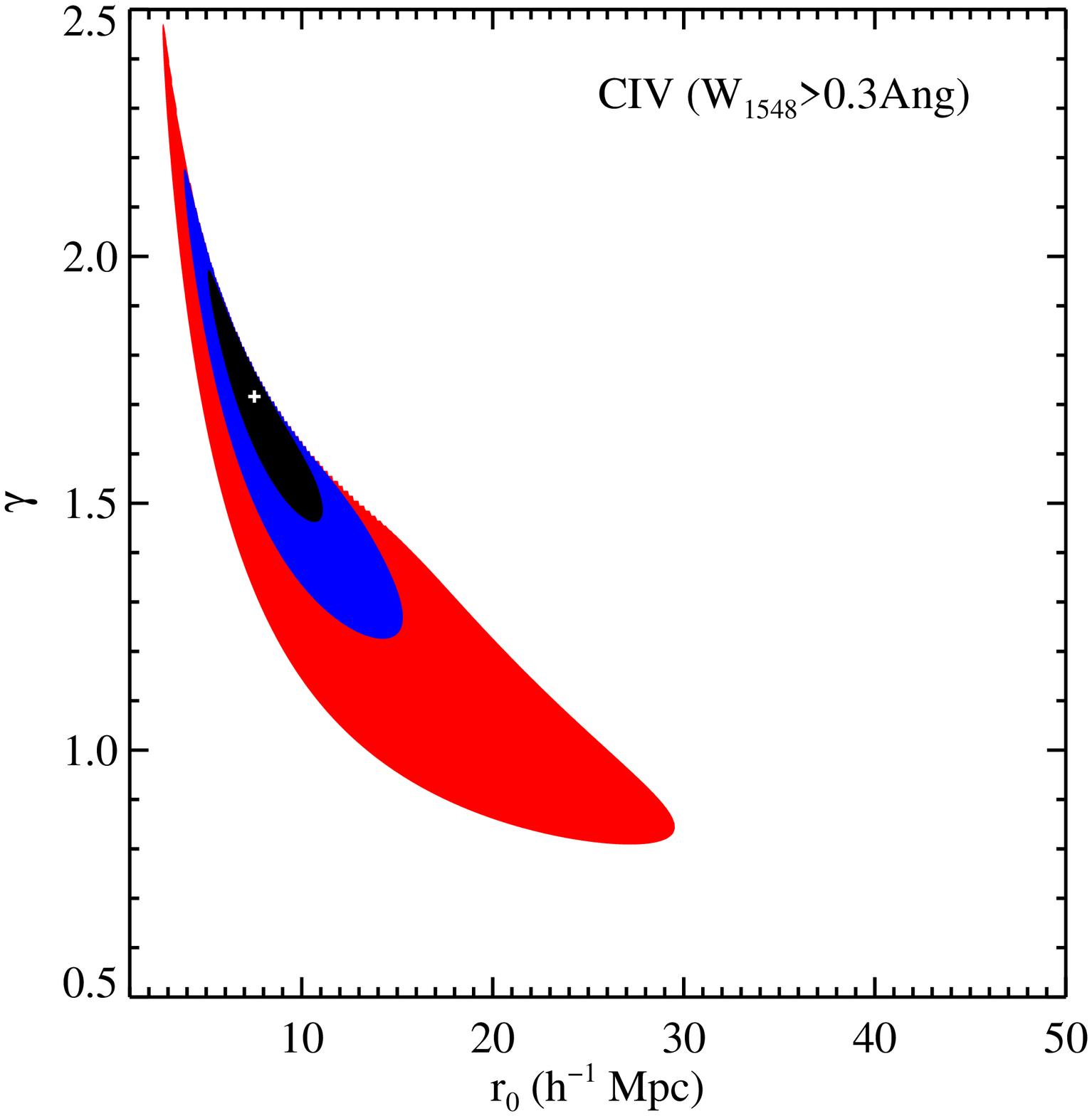}
\caption{Constraints on the cross-correlation function for strong
  ($\mwciv > 0.3$\AA) \civt\ absorption and quasars at $z \approx
  2.5$, parameterized as $\mxic(r) = (r/r_0)^{-\gamma}$.  The best-fit
  values are $\gamma = \vgmm$ and $r_0 = \vro$, with
  degeneracy between the two.  For $\gamma=1.6$, typical of
  galaxy-galaxy and galaxy-quasar clustering, the data demand $r_0 >
  5 \mhMpc$ at 95\%\ confidence.  The colors (black,blue,red) describe
  contours of (1,2,3)$\sigma$ confidence levels.
}
\label{fig:contour}
\end{figure}

\begin{figure}
\includegraphics[width=3.7in,angle=90]{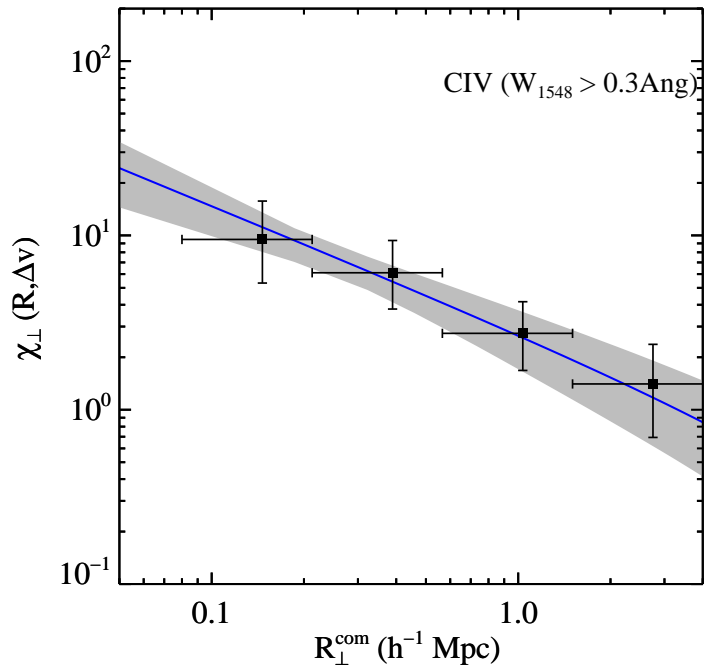}
\caption{Projected cross-correlation function between strong
  \ion{C}{4} absorbers and quasars, estimated in arbitrary bins of
  \rcom.
  Overplotted on these values is the best-fit model from a maximum
  likelihood analysis 
  $\mxic(r) = (r/r_0)^{-\gamma}$ (blue line; $r_0 = \vro$, $\gamma =
  \vgmm$), and our estimate of the
  $1\sigma$ uncertainties.  This model provides a good description of
  the observations.  
}
\label{fig:trans_corr}
\end{figure}

A key consideration for the analysis is the limiting equivalent width
$\mwlim_{\rm CIV}$ for systems in the correlation analysis.  
In the previous section, we presented covering fraction
results for $\mwlim_{\rm CIV} = 0.2$\AA.
For the following, we take $\mwlim_{\rm CIV} = 0.3$\AA\ and require a
$3 \sigma$ detection at this equivalent width limit.  This is a
stricter definition and is also motivated by our requirement to have a
well-measured incidence of `random' \ion{C}{4} absorbers.
In addition, we
restrict to pairs where $\mzfg < 3.5$.  The sample encompasses
\ncpair\ pairs with $\langle \mzfg \rangle = 2.34$ and comoving impact
parameter $\mrphys^{\rm com}$ ranging from 0.9 to 2.6\,\hMpc.

The results of our maximum likelihood analysis\footnote{See QPQ2 and QPQ6 for
  details on the methodology.} 
for $r_0$ and $\gamma$
are presented in Figure~\ref{fig:contour}.  We recover $r_0 = \vro$
and $\gamma = \vgmm$ with degeneracy between the two
parameters.  The contours exhibit a sharp cutoff on
the right-hand side (larger $r_0$) where the models predict the
detection of at least one \ion{C}{4} absorber at an impact parameter
where none was found.  
Our best-fit $\gamma$ values follow results
reported previously for galaxy-galaxy clustering,
as predicted from analysis of dark matter
halo clustering in numerical simulations \citep[e.g.][]{mw96}.
This suggests that the \ion{C}{4} gas may be associated to 
dark matter halos in the quasar vicinity, i.e.\ the CGM of neighboring
galaxies clustered to the quasar host.
The large correlation length, meanwhile, reflects both the mass of
the quasar host galaxy but also implies the \ion{C}{4} gas occurs
preferentially near massive halos (similar to our inferences on LLS
in QPQ6).  We explore this assertion further in
the next section.

Figure~\ref{fig:trans_corr} presents a binned
evaluation of the transverse, projected cross-correlation function
$\chi_\perp(R,\Delta v)$ for the quasar pair sample and the best-fit
model with a band illustrating the uncertainty in $r_0$.  Clearly,
this model provides a good description of the data.
We conclude that strong \ion{C}{4} systems are highly clustered to
quasars at $z \sim 2$ with a clustering amplitude $r_0 \approx 7.5 \mhMpc$. 

\begin{figure}
\includegraphics[width=3.5in]{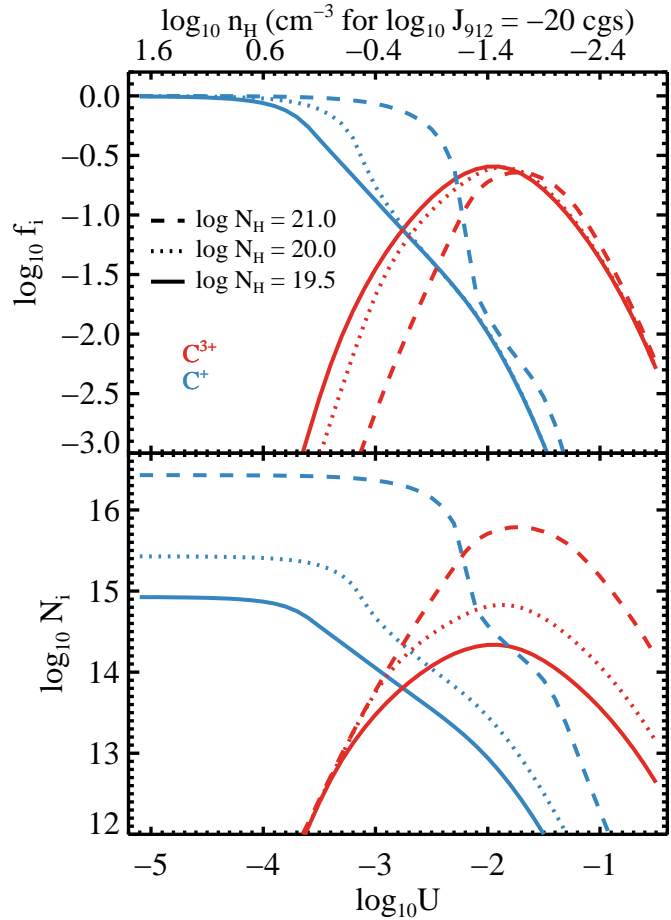}
\caption{(top): Ionization fractions $f_i = N_i/N$ for C$^+$ (blue) and
  C$^{3+}$ (red) as a function of ionization parameter $U$ and total
  hydrogen column density $N_{\rm H}$.  All models assume a standard
  EUVB radiation field at $z=2$ and a gas metallicity [C/H]~$=-1$.
  At $\log U = -3$, which we believe is representative of the cool
  CGM (QPQ3, QPQ8), C$^+$ has $f_i = 0.1 - 0.3$ and C$^{+3}$ shows
  $f_i < 0.1$.  This is consistent with the stronger \ciit\ absorption
  observed relative to \civt\ (Figure~\ref{fig:ewscatt}; QPQ5).
  The top axis is labeled by the number density $n_{\rm H}$ assuming
  an intensity for the EUVB at the Lyman limit $J_{912} = 10^{-20} \,
  {\rm erg \, s^{-1} \, cm^{-2} \, Hz^{-1}}$.
  (bottom): The predicted column densities $N_i$ as a function of
  $\log U$ and $N_{\rm H}$.  These photoionization models predict
  $\N{C^+} > \N{C^{3+}}$ for $\log U < -2.5$.  
  From the C$^+$ column densities estimated, one requires $N_{\rm H} >
  10^{20} \cm{-2}$.  The figure also demonstrates that in gas where
  only \civt\ is detected, one expects $\log U \gtrsim -2$.
}
\label{fig:cldy}
\end{figure}

\section{Discussion}
\label{sec:discuss}

In this section, we consider the implications of our results for the
ICM, the origin of the CGM, and the nature of strong metal-line
absorbers. 
Before proceeding, we present the results from a simple series of
photoionization calculations 
made with the Cloudy software package \citep[v10;][]{ferland13}, 
to offer physical insight into our observations.
Figure~\ref{fig:cldy} shows the predicted ionic fractions $f_i \equiv
N_i/N$ and ionic column densities $N_i$ of C$^+$
and C$^{3+}$ for a plane-parallel gas slab with total hydrogen column
densities $N_{\rm H} = 10^{19.5}, 10^{20}$ and $10^{21} \cm{-2}$,
metallicity [C/H]~$= -1$, and an assumed EUVB background
from the CUBA package \citep[$z=2$;][]{hm12} with a range of ionization
parameters $U \equiv \Phi/cn_{\rm H}$.
At low $U$ values ($\log U < -2.5$), the lower ionization state
dominates and one predicts a modest C$^{3+}$ column density, consistent
with our observations.  From a \ciit\ equivalent width of $\mwcii =
1$\AA, one conservatively estimates a column density $\N{C^+} >
10^{14.7} \cm{-2}$ from the linear curve-of-growth approximation.
For [C/H]~$= -1$, this implies $N_{\rm H} > 10^{20}
\cm{-2}$.  Such models also reproduce the large \nhi\ values observed
in the QPQ sample (QPQ6).
The figure also demonstrates that one can reproduce systems
with strong \civt\ absorption and negligible \ciit, characteristic of gas
at large impact parameters from the quasar hosts, provided $N_{\rm H}
\approx 10^{20} \cm{-2}$ and $\log U \gtrsim -2$.
This implies a medium whose volume density decreases with radius but
with a roughly constant surface density.

\begin{figure}
\includegraphics[width=3.7in,angle=90]{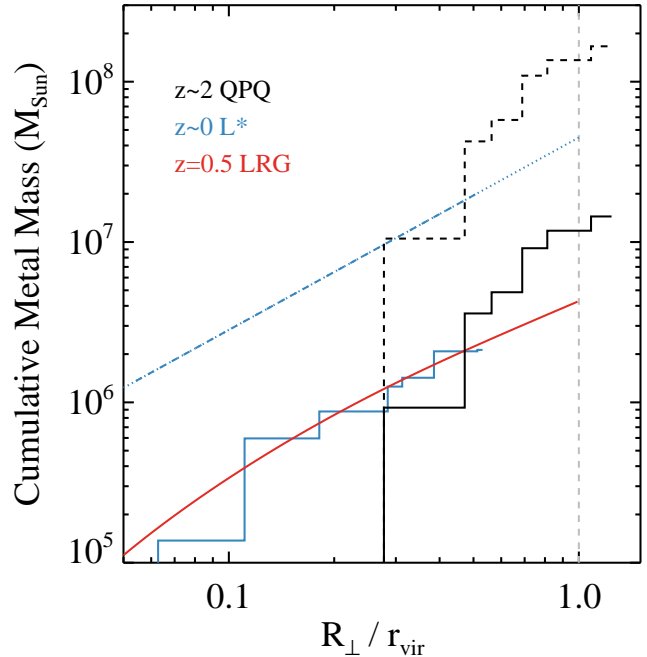}
\caption{Estimated cumulative mass profiles for the cool CGM of
  several galaxy populations:
  $z \sim 2$ massive galaxies (black; QPQ),
  luminous $z \sim 0$ galaxies \citep[red;][]{werk+14},
  and LRGs (blue; Z14).
The solid curves offer very conservative (i.e.\ lower limit) estimates
based on saturated absorption and no ionization corrections (see
$\S$~\ref{sec:ICM}
for details).  The dashed curves present estimates that include
ionization corrections \citep{werk+14} and the dotted curve represents
the extrapolation of such analysis.  Comparing results that adopt the same
set of assumptions (i.e.\ restricting comparison to curves with the same line-style), we
find that the total mass at \rvir\ is highest in the halos of $z
\sim 2$ massive galaxies.
}
\label{fig:Cmass}
\end{figure}

\subsection{Constraining Enrichment Models for the ICM}
\label{sec:ICM}

One motivation for our analysis was to test the prediction that
massive galaxies expel a large mass in metals at early times to
reproduce $z<1$ observations of the ICM
\citep{renzini93,mg95,arrigoni10b,yates14}.
Taking one recent calculation as a fiducial example, the preferred
model of \cite{yates14} predicts that a $10^{12.5} \msol$ galaxy
at $z=2$ will have ejected several $10^9 \msol$ in metals by $z=2$. 
In this model, the early enrichment and feedback is dominated by
massive stars.
We may then compare these predictions to estimates for the metal mass
surrounding the massive, $z \sim 2$ galaxies of the QPQ survey. 

The results presented in
Figures~\ref{fig:ewscatt}~and~\ref{fig:fC} imply a large
mass of metals traced by \ciit\ in the halos of $z \sim 2$, massive
galaxies.  To provide a preliminary and rough estimate for this mass, 
consider the following calculation which gives
a conservative, lower limit to the metal mass:
 (i) convert the individual \wcii\ values into C$^+$ column densities
 $\N{C^+}$ conservatively assuming the linear curve-of-growth (COG) approximation
 \citep[e.g.][]{spitzer78}.  This includes negative values (i.e.\ all
 non-detections); 
 (ii) average these $\N{C^+}$ values in bins of \rphys\ with
 \ncbin~measurements in each bin, 
 $\left <\N{C^+} \right >$;
 (iii) calculate the mass in the $i$th annulus defined by the $i$th set of
 \ncbin~pairs, $M_i = \left< \N{C^+} \right>_i \, m_{\rm C} \, \pi \ltk R_{i,max}^2
 - R_{i,min}^2 \rtk$ and;
 (iv) assume that carbon represents 20\%\ of the total estimated
 metal mass, as in our Sun.  

The cumulative mass profile, plotted relative to the
virial radius,  is shown as the solid black curve in
Figure~\ref{fig:Cmass}.  
This represents a very conservative lower
limit to the metal mass of the cool CGM phase because we have ignored
line-saturation and because C$^+$ is likely only a small fraction
of all C ($<50\%$; Figure~\ref{fig:cldy}), even in this cool phase.  
Nevertheless, the total mass \mmetal\ is
substantial: approximately $5 \sci{6} \msun$ at the estimated virial
radius and nearly $10^7 \msun$ at $\mrphys = 200$\,kpc.
This is comparable to the metal mass in a $10^{10} \msun$ gas with
1/10 solar metallicity\footnote{One may also infer this mass by the fact that
  strong \ion{C}{2} absorption implies $\mnhi > 10^{18} \cm{-2}$,
  e.g. QPQ5.}, and we stress again that this estimate gives a
very conservative lower limit to \mmetal.  
While in principle the ISM of satellite galaxies will
contribute to this estimate, none of these sightlines exhibit \ion{H}{1}
column densities characteristic of a galactic ISM (i.e.\ $\mnhi >
10^{20} \cm{-2}$).  
And, the high covering fraction of \ciit\ absorption implies a
widespread CGM (Figure~\ref{fig:CII_map}) that cannot be
easily explained by satellites \citep[e.g. QPQ3][]{tumlinson+13}.

To provide a better estimate for the metal mass, 
we further assume the following:
 (v) the QPQ sightlines with $\mwcii>0.2$\AA\ require a (conservative)
 0.3\,dex correction to their COG-estimated $\N{C^+}$ values.  This is
 based on analysis of ions that exhibit multiple transitions with a
 wide range of oscillator strengths (e.g.\ \ion{Fe}{2});
 (vi) C$^+$ represents only $1/3$ of the total C in this cool phase
 (i.e.\ $\log U \approx -3$; Figure~\ref{fig:cldy}); 
and
 (vii) C/O is sub-solar by a factor of 2, as observed in metal-poor stars
 \citep{acn+04}, such that C represents only 10\%\ of the total mass
 in metals.  
The second correction (C$^+$/C = 1/3) is supported by
detailed photoionization modeling of the gas from observed ionic
ratios (QPQ3, QPQ8).
Meanwhile, the saturation correction is still conservative
given the very large \wcii\ measurements.  
Together, these corrections imply $\approx 10\times$ more metal mass
in the cool CGM, i.e.\ $M_{\rm CGM}^{\rm cool}$ exceeds $10^8 \msun$ at $\approx \mrvir$, and {\it we
still consider this a conservative estimate}.

Comparing to the predictions from chemical evolution models
\citep[e.g.][]{yates14}, the cool CGM represents $\gtrsim 10\%$ of the
metals predicted to in galactic halos with $M \approx 10^{12.5} \msun$
at $z \approx 2$.  We conclude that the cool CGM represents a small
but non-negligible fraction of the metals of the incipient
intragroup/intracluster medium (IGrM/ICM).  
We further emphasize that a massive reservoir of 
hot, enriched gas could exist within the QPQ halos.  
The observed cool gas almost certainly requires an ambient warm/hot
medium to provide pressure support (e.g.\ QPQ3) and
such hot halos are predicted to be already
ubiquitous at $z \sim 2$ in in massive halos \citep[e.g.][]{fhp+14}.
Unfortunately, the far-UV diagnostics
provided by the absorption spectra are insensitive to material with $T > 10^6$\,K.  
Even the apparent absence of strong \ion{N}{5} and \ion{O}{6}
absorption (QPQ3,QPQ8) would only rule out
metal-enriched gas with $T \approx 10^{5-6}$\,K.  
Perhaps, one may offer model-dependent estimations of the hot gas by
considering predictions for such ions when cool gas interacts with a
hot medium \citep[e.g.][]{kwak11}.

Returning to the cool CGM and its relation to the IGrM/ICM, we are
motivated to examine the relative abundances within this medium (e.g.,
O/Fe, Si/Fe) to test competing scenarios of early enrichment and
feedback \citep[e.g.][]{arrigoni10b,yates14}.  This will require,
however, observations with high spectral resolution and a careful
treatment of ionization corrections.  We defer such analysis to future
work.




\begin{figure}
\includegraphics[width=3.6in]{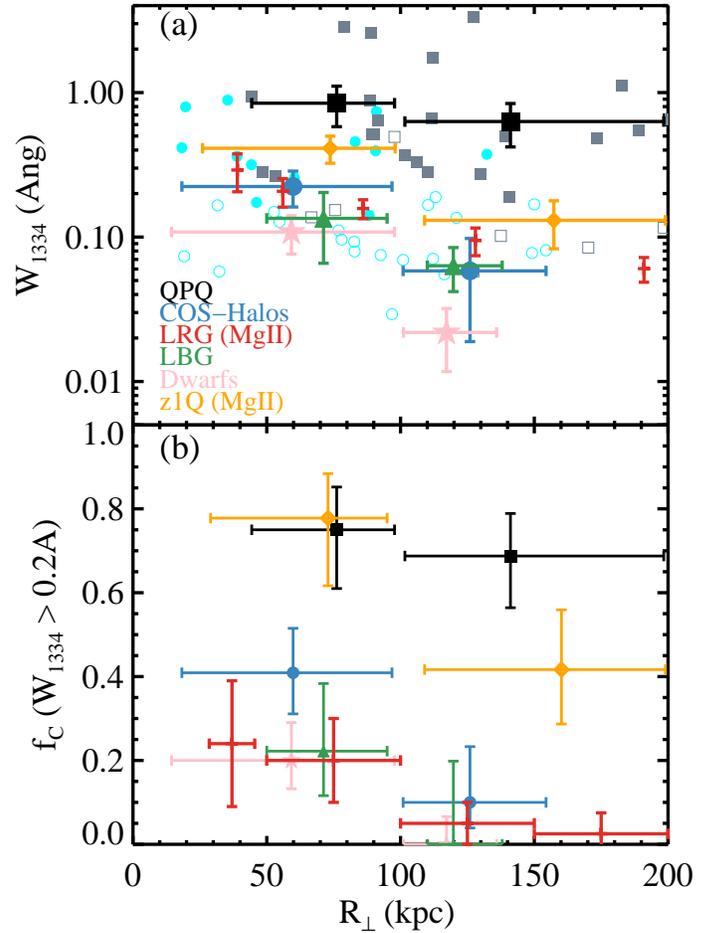}
\caption{ (a) Scatter plot of the rest-frame equivalent widths of \ciit\ along
  quasar sightlines with impact parameters \rphys\ to the $z \sim 2$
  massive galaxies hosting quasars (black squares; this paper) and 
  present-day \lstar\ galaxies \citep[blue circles;][]{werk+13}.
  Although the loci of these samples overlap, the QPQ data exhibit
  larger values on average and a higher incidence of positive
  detections. The figure also shows a set of binned evaluations
  \avgcii, including estimates for $z \sim 0$ dwarf galaxies (pink
  stars; Bordoloi et al., in prep.), 
  $z \sim 0.5$ LRGs (red symbols; scaled
  from \wmgii, see $\S$~\ref{sec:compare}; Z14), and 
  the $z \sim 2$ LBGs
  \citep[green triangles;][]{ass+05,simcoe06,rudie13,crighton13}. 
  It is evident that the CGM of the QPQ sample exhibits the strongest
  low-ion absorption of any galaxy population.  This conclusion is further emphasized in panel
  (b) which gives the covering fraction in each sample for
  $\mwcii > 0.2$\AA.
}
\label{fig:CII_compare}
\end{figure}

\subsection{Comparing the Cool CGM to other Galaxy Populations}
\label{sec:compare}

By drawing comparisons between the QPQ observations and other
CGM measurements in galaxies with a wide range of halo mass and
age, we may gain insight into the physical processes that dominate
the CGM.  This sub-section performs the empirical comparison and we
discuss implications for physical processes in $\S$~\ref{sec:origin}.

A principal result of our analysis is the predominance of strong,
low-ion metal absorption to $\mrphys \approx 200$\,kpc 
in the halos of galaxies hosting quasars.
Strong low-ion absorption from the CGM of galaxies (aka halo gas) has
been recognized previously at $z<1$
\citep[e.g.][]{bergeron86,steidel93,lzt93}.  
Indeed, there is an extensive, and
still growing, literature on the incidence and nature of \ion{Mg}{2}
absorption in the CGM of luminous galaxies at $z<1$
\citep[e.g.][]{chg+10,nck+13,werk+13}, and also the hosts of quasars at
$z \lesssim 1$ \citep{bhm+06,farina13}.  A primary
conclusion of this literature
is that strong \ion{Mg}{2} absorption occurs at impact parameters
$\mrphys \lesssim 75$\,kpc for $L^*$ galaxies. 
There is also the indication that \ion{Mg}{2} absorption strength
correlates with galaxy luminosity and/or halo mass
\citep{cwt+10,churchill13,farina+14}. 
We turn to our results in the context of this previous work,
restricting the comparison to scales of the dark matter halo and
presented in physical units (kpc).

Figure~\ref{fig:CII_compare}a presents the \ion{C}{2}~1334
equivalent width
distribution for the QPQ7 sample (black squares) compared against the
set of such measurements for $L \approx \mlstar$ galaxies at $z \sim 0.2$
taken from the COS-Halos survey \citep[red
circles;][]{werk+13,tumlinson+13}.
The loci of the measurements overlap, but the
detection rate and typical \wcii\ values are much higher for the QPQ7
sample.  These differences are emphasized in the binned evaluations \avgcii,
and the estimated covering fractions \fc\ for $\mwcii >
0.2$\AA\ (Figure~\ref{fig:CII_compare}b).
The contrast is most striking at $\mrphys = 100-200$\,kpc
where positive detections are very rare for \lstar\ galaxies but are
common in the outer halos of the $z \sim 2$ quasar hosts.  
These differences would be further accentuated if one scaled the impact
parameters by estimates of the virial radii for these two
populations ($\mrvir^{\rm QPQ} \approx 160$\,kpc, $\mrvir^{L^{*}}
\approx 290$\,kpc).  
We conclude, at very high confidence, that the halos of
massive, $z \sim 2$ galaxies contain a much greater reservoir of
cool metals than modern \lstar\ galaxies. 

\begin{figure*}
\includegraphics[scale=0.67,angle=90]{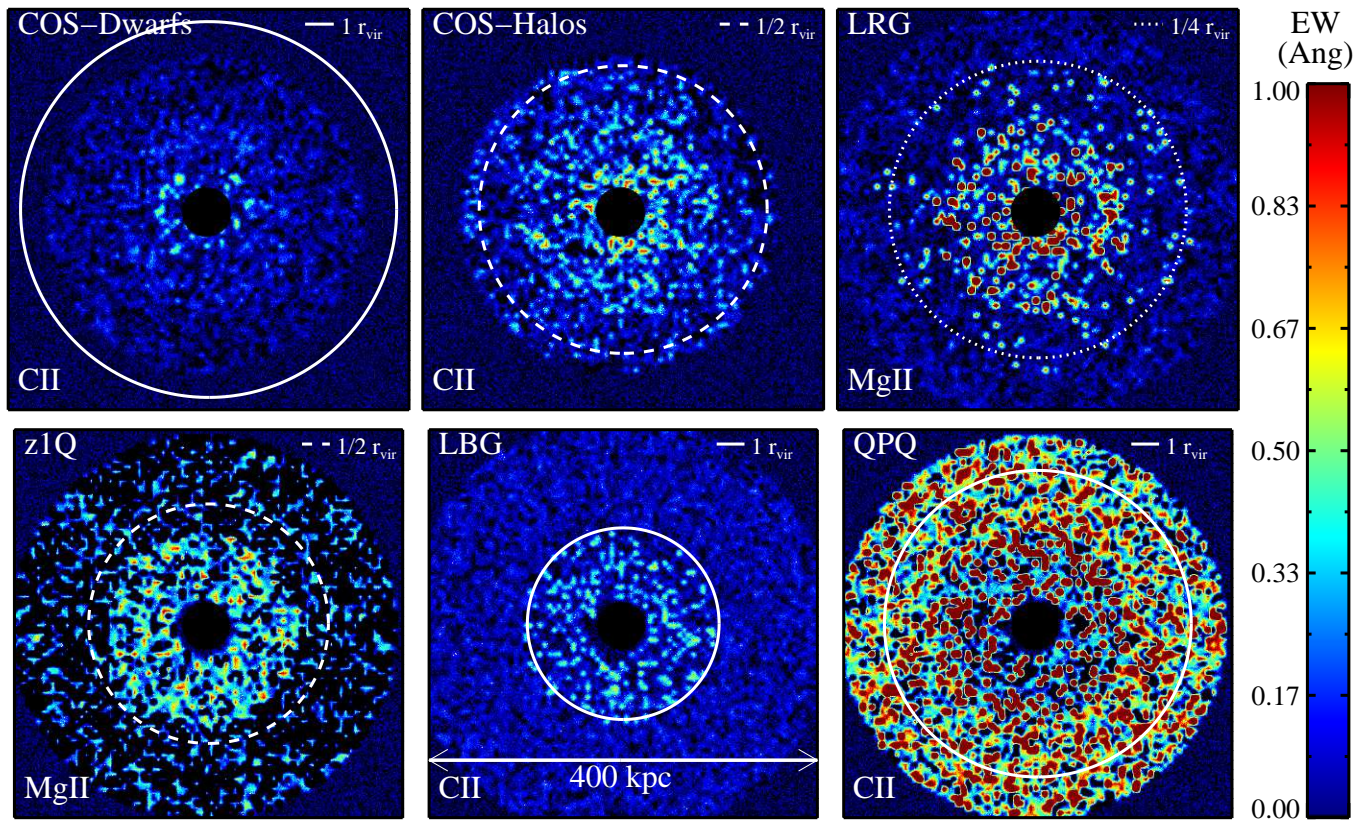}
\caption{
 A visualization of the \ciit\ absorption measured
 statistically for $z \sim 0$ luminous galaxies
 \citep[COS-Halos;][]{werk+13}, $z \sim 2$ star-forming galaxies
 (LBGs), [LRGs], and
 the massive $z \sim 2$ galaxies hosting quasars (QPQ; this paper).
 The color bar indicates the \wcii\ values in \AA.  Each cut-out is
 400\,kpc on a side and the solid (dotted/dashed) white circle indicates the
 estimated full (quarter/half) virial radius of each galaxy population.  
 This exercise ignores the inner 20\,kpc where the ISM of the galaxy 
 may dominate and little data exists.
 The outer edges of the maps are also arbitrarily defined, i.e.\ not physical.
 The maps are constrained to give the observed distribution of
 \wcii\ and/or \avgcii\ values in several annuli but we emphasize
 that the size of the clumps (taken to be 5\,kpc) and their
 distribution (taken to be random) need not hold in the real universe.
}
\label{fig:CII_map}
\end{figure*}

The figure also presents estimations of \avgcii\ for the COS-Dwarfs
survey \citep[][Bordoloi et al., in prep]{bordoloi14}.  These are
based on 39~measurements of \wcii\ from $L \lesssim 0.1 \mlstar$
galaxies at $z<0.1$ drawn from the SDSS.  The measured values are smaller still
than the (nearly) coeval \lstar\ galaxies, suggesting that the cool CGM depends on
stellar/halo mass.

Figure~\ref{fig:CII_compare} also presents estimations for the average
equivalent widths of low-ion metals from 
the halos of $z \sim 0.5$, luminous red galaxies (LRGs).
This estimate is derived from 
the measurements of \citet[][hereafter Z14]{zhu+14} for the average
equivalent widths of \mgiit, $\left < \mwmgii \right >$. 
Because of their similar ionization potentials, we expect C$^+$ to be
roughly co-spatial with Mg$^+$-bearing gas.\footnote{Each element is
  ionized to this first state by $h\nu < 1$\,Ryd photons.  One does
  note, however, that Mg$^+$ has a significantly lower ionization
  potential than C$^+$ (15.0\,eV vs.\ 24.4\,eV).}
Furthermore, we expect
comparable equivalent widths for \ciit\ and \ion{Mg}{2}~2796,
especially when the transitions are both saturated. 
Empirically, the $\approx 10$ systems from the COS-Halos
dataset with strong detections\footnote{Weaker \mgiit\ systems show a
  nearly the same equivalent width as \ciit\ \citep{narayanan08}.} 
of both \ciit\ and \mgiit\
show an average ratio of \wmgii/\wcii=1.7 \citep{werk+13}.  
This scaling follows from the difference in rest wavelengths
between the two transitions under the expectation that
$W_\lambda/\lambda$ is roughly constant.
Therefore, we have scaled
down the Z14 measurements accordingly to estimate
\avgcii\ for the LRGs.
Surprisingly, the LRG results trace (at least roughly) the 
values for the \lstar\ galaxies. 
This indicates that the massive halos of $L \gg \mlstar$ galaxies
contain substantial, cool and enriched gas at all epochs, 
independent of on-going star-formation in the central galaxy \citep[as
found for red-and-dead galaxies at $z \sim 0.2$;][]{thom12}.  
This even includes galaxy clusters which exhibit an enhanced incidence
of strong \ion{Mg}{2} systems within \rvir\ \citep[$\mwmgii >
2$\AA;][]{lopez08,andrews13}.
Regarding our study of $z\sim 2$ massive galaxies, the QPQ values
systematically exceed the estimates for the LRGs at all impact
parameters.  Estimations of the covering fraction of \ion{Mg}{2}
absorption for LRGs also give much lower values than observed for the
quasar hosts (Figure~\ref{fig:CII_compare}b).
And, again, scaling by the virial radii would only further accentuate
these differences ($\mrvir^{\rm LRG} \approx 600$\,kpc).


Pushing to $z > 1$, 
we compare our results against the coeval population of
star-forming galaxies at $z \sim 2$, known as LBGs.
Clustering analysis of the galaxies provides halo mass estimates
of $M_{\rm halo} \lesssim 10^{12} \msun$ \citep{adel05,bielby+13}, 
i.e.\ a factor of three to five lower mass than the halos hosting luminous quasars.
Figure~\ref{fig:CII_compare} shows estimations for \avgcii\ 
and its covering fraction from the modest
set of measurements in the literature
\citep{adel03,ass+05,simcoe06,rudie13,crighton13}.\footnote{We 
  discuss results from the LBG stacked spectra of
  \cite{steidel+10} in the following section.}
Consistent with the low covering fraction of optically thick gas
reported by \cite{rudie13}, the \avgcii\ and $f_C$
values are small and significantly lower than the QPQ measurements. 
Even if we scale by the (likely) smaller virial radius for LBGs
($\mrvir^{\rm LBG} \approx 100$\,kpc), the massive galaxies hosting
quasars exhibit far greater low-ion absorption on all scales within \rvir.
Lastly, the figure shows recent results on the measurements of
\ion{Mg}{2} around $z \sim 1$ quasars \citep[z1Q;][]{farina+14}.
As with the LRGs, we scale these \ion{Mg}{2} measurements 
down by a factor of 1.7 to compare them to the \ciit\ results.
It is evident that the z1Q results lie between the QPQ and all other
studies.  This comparison indicates modest evolution in the strength
and covering fraction of cool gas around galaxies hosting quasars,
consistent with the expected mild evolution in the masses of halos
hosting quasars with decreasing redshift \citep{richardson12,yshen13}.

To summarize the results of Figure~\ref{fig:CII_compare}, the CGM of
massive galaxies hosting $z \sim 2$ quasars exhibits a higher incidence
and larger average equivalent widths of low-ion absorption than that
measured for any other galaxy population at any epoch.
Previously, we reached the same conclusion for \ion{H}{1}~\lya\
absorption (QPQ5, QPQ6).
We conclude that
{\it the CGM of these massive $z \sim 2$ galaxies 
represents the pinnacle of cool halo gas in the Universe.  
}

\begin{figure}
\includegraphics[width=3.0in,angle=90]{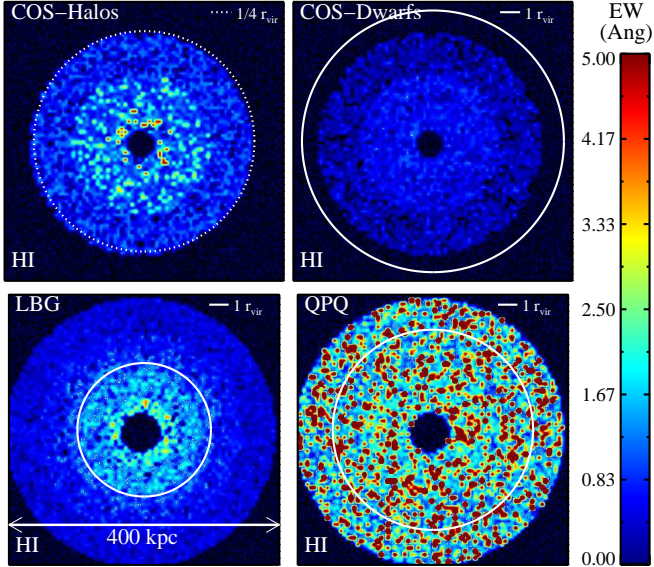}
\caption{Visualization of the CGM in \ion{H}{1} \lya\ with
  rest-frame equivalent width given by the color bar (in \AA;  see
  Figure~\ref{fig:CII_map} caption and the text for details).
  The figure illustrates clearly the much higher incidence of strong
  \ion{H}{1} \lya\ absorption associated with the $z \approx 2$
  massive galaxies hosting quasars.  This population marks the peak in
  the cool CGM.
  The inner/outer edges of the maps are arbitrarily defined by data
  coverage, i.e.\ not physical.
  The maps are constrained to give the observed distribution of
  \wcii\ and/or \avgcii\ values in several annuli but we emphasize
  that the size of the clumps (taken to be 5\,kpc) and their
  distribution (taken to be random) need not hold in the real universe.
}
\label{fig:HI_map}
\end{figure}

\begin{figure}
\includegraphics[width=3.5in]{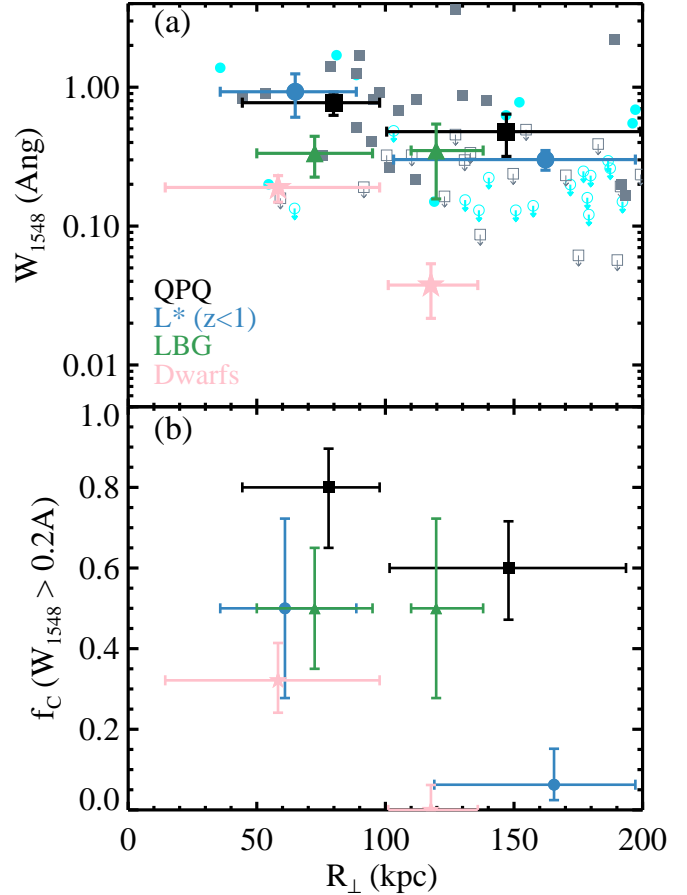}
\caption{Same as Figure~\ref{fig:CII_compare} but for the \civt\
  transition.  For the $L^*$ galaxies at $z<1$, we adopt the
  measurements of \cite{borthakur+13} and \cite{clw01}.
  The QPQ sample exhibits the highest rest-frame equivalent widths and covering
  fractions than any other galactic population, but the differences are smaller
  than those observed for lower ionization states.
}
\label{fig:civ_compare}
\end{figure}

To further illustrate this primary result, we constructed 
``maps'' of the CGM.
For \ciit\ in QPQ, we average the \wcii\ measurements in several
annuli for $\mrphys \le 200$\,kpc.  
Within each annulus, we drew randomly from the 
observed \wcii\ distribution,
assumed random gaussian noise with a deviate of 50\,m\AA, and
generated ``clumps'' with an arbitrary
size of 5\,kpc. 
These clumps were randomly placed within each annulus,
constrained to give approximately the observed \wcii\ distribution of
\ciit\ absorption including non-detections (i.e.\ reproducing the
observed covering fraction). 
Explicitly, for the annulus $\mrphys = [30., 100]$\,kpc we inserted
1152 clumps with an \wcii\ distribution matching the 12 pairs from the
QPQ sample at these impact parameters with an additional dispersion of
40\,m\AA.
We ignored the inner 25\,kpc which our pairs do not sample and
where the galactic ISM is expected to
contribute/dominate.  The maps are also arbitrarily cut at $\mrphys \approx
200$\,kpc or less (e.g.\ the COS-Halos and COS-Dwarfs samples extend 
to only $\approx 150$\,kpc).

Figure~\ref{fig:CII_map} shows the results in 400$\, \times 400$\,kpc
cut-outs for the COS-Halos survey \citep{werk+13}, the COS-Dwarfs
survey (Bordoloi et al., in prep), 
the LRGs (Z14; see the Appendix for details),
the $z \sim 1$ quasars \citep[z1Q;][]{farina+14},
the LBGs, and our QPQ results.
This exercise illustrates that the
low-ion absorption from the CGM of massive, $z \sim 2$ quasars is
qualitatively stronger than that of any other galactic population. 
We have repeated this exercise for the \ion{H}{1} \lya\ measurements
where available, as presented in Figure~\ref{fig:HI_map}. 
For the LBGs, we take the average \wlya\ values reported by
\cite{steidel+10} and \cite{rakic12}, assuming a 50\%\ scatter,
and a unit covering fraction.
All of these equivalent widths were corrected for absorption by the
background IGM. 
The results further emphasize
that cool gas absorption is systematically
larger for the QPQ dataset on all physical scales and also when
considered as a fraction of the virial radius.
At the same time, the CGM of the \lstar\ galaxies at $z \sim 0$ 
is qualitatively similar to the LBGs (the latter 
is stronger when scaled by \rvir).

\begin{figure}
\includegraphics[width=3.0in,angle=90]{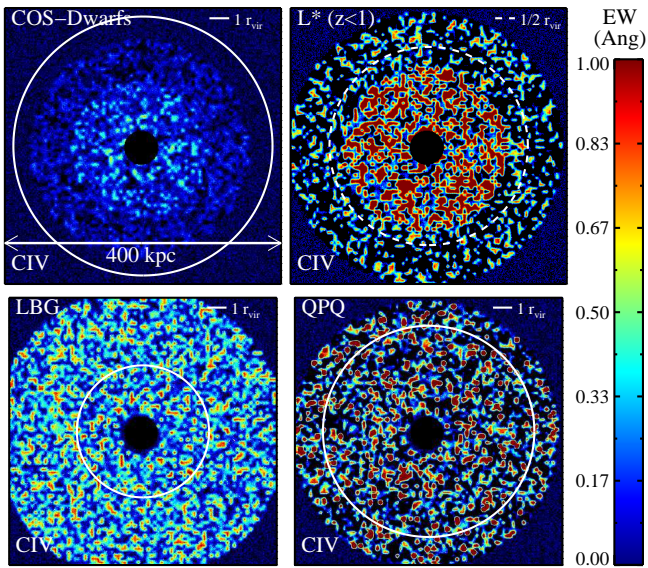}
\caption{Illustrative map of the CGM in \civt\ absorption with
  equivalent width given by the color bar (in \AA;  see
  Figure~\ref{fig:CII_map} caption and the text for details).
  The figure illustrates a high incidence of strong
  \civt\ in all populations except the dwarf galaxies. 
}
\label{fig:CIV_map}
\end{figure}

To complete the comparisons, we present results on \civt\ 
where available (Figures~\ref{fig:civ_compare}, \ref{fig:CIV_map}).
Similar to the transitions known to trace the cool CGM (\ion{H}{1}
\lya, \ion{Mg}{2}, \ion{C}{2}), the \ion{C}{4} absorption strength for
QPQ exceeds that for all of the $z<1$ populations.  One notes, however,
that the LBG measurements more closely follow those of the $z \sim 2$
quasar hosts.  The differences between the populations are far more
apparent in the \ciit\ and \ion{H}{1} gas, implying that \civt\ (which likely traces
a more highly-ionized and possibly warmer gas) may be associated to
a medium that is less sensitive to the underlying halo mass.

We may also compare mass estimates for the cool CGM metals 
around these galaxies with our own QPQ estimates, 
as presented in Figure~\ref{fig:Cmass}. We remind the reader that
these estimates are strictly lower limits.
Overplotted on the QPQ result is an estimate for the cumulative mass in
metals traced by \ion{Mg}{2} absorption in the halos of LRGs
(Z14). The curve shows the best-fit halo model from their
analysis, where we have assumed
that Mg represents 6\%\ of the total metal mass (as in the Sun) and 
no correction for ionization, but note that Z14 did impose a modest
correction for line-saturation.  
The figure also shows the cumulative mass profile from the COS-Halos
sample ($z\sim0, \mlstar$ galaxies) generated in
the same manner
as the QPQ data (their survey, however, only probes to 
$\mrphys \approx \mrvir/2$).
Remarkably, the COS-Halos and LRG profiles are
very similar, suggesting a similar mass in cool gas within
\rvir\ despite the order-of-magnitude difference in their halo masses
(although formally these estimates are lower limits).
These curves intersect the profile for the $z \sim 2$ massive
galaxies\footnote{These begin at $\mrphys/\mrvir \approx 0.3$ because
  of the lack of sightlines that probe smaller scales.}, 
but the latter rises to a value several times larger than that
derived for LRGs at \rvir.  And, we emphasize that a majority of the
metal mass in the halos of LRGs occurs at $\mrphys > 250$\,kpc where 
satellite galaxies may
dominate, as suggested by the low covering
fraction of strong \ion{Mg}{2} absorption at such impact parameters
\citep[][see also the Appendix]{bc11,gc11}.

The figure also shows the total, cumulative metal mass estimated for
the cool CGM of
\lstar\ galaxies, taken from the analysis of \citet[][see also Peeples
et al.\ 2014]{werk+14}.
Specifically, we scaled their estimate for the surface density
profile of silicon (their Equation~6) by assuming silicon represents 7\%\
of the total mass in metals. 
Unlike the mass estimations that we provided first (solid curves), 
the \cite{werk+14} analysis
includes significant ionization corrections based on their detailed
modeling of the absorption lines and line-saturation has a more modest
effect. For $\mrphys > 0.55 \mrvir$, we have extrapolated 
their best-fit and show that curve as a
dotted line.  
This mass profile exceeds the QPQ profile based on our simple and
highly conservative prescription (solid black line).
The more realistic estimate for the QPQ mass profile (dashed
black line), however, exceeds the values for \lstar\ galaxies at $z \sim 0$.
These results further supports the conclusion
that the cool CGM of galaxies peaks in
$z \sim 2$, massive galaxies which host luminous quasars.  
In $\S$~\ref{sec:origin}, we speculate on possible physical
explanations.

\subsection{The Incidence of Strong \ion{Mg}{2} Absorption}
\label{sec:mgii}

The concept that the cool CGM peaks at $z \sim 2$ was previously
suggested from analysis on the incidence of strong \ion{Mg}{2}
systems \lxmg\ \citep{ppb06,menard+11,ms12}.  
These authors emphasized that the redshift evolution of \lxmg\ with
redshift tracks the cosmic star formation history (SFH)
of the universe.  We
discuss connections between SF and the CGM in the following
sub-section, but consider here the contribution of massive $z \sim 2$
galaxies to the strong \ion{Mg}{2} absorbers.   

To perform the comparison, we must first estimate the covering
fraction of strong \ion{Mg}{2} absorption around quasar halos.
Unfortunately, our dataset provides only a small sample of pairs where we
can analyze \mgiit\ directly.  An alternate approach is to adopt the
$f_C = 70\%$ covering fraction of strong \ciit\ absorption ($\mwcii >
0.2$\AA) and scale this equivalent width limit by 1.7 (see the
previous section).  We conservatively adopt $f_C = 0.7$ to $\mrphys =
200$\,kpc for $\mwmgii \ge 0.3$\AA.
Our sample does include 13~pairs with $\mrphys < 200$\,kpc and good
spectral coverage of \mgiit.  As expected,
all of the systems with strong \ciit\ absorption also
exhibit strong \mgiit\ absorption.
Of the 13 pairs, five have $\mwmgii > 1$\AA\ for a covering fraction
of $0.4 \pm 0.1$.
For the following we conservatively assume that these massive galaxies have 
$\mfc^{\rm MgII}(>1{\rm \AA}) = 0.3$ to $\mrphys = 200$\,kpc.

\begin{figure*}
\includegraphics[height=6.9in,angle=90]{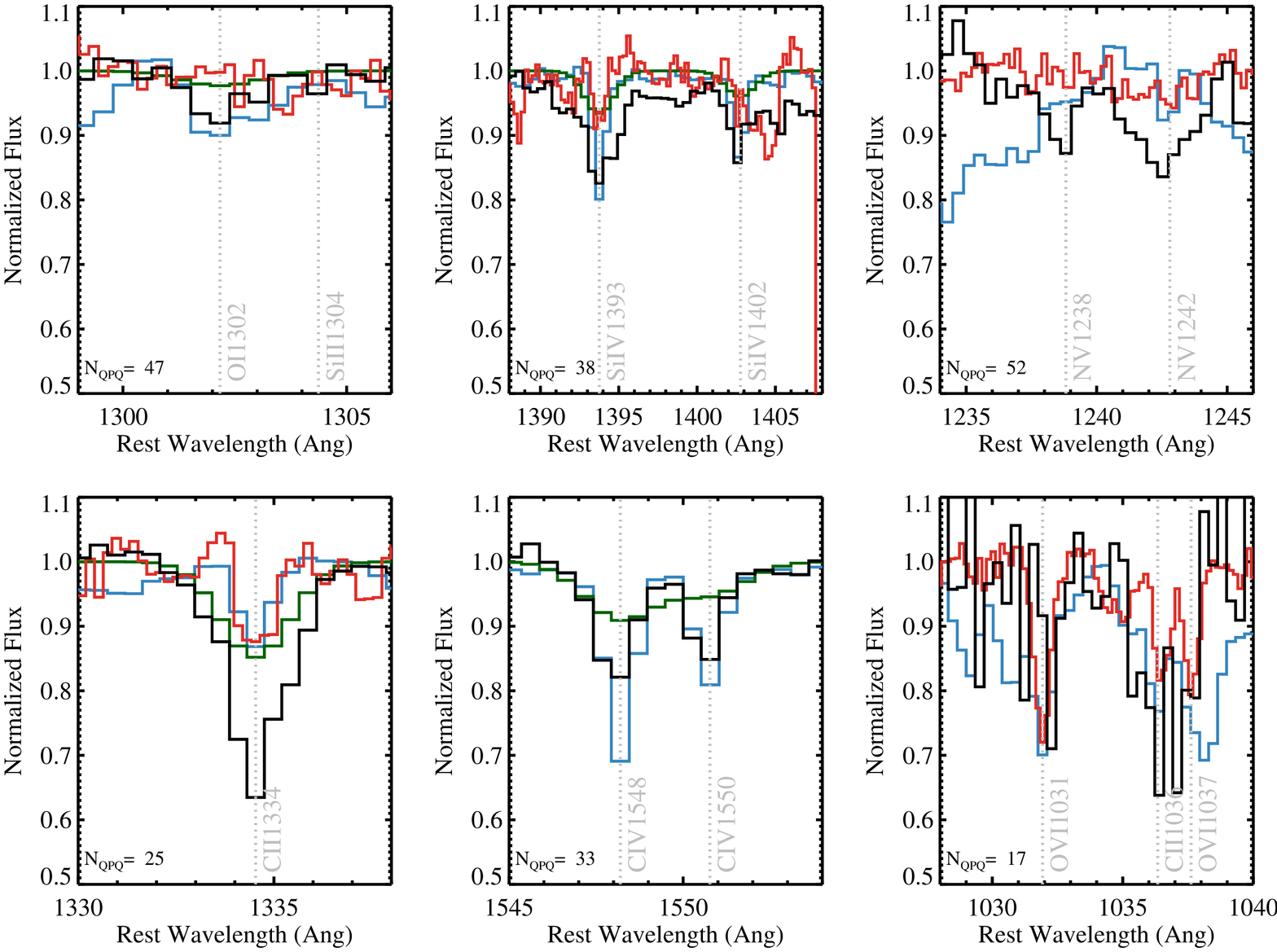}
\caption{Stacked spectra for a series of low and high-ion metal
  transitions tracing the CGM of galaxies across cosmic time: 
  QPQ (black; this program), 
  $z \sim 2$ LBGs \citep[green;][]{steidel+10}, 
  $z \sim 0$ \lstar galaxies \citep[red;][]{werk+13}, 
  and a set of LLS at $z \sim 3$ with $\mnhi < 10^{17.5-19} \cm{-2}$
  \citep[cyan;][]{fop+13}.  It is evident that the $z \sim 2$
  massive galaxies hosting quasars exhibit the strongest metal-line
  absorption at $\mrphys < 200$\,kpc of any galactic population.  
  There are indications of \ion{N}{5} and \ion{O}{6} absorption in the
  QPQ dataset, although these profiles are significantly affected by
  coincident IGM absorption. The profiles from QPQ 
  are relatively similar to that observed in the
  $z \sim 3$ LLS, suggesting the latter trace a similar CGM.
}
\label{fig:stack_compare}
\end{figure*}

The incidence of \ion{Mg}{2} absorption that 
one ascribes to the CGM of galaxies scales with their comoving
number density $\mncom^{\rm QPQ}$ and the projected effective area of
absorption from the CGM \citep{bp69},

\begin{equation}
\mlxmg = \frac{c}{H_0} \, n_{\rm com}^{\rm QPQ} \, \mfc^{\rm MgII} 
\, \pi \, R_{\rm max}^2 \;\;\; .
\end{equation}
A very conservative estimate for
$\mncom^{\rm QPQ}$ is to adopt the number density of luminous quasars
\citep[$10^{-5} \, \rm Mpc^{-3}$;][]{hrh07}, giving $\mlxmg = 0.0016$
for $\mwmgii > 1$\AA.  
Because the duty cycle of quasar activity is believed to be short
($10^{-3}$), this is a very conservative limit.  
If we instead assume that the CGM profiles
of massive galaxies are independent of whether a quasar is
shining\footnote{We also stress that the radiation field from the
  quasar is very likely to {\it reduce} the covering fraction by
  ionizing Mg to a higher ionization state than Mg$^{+}$.}, 
a more
reasonable estimate for $\mncom^{\rm QPQ}$ is the number density of
halos with mass consistent with the observed clustering, i.e. $M_{\rm
  halo} \sim
10^{12.5} \msun$ \citep{white12}.  
Here we adopt a minimum mass $\mmminq = 10^{12} \msun$
giving $\mncom^{\rm QPQ}(M_{\rm halo}>\mmminq) = 7 \sci{-4} \, \rm Mpc^{-3}$ 
at $z=2.5$, and emphasize that this number density has a steep
dependence on \mminq.   Altogether, we recover 
$\mlxmg^{\rm QPQ}(>0.3{\rm \AA}) = 0.25$ and
$\mlxmg^{\rm QPQ}(>1{\rm \AA}) = 0.11$.
The observed incidence of strong \ion{Mg}{2} systems at $z \gtrsim 2$
has been measured along quasar sightlines to be:
 $\mlxmg^{\rm IGM} (>0.3{\rm \AA}) = 0.27$ and
 $\mlxmg^{\rm IGM} (>1{\rm \AA}) = 0.11$ 
   \citep{ntr05,ppb06,ms12,seyffert+13}.
We conclude that a significant fraction and very possibly the
overwhelming majority of strong \ion{Mg}{2} systems at $z \sim 2$
occur within the halos of massive galaxies. 

One test of this conclusion is to search for the galactic counterparts
of strong \ion{Mg}{2} systems at $z=2$.  \cite{bouche12} have
performed a survey of 20~\ion{Mg}{2} systems with $\mwmgii \gtrsim
2$\AA\ at $z \approx 2$ using integral field unit (IFU) spectroscopy tuned to
the H$\alpha$ emission line.  They report an unobscured-SFR sensitivity limit 
of $\approx 3 \msun \, {\rm yr^{-1}}$
and find only 4 galaxy counterparts within their $\approx 40$\,kpc 
search radius.  At first glance their results appear inconsistent with
our assertion that strong \ion{Mg}{2} absorption is dominated by
massive halos, whose SFRs we expect to exceed $5 \msol \, {\rm
  yr^{-1}}$.  We stress, however, that the cross-section is dominated
by large impact parameters, i.e.\ at $\mrphys > 40$\,kpc which exceeds their
IFU field-of-view.  Therefore, we encourage a similar
search for galaxy counterparts to larger impact parameters.

Before concluding this sub-section, 
we also note
that one recovers a similar estimate for \lxmg\ from the CGM of
$z \sim 1$ quasars where \cite{farina+14} measure
$f_C(\mwmgii>0.6{\rm \AA}) \approx 0.2$ to $\mrphys = 200$\,kpc.
Quasar clustering at $z \approx 1$ implies a halo mass of 
$\approx 4 \sci{12} h^{-1} \msol$ \citep{richardson12,yshen13}.
Therefore, we adopt $M_{\rm min}^{\rm z1Q} = 10^{12} \msun$
and calculate $\mncom^{\rm z1Q}(M>M_{\rm min}^{\rm z1Q}) 
= 1.5 \sci{-3} \, \rm Mpc^{-3}$ at $z=1$. 
Altogether, this gives $\mlxmg^{\rm z1Q} = 0.16$ which is an
appreciable fraction ($\sim 70\%$)
of the incidence observed along quasar sightlines:
$\ell_{\rm MgII}^{IGM}(X; z=1, \mwmgii \ge 0.6\,{\rm \AA}) = 0.22$ 
\citep{seyffert+13}.

\subsection{Insights from Stacked Spectra}
\label{sec:stack}

In Figures~\ref{fig:CII_stack} and \ref{fig:CIV_stack}
($\S$\ref{sec:ew}), we presented the
averaged \ciit\ and \civt\ profiles as a function of pair separation.  These
stacked profiles illustrated the results apparent in the individual
measurements: a high incidence of strong \ciit\
absorption at $\mrphys < 200$\,kpc, the steep decline beyond, and the
sustained incidence of strong \ion{C}{4} absorption to $\mrphys
\approx 1$\,Mpc.  We now extend this exercise to additional
transitions and offer qualitative comparison to stacked spectra
generated from other CGM datasets.  Our primary interests are to assess
additional ionization states and to further accentuate aspects of the
CGM relative to other galaxy populations.  We restrict all
of the following analysis and discussion to $\mrphys < 200$\,kpc.

We have generated stacked QPQ spectra at four additional transitions -- 
\ion{O}{1}~1302, \ion{N}{5}~1238, \ion{Si}{4}~1393, and
\ion{O}{6}~1031 -- permitting an assessment of those and a few
additional, neighboring transitions.  Except for \ion{Si}{4},
these new stacks include spectral regions within the \lya\ forest
of the b/g quasar and therefore are significantly contaminated by the
coincident $z \sim 2$ IGM.  With sufficient sample size, the IGM does
average down to a relatively smooth effective opacity
\citep[QPQ6][]{bhw+13}, but for the bluest transitions considered here
the sample is small (several tens of pairs) and both \lyb\ and \lya\
opacity contributes.  These data are compromised by the IGM.

Aside from \ion{Si}{4}, we utilize the same algorithm employed in QPQ6
for continuum estimation in the \lya\ forest and to stack the data 
at the transitions of interest.  Each pair has equal weighting and each
spectrum was resampled to $\Delta v = 100\mkms$ pixels centered on the
transition before performing a straight average.  
We have also normalized each stacked spectrum to give
approximately unit flux at large offsets from the expected transitions.
The results are presented in Figure~\ref{fig:stack_compare}.

The spectra exhibit strong detections of the \ion{Si}{4} doublet
and weak but significant absorption at \ion{O}{1}~1302.
The \ion{Si}{2}~1304 transition is not positively detected which is
somewhat surprising given the detection of \ion{Si}{2}~1526
in the \civt\ stacks (Figure~\ref{fig:CIV_stack}).  
We attribute this to systematics from IGM absorption in the
\ion{Si}{2}~1304 profile and a significantly poorer S/N.
Meanwhile, 
the \ion{N}{5} and especially the \ion{O}{6} doublets suffer from
stochastic variations in the IGM absorption.  
The data suggest positive detections but we only
set generous upper limits
to the average equivalent widths of $W_{1238} < 0.2$\AA\ and 
$W_{1031} < 0.5$\AA.\footnote{ 
  Taking all of the QPQ7 sightlines that cover \ion{N}{5} (i.e.\ to
  $\mrphys = 1$\,Mpc), the stacks suffer much less from IGM
  stochasticity and we measure $\mwnv < 0.1$\AA.}
One will require higher spatial resolution and/or a much larger dataset
to more effectively probe this highly ionized gas.

Overplotted on these data are the average absorption-line profiles for
20~LLS with $\tau > 2$ 
taken from the $z \sim 3$ survey of optically thick gas by \cite{fop+13}. 
We have restricted their sample
 to systems without strong damping at
\ion{H}{1}~\lya\footnote{
  Including the LLS with larger \nhi\ values increases the average
  absorption, especially \ion{O}{1}~1302.},
corresponding to $\mnhi \lesssim 10^{20} \cm{-2}$,
because very few of the QPQ sightlines exhibit such 
high \ion{H}{1} column densities (QPQ6).
We have also smoothed stacks of their MagE spectra by 4~pixels
and resampled to 50\,\kms\ pixels.
With the exception of \ciit, where the QPQ sample shows stronger
absorption, the sets of stacked profiles from QPQ and these LLS
are qualitatively similar.
This suggests that the CGM of massive galaxies is a major contributor to
at least the set of strong metal absorption systems
in the LLS cohort.  We reached a similar conclusion in QPQ6 based on
the high covering factor and an estimated abundance for the halos
hosting quasars.

Figure~\ref{fig:stack_compare} also shows pseudo-spectra for the
CGM surrounding LBGs using the results of \cite{steidel+10}. 
These were generated by averaging the reported equivalent widths from
their stacked spectra for the $\mrphys \approx 63$ and 103\,kpc bins
(their Table~4) and representing the profiles as Gaussians with
$\sigma = 200\mkms$.\footnote{
  We have digitized their Figures~17-20 and confirm that our
  pseudo-spectra are a reasonable match to the data when compared at
  the same spectral resolution.}
Consistent with the results for the individual measurements of \ciit\
(Figure~\ref{fig:CII_compare}; QPQ5) 
and \ion{H}{1}~\lya\ (QPQ6), the absorption strength of LBGs for
$\mrphys < 120$\,kpc is weaker than the average absorption observed
for the halos hosting quasars averaged over $\mrphys \le 200$\,kpc.  
This includes both the low and high-ionization species.
These stacked spectra
confirm that the CGM of the massive galaxies hosting
quasars exceeds that of the coeval, star-forming LBG population.

Lastly, we include a set of stacked profiles from the COS-Halos
survey, generated with the same QPQ algorithms but sampled with
50\,\kms\ pixels.  The COS-Halos sample includes
all 44~galaxies studied in \cite{werk+13}.  
Consistent with Figure~\ref{fig:CII_compare},
the average \ciit\ absorption is significantly stronger in the 
CGM of massive $z \sim 2$ galaxies.  
Similarly, the COS-Halos sightlines exhibit negligible absorption from
the other low-ion transitions.  In fact, the absorption in the $z \sim
0$ CGM of \lstar\ galaxies is dominated by intermediate ions (C$^{++}$,
Si$^{++}$; not shown here) and \ion{O}{6}, in addition to the 
\ion{H}{1} Lyman series \citep{ttw+11,werk+13,tumlinson+13}.

To summarize, the average profiles of the QPQ sample exhibit 
metal absorption with systematically larger equivalent
widths than any other CGM, especially for the lower ionization states.
These stacks more resemble, at least qualitatively, the profiles
exhibited by strong LLS at similar redshift suggesting a significant
fraction of the LLS may be associated to massive halos (see also
QPQ6).

\subsection{Inferences on the Cool CGM}
\label{sec:origin}

The results presented in the previous sub-sections
demonstrate that the gas
surrounding massive, $z \sim 2$ galaxies hosting quasars represents
the pinnacle of the cool CGM.  In terms of the strength of 
\ion{H}{1} and low-ion metal absorption, the radial extent -- physical
and scaled to \rvir\ -- of this
cool gas,  and the estimated metal mass, 
the CGM of the QPQ sample represents the greatest reservoir of cool gas. 
We now explore and speculate on the conditions that favor the growth
of this massive reservoir in this environment and at this epoch.  

Fundamentally, there are two factors that set properties of the cool CGM: 
  (1) the total mass in gas and metals within the galactic halo; and 
  (2) the fraction of this medium that is in a cool phase ($T \sim
  10^4$\,K).
The first factor, we believe, is set by the mass of the dark matter
halo; the baryonic mass scales with dark matter and, presumably, the
metal mass tracks the stellar mass which is also proportional to halo
mass \citep[e.g.][]{mandelbaum05,moster+10}.
The second aspect -- cool gas fraction -- is determined by a
complex set of competing physical processes: the flow of cool gas into/out of
galaxies (processes that strip the ISM from galaxies, 
accretion of cool material from the IGM), shock heating, cooling of
warm gas in the halo, interactions of the cold phase with the predicted hot phase,
via processes like conduction, turbulent mixing/ablation, etc. 
We begin by considering several of these processes and argue that
individually they
are sub-dominant to the halo-mass dependence.

Is the cool CGM driven by quasar feedback? 
Given that the QPQ experiment uses quasars as signposts for
the locations of
massive galaxies, one might speculate that the AGN directly impacts 
the results, especially since quasar activity 
peaks at $z \sim 2-3$.  
We also note that both samples probing quasars, ours and the
experiment at $z \sim 1$ \citep{farina+14} recover the two dominant
populations regarding the cool CGM.
Indeed, quasar feedback is frequently invoked as an
effective means to transport cool, dense gas from the ISM of galaxies
\citep[e.g.][]{SilkRees98,Sijacki07,choi+14}.  
Furthermore, we have identified
examples in our own QPQ survey of extreme kinematics suggestive of
non-gravitational flows (QPQ3).  Such systems are relatively
rare, however, and are not uniquely explained 
by quasar feedback (QPQ3, QPQ8).  

While quasars undoubtedly play a role in the galaxy formation process, 
the body of data presented in the previous sub-sections
(Figures~\ref{fig:Cmass}-\ref{fig:CII_compare}) indicates that the AGN
itself has a minor role in producing the cool CGM. 
First, and most obvious, we recognize that many galaxy populations
exhibit a substantial cool CGM without a quasar
(e.g.\ Figure~\ref{fig:CII_compare});  an active galactic
nucleus is not required.
Second, as regards the QPQ measurements, our results indicate
substantial mass in \ion{H}{1} and metals to at least 200\,kpc.
Even if quasars are active for $10^8$\,yr and accelerate material to
500\kms, this gas would only reach 50\,kpc upon the termination of 
that quasar cycle.
One would need to invoke multiple quasar episodes to reach 200\,kpc. 
Third, we question whether quasar feedback could expel the total mass inferred 
(nearly that of an entire ISM) and provide a nearly unit covering fraction of cool
material (QPQ3,QPQ6; Figure~\ref{fig:Cmass}).  
Fourth, quasars are more likely
to {\it suppress} the presence of cool halo gas because their ionizing radiation field
easily over-ionizes gas to very large distances 
\citep[$\sim 1$\,Mpc; e.g. QPQ2,][]{cmb+08}. 
While we have argued that quasars emit their radiation
anisotropically, we still expect suppression within the nearby
environment. 
Fifth, while kpc-scale jets from radio-loud quasars could play a role,
only a small fraction of quasars exhibit such emission \citep[$\sim
10\%$][]{ivezic02}. 
Sixth, some models for triggering quasar activity envision
galaxy-galaxy mergers are required to funnel gas to the galaxy
centers.  While this could enhance the incidence of cool gas on scales
of tens kpc, we question whether such interactions would influence the
CGM at $\mrphys > 100$\,kpc.
We conclude that 
at most, quasar episodes help shape the nature of the CGM but that
they do not define it.


Is the cool CGM generated from flows driven by star-formation feedback, i.e.\ the
outflow of cool gas from the ISM?
The presence of heavy elements within the CGM has led many researchers
to link this gas to processes of star-formation feedback
\citep[e.g.][]{ass+05,od06,steidel+10,menard+11,sbp+12,smg+13}.
Figure~\ref{fig:Cmass} emphasizes the nearly ubiquitous presence
of metals, indicating a fraction of the observed medium has previously cycled
through a galaxy and has then been transported into the halo.
Such associations are supported by the observation of 
cool gaseous outflows from star-forming galaxies across cosmic time
\citep[e.g.][]{pks+98,martin05,wcp+09,rubin+14}.
One notes further that $z \sim 2$ corresponds to the approximate peak
in the cosmic star-formation history (SFH).  Perhaps this explains
the remarkable CGM of our $z \sim 2$ massive galaxies, i.e.\ 
one could associate the peak in SF to a peak in the cool CGM.  
By the same token, of course, a peak in the cool CGM of galaxies may
drive (i.e.\ fuel) a peak in SF activity.
So, is it the chicken or the egg?
Perhaps it is neither but both.  
To maintain even a modest SFR for Gyrs, one requires a fresh fuel
supply and, in turn, SF feedback enriches the surrounding medium.  In
this regard, elevated SF may be a natural outcome of a massive, 
cool CGM and vice-versa.  

Turning to the results presented here, the majority of
galaxies known
to exhibit a cool CGM also are actively forming stars and, presumably,
supernovae with associated feedback.\footnote{Currently, there is no galaxy population
  without a cool CGM, although such gas appears suppressed in the
  cluster environment \citep[][but see \citet{lopez08}]{yp13}.}
This is not universal, however.
The obvious exceptions
are the LRGs whose large covering fraction to strong \ion{Mg}{2}
absorption at small \rphys\ indicates gas related to the central
galaxy \citep[Figure~\ref{fig:CII_map};][Z14]{bc11}.
Furthermore, 
a cool CGM is also present in the halos of present-day, red-and-dead galaxies 
\citep{thom12}.  Furthermore, within the star-forming population, there
is little dependence of the strength of absorption
on SFR or specific SFR. For example,
Figure~\ref{fig:CII_compare} demonstrates that the present-day \lstar\
galaxies and $z \sim 2$ LBGs have CGM with similar characteristics
\citep[see also][]{chen12}.
Both populations have
comparable halo mass, yet an
order-of-magnitude difference in active SFR.  
Similarly, the galaxies hosting quasars do not exhibit evidence for
elevated SFRs but instead lay along the so-called ``main sequence'' of SF
at $z \sim 2$ \citep{rtl+13}.
Lastly, the excess in cool gas extends to
many hundreds kpc, i.e.\ too great a distance to be directly
influenced by the host galaxy. 
We conclude that the instantaneous SFR is unrelated to the 
current properties of the cool CGM.  
While SF feedback is an absolutely critical ingredient to enriching
the cool CGM, we suspect its integrated impact only contributes over long
time-scales. 
More likely, the presence of a cool CGM is a prerequisite -- but not
a necessary condition -- for active SF.

Is the cool CGM driven by the accretion of fresh, cool material from the
IGM (i.e.\ ``cold flows'' or ``streams'')?  Several lines of argument
disfavor this scenario.  First, the CGM gas is significantly enriched,
at all epochs.  Metal enrichment by the first stars and/or low mass
galaxies infalling with the streams undoubtedly generate some metals
\citep[e.g.][]{mfr01,shen+11}, but current models predict lower metallicity
flows \citep[$\approx 1/100$ solar;][]{fpk+11,freeke12,smg+13}
than observed in the CGM \citep[with important exceptions;][]{ribaudo11,fop11}.
Second, cosmological simulations predict that present-day \lstar\
galaxies have ceased to accrete dark matter 
\citep{dkm07,dmk13}. Therefore, the cool CGM of modern
galaxies is unlikely to arise primarily from ongoing cold gas accretion. 
Third, none of the existing models predict a covering fraction of cool
gas comparable to that observed in the massive galaxies hosting $z
\sim 2$ quasars
\citep[QPQ6;][]{fhp+14}. 
In short, we conclude that cold accretion alone cannot
reproduce the observed cool CGM.
We may speculate, however, that the extreme CGM exhibited by
massive $z \sim 2$ galaxies does indicate a contribution from the elevated
accretion of cool gas onto these halos.
Indeed, a supply of cool \ion{H}{1} gas may be required 
to fuel star-formation in these massive
galaxies and, especially, the extreme examples among the 
population (sub-mm galaxies).
Given the properties
of the CGM illustrated in this manuscript, one may speculate whether
the galaxies hosting quasars are poised to undergo a major burst of SF.  
We return to this point in $\S$~\ref{sec:quench}.

In lieu of quasar activity, SF feedback, cold accretion and any other
astrophysical mechanism\footnote{We add that tidal disruption of
  satellite galaxies is very unlikely to generate the entirety of the
  observed CGM but may certainly contribute.}
as the obvious dominant factor for the cool
CGM, we posit that its properties are most fundamentally driven by the halo
mass.  This conclusion follows from several of the comparisons
presented in the previous subsections
(Figures~\ref{fig:CII_compare}-\ref{fig:stack_compare}). 
Consider first the results for LBGs against those for the $z \sim 2$ massive
galaxies hosting quasars.  Aside from the luminous quasar (discussed
above), these coeval galaxy populations differ primarily in one
characteristic: halo mass, by a factor of $\approx 3-5$.  
The SFRs estimated from far-IR observations of galaxies
hosting quasars indicate they lie along the
``main-sequence'' of star-formation at $z \sim 2$ \citep{rtl+13}.  
Owing to the higher halo mass of these systems, the SFR may be
higher than the LBGs but the dependence on mass is
modest \citep[SFR~$\propto M_*^{0.57}$;][]{whitaker12}.
The difference in halo mass, meanwhile, is well established
through the clustering strength of LBGs and quasars
\citep{ass+05,white12}.  It is further confirmed 
by the difference in \ion{H}{1}
absorption strength on large scales
\citep[QPQ6,][]{rakic12,rakic+13,font13}.  
Excess \ion{H}{1} absorption extends down to the smallest
scales probed in each sample, i.e.\ well within the dark matter
halos.  And the same holds for every other ion examined in both
populations (Figure~\ref{fig:stack_compare}).  We 
conclude that the distinct CGM properties are 
manifested by the difference in halo mass.

The dominant role of halo mass is further supported through comparison
of the nearly coeval, $z \sim 0$ \lstar\ and dwarf galaxies.  One
finds that the latter exhibit a much weaker, cool CGM
(e.g.\ Figure~\ref{fig:CII_compare}).
This holds despite the fact that the latter are predicted to be
have expelled a much higher fraction of metals and gas from their
central galaxy \citep[e.g.][]{shen+14}.
In addition, comparing the cool CGM of
the \lstar\ galaxies at $z \sim 0$ with $z \sim 2$ LBGs one observes
very similar CGM properties despite
substantial differences in the active SFR and SFH.
Their halo and stellar masses, however, are comparable.
Lastly, studies that associate \ion{Mg}{2} absorption to galaxies
suggest a scaling in absorption strength with halo mass
\citep{cwt+10,churchill13}.  Therefore,
we conclude that {\it halo mass is the dominant factor in establishing
  the properties of the cool CGM.}

This final conclusion poses an immediate question:
  What about the LRGs which are the most massive halos and galaxies
  considered? 
Despite an order-of-magnitude higher halo mass, the absorption strength and
estimated metal masses for the cool CGM of LRGs do not greatly exceed
(or even match) that observed in the lower mass halos of $z \sim 0,
L^*$ galaxies, the LBGs, and quasar hosts.
This result is very unlikely to reflect lower gas
and metal masses within the LRG halos.
Indeed, X-ray
observations of small galaxy groups with virial temperatures 
characteristic of the halos of LRGs 
($k T \sim 1$\,keV)
exhibit a metal mass in Fe
{\it alone} that exceeds $10^9 \msun$ \citep{sms14}.  
Therefore, the
cool gas around LRGs likely represents a tiny fraction of the
total gas and metal budget within their dark matter halos.
One draws similar conclusions for the cool gas related to the
ICM of modern galaxy clusters \citep{lopez08,yp13}.
Within these most massive halos,
the cool CGM must be suppressed. 
An obvious explanation is that the
plasma is too hot to support a major reservoir of cool gas.  
In fact, it is very possible that nearly all of the cool gas detected
in the outer regions of these massive halos is related to material
recently stripped from or within satellite galaxies (see also
Appendix~B).

So is halo mass the only factor?  Certainly not.  Better stated, what astrophysical
processes that scale with halo mass most influence the cool CGM?
One aspect must be the greater production of metals, i.e.\ a higher
stellar mass within higher mass halos yields a higher output of
metals.  Only a fraction of these metals, however, are dispersed within the halo
and it is possible that a smaller fraction is output for higher mass
galaxies given their larger potential wells.  We suspect, therefore,
that another astrophysical factor is required.  We propose that the
principal factor is the characteristic density of halo gas \nhalo.  
The neutral fraction is sensitive to $\mnhalo^2$ via recombination
and there may be additional non-linear effects, e.g. the
self-shielding of ionizing radiation, higher cooling rates, a greater
probability for instabilities that generate cool clouds. 
One further expects that for a given mass halo that \nhalo\ is higher 
at higher redshift because the universe has a higher mean density.
In fact, with the QPQ (and to a lesser extent zQ1) samples, we may
have all of the astrophysical processes that contribute to the CGM
working together to yield the observed pinnacle of the CGM.



\begin{figure}
\includegraphics[width=3.7in,angle=90]{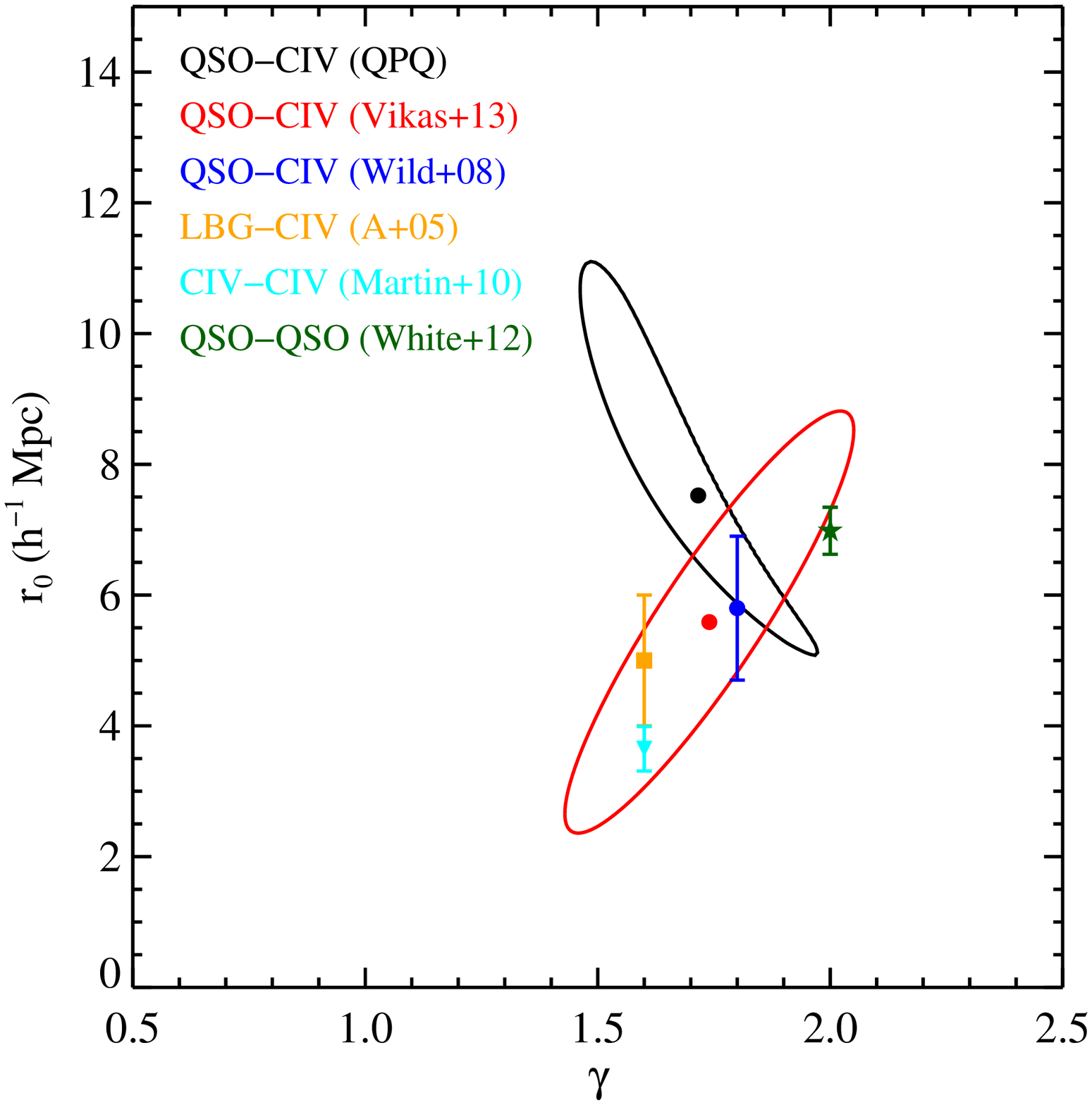}
\caption{Comparison of constraints on the two-point 
  quasar-quasar \citep[green;][]{white12},
  \ion{C}{4}-\ion{C}{4} \citep[cyan;][]{martin+10}, 
  and quasar-\ion{C}{4} \citep[blue,black,red; this
  paper,][]{vikas+13,wkw+08}
  cross-correlation functions.  
  The latter exhibit good agreement at $r_0 \approx 6 \mhMpc$ and
  $\gamma \approx 1.8$.
  We also present the LBG-\ion{C}{4}
  cross-correlation function \cite{adel05} for $N_{\rm CIV} \ge
  10^{12} \cm{-2}$ (yellow). 
}
\label{fig:civ_cluster}
\end{figure}

\subsection{The Clustering of \ion{C}{4}} 
\label{sec:CIV}

In $\S$~\ref{sec:cross}, we presented analysis on the clustering of
strong ($\mwciv > 0.3$\AA) \ion{C}{4} absorption 
with quasars, as
estimated from the projected cross-correlation on scales $\mrcom < 3
\mhMpc$.   Our observations are well-described by a power-law, cross-correlation
function $\mxic(r) = (r/r_0)^{-\gamma}$ with 
$r_0 = \vro$ and $\gamma = \vgmm$ (Figures~\ref{fig:contour},
\ref{fig:trans_corr}).  The large clustering amplitude, comparable to
that measured from quasar clustering, impies that the \ion{C}{4} gas
traces the same large-scale over-densities as the halos 
manifesting $z \sim 2$ quasars.
We now explore further the implications of this result. 

Previous studies have examined the clustering of \ion{C}{4} gas with
quasars on larger scales 
\citep[$\mrcom > 10 \mhMpc$;][]{wkw+08,vikas+13}, 
the clustering with LBGs on comparable scales to our experiment \citep{ass+05,cwg+06}, 
and also the \ion{C}{4} auto-correlation function
\citep{qb98,martin+10}.   All have reported significant clustering
amplitudes, as summarized in Figure~\ref{fig:civ_cluster}.  
These results have been interpreted as evidence that the
\ion{C}{4} gas is
physically associated to galaxies, i.e.\ the \ion{C}{4} gas resides
within the dark matter halos of galaxies whose clustering amplitude
scales with mass. 
Indeed, \cite{vikas+13} estimated that strong \ion{C}{4} ($\mwciv >
1$\AA) at $z \sim 2.5$ occurs primarily within galaxies hosted by 
dark matter halos with $M_{\rm halo} > 10^{12} \msun$.  This estimate follows from
the cosmological paradigm which predicts that baryons trace the
gravitational potential generated by dark matter and may be related to
the latter in the regime of linear bias by a bias factor $b$,
$\xi(r) = b^2 \xi_{\rm DM} (r)$ \citep[e.g.][]{dp77,sw98}, which is
usually valid at large scales $\gg 1$\,Mpc \citep[but see][]{tejos+14}.
When combined with 
independent estimates of the quasar auto-correlation function, one may
assess the bias factor and estimate the halo mass.
Our results on the quasar-\ion{C}{4} clustering, 
obtained on smaller (and presumably non-linear) scales
serve to confirm the \cite{vikas+13} inferences.  
The results imply that the \ion{C}{4} absorption associated to quasars
is characteristic of all massive halos.

One may also gain insight into the nature of absorption-line systems
through comparisons of the cross-correlation functions with quasars.
Taking the results presented here on \ion{C}{4} and those from QPQ6,
we have (in order): 
 $r_0^{\rm DLA} = 3.9 \pm 2.3 \, \mhMpc$, 
 $r_0^{\rm CIV} = \vro$, 
 $r_0^{\rm LLS} = 12.5^{+2.7}_{-1.4} \, \mhMpc$, 
and
 $r_0^{\rm SLLS} = 14.0^{+7.6}_{-2.7} \, \mhMpc$.
It is astonishing (and possibly of concern) that the absorption
systems known to trace the highest gas densities (DLAs) exhibit the
smallest clustering amplitude.
These amplitudes, which we emphasize are generally
evaluated from clustering in the non-linear regime, suggest that the
strongest \ion{H}{1} absorption is less biased to the massive
halos.  Furthermore, our new measurements on \ion{C}{4} lend
additional support to the conclusion of QPQ6 that optically thick gas
with $\mnhi < 10^{20} \cm{-2}$ is highly biased to the environment of
massive halos.  These results demand greater inspection within
cosmological simulations, ideally incorporating a proper treatment
of radiative transfer.

The excess of strong \ion{C}{4} absorbers around quasars implies that
massive galaxies may contribute a significant fraction of the $\mwciv
\ge 0.3$\AA\ systems discovered along random quasar sightlines.
Indeed, we estimated in $\S$~\ref{sec:mgii} that galaxies with
$M_{\rm halo} \ge 10^{12} \msun$ can reproduce the observed incidence
of strong \ion{Mg}{2} absorbers.  Integrating our cross-correlation
function to $\mrphys = 200$\,kpc (and never allowing the covering fraction
to exceed unity), we recover $\mlxciv = 0.22$ or only $\approx 10\%$ of the
random incidence.  Therefore, we find that if strong \ion{C}{4} occurs
primarily within the CGM of galaxies than lower mass halos must
dominate.  This implies some tension with the clustering estimates
described above, although the amplitude measured on large-scales
(i.e.\ in the linear regime) corresponds to $\mwciv > 1$\,\AA.

\subsection{Do Quasars Quench Star Formation?}
\label{sec:quench}


A principal focus of galaxy formation studies, possibly the primary
focus, has been to identify and understand the physical process(es)
that convert a blue, star-forming galaxy into a `red-and-dead' system
\citep[e.g.][]{faber07}.
In recent years, it has become fashionable to invoke feedback from AGN
-- thermal, radiative, mechanical -- as the primary mechanism 
\citep[e.g.][]{Springel05,choi+14}.
Theorists have been motivated, at least in part, by the empirical
revelation that all massive galaxies host super-massive black holes.
One further recognizes that the astrophysics of gas accretion onto a
supermassive black hole produces a tremendous output of energy that,
if tapped, could easily influence the surrounding galaxy.
A full study of AGN quenching in the context of QPQ may be the focus
of a future manuscript;  here, we offer a few comments and
speculations.

Directly, we contend that the principal results of our QPQ experiment
-- the omnipresence of cool, enriched gas transverse to the sightline
to quasars on scales to $\approx 200$\,kpc -- {\it directly contradict
  the thesis that quasars quench star-formation}.
On the contrary, this cool CGM represents a tremendous reservoir for
current and future star-formation.
Our conservative estimate gives $\mmccgm > 10^{10} \msun$ of cool gas
within these halos.  At a temperature of $T \approx 10^4$\,K, this gas
need cool no further to be accreted onto the central galaxy, and 
following \cite{mb04} we may estimate the infall time\footnote{The
  shortest timescale of interest is the dynamical, infall time of
  $\sim 10^8$~yr.}.  Those authors
considered two processes:  
  (i) ram pressure drag as the gas moves through a hot gas halo 
  and (ii) cloud-cloud collisions \citep[see also][]{mcd99}.
Their calculations for a low-$z$ halo with $M_{\rm halo} \approx
10^{12} \msun$ and $\mmccgm = 2\sci{10} \msun$ imply 
that cloud-cloud collisions dominate and that the infall time is
approximately $\tau_{\rm in} \approx 1$\,Gyr.  For our higher mass
and higher redshift halos, the
infall times should be smaller.  Therefore, we estimate a cool gas inflow rate
of ${\dot M}_{\rm CGM}^{\rm cool} \approx \mmccgm / \tau_{\rm in} \approx 10 \msun \,
{\rm yr^{-1}} (\mmccgm/10^{10} \msun) (\tau_{\rm in}/1\,{\rm
  Gyr})^{-1}$.  
Allowing for a considerably higher cool gas mass
(probable) and/or a shorter infall time (also probable), we infer a
mass infall rate that likely exceeds 100\,$\msun \, {\rm yr^{-1}}$.
This is comparable to the SFR estimated for galaxies hosting quasars
\citep{rtl+13}.   We conclude that quasars are not
quenching\footnote{It is, however, without a doubt that quasars
  significantly affect the gas that they shine upon;  this feedback,
  however, must be anisotropic.}
star-formation at $z \approx 2$ and that their halos contain a
sufficient reservoir of cool gas to fuel the observed SFRs for at
least 1\,Gyr.

We may also offer a direct test
to a recent prediction on AGN quenching from the literature.
\cite{choi+14} have implemented several forms of AGN feedback into a
set of cosmological simulations studying the formation of early-type
galaxies.  Their analysis favors a scenario of radiative+mechanical
feedback and they emphasize that thermal feedback over-predicts the
observed X-ray luminosities of modern, early-type galaxies.
From these models, they predicted the total gas surface density
profiles $\Sigma_{\rm gas}(r)$ for the galaxies at $z=1.5$, estimating
$\Sigma_{\rm gas} \approx 10^{5.5} \msun \, {\rm kpc^{-2}}$ 
at $r \approx 100$\,kpc (their Figure~7).  
This corresponds to $N_{\rm H} \approx 10^{19} \cm{-2}$ 
assuming that helium contributes 25\%\ of the mass.
In QPQ3, we measured $\mnhi = 10^{19.7} \cm{-2}$ for an optically
thick sightline at $\mrphys = 108$\,kpc and 
$N_{\rm H} = 10^{21} \cm{-2}$ based on photo-ionization modeling.
This exceeds the \cite{choi+14} estimate by two orders-of-magnitude.
In QPQ5, we argued that $N_{\rm H} > 10^{20} \cm{-2}$ based on the
very strong \ion{C}{2} absorption.  The results of this manuscript
only strengthen the result.   We conclude that the \cite{choi+14}
prescription removes too much gas from galactic halos and caution
against similar AGN models for quenching star-formation.


\section{Summary and Concluding Remarks}
\label{sec:summary}


From a sample of \npair\ projected quasar pairs, we have analyzed the
incidence and absorption strength of the \ciit\ and \civt\ transitions
associated to the environment of the f/g quasar.  These pairs have
physical separations $\mrphys \approx 39$\,kpc to 1\,Mpc at the f/g
redshift \zfg, an average f/g quasar redshift of $\langle \mzfg
\rangle = 2.41$, and b/g quasar spectra with S/N~$\ge 9.5$\,per \AA\ at
the f/g \ion{H}{1} \lya\ transition.    
Measurements of the two-point correlation function of $z \sim 2$
quasars imply they inhabit dark matter halos with mass $M_{\rm halo} =
10^{12.5} \msun$ and a characteristic virial radius of $\mrvir \approx
160$\,kpc. 
Following our work on neutral hydrogen (QPQ6),
we adopt an analysis window $\delta v = \pm 1500\mkms$ to account for
uncertainties in \zfg.

We summarize our primary results and conclusions as:

\begin{itemize}
 \item The galactic halos of luminous, $z \sim 2$ quasars frequently
   exhibit strong \ciit\ absorption.  For $\mrphys \le \mrvir$, we
   measure the covering fraction for $\mwcii \ge 0.2$\AA\ to be
   $f_C^{1334} = \vfct$.  Beyond 200\,kpc, the incidence rapidly
   declines and we measure $f_C^{1334} = 0.08 \pm 0.03$.  
   We associate the observed \ciit\ absorption to the cool, CGM
   surrounding $z \sim 2$ galaxies hosting quasars. 

 \item The sightlines also frequently exhibit strong 
   \civt\ absorption, with an excess incidence relative to random
   sightlines to at least 1\,Mpc.  We measure the cross-correlation
   between strong \civt\ absorption ($\mwciv \ge 0.3$\AA) and quasars
   for $R < 3 h^{-1}$ comoving Mpc.  Adopting a standard power-law
   description,$\mxic(r) = (r/r_0)^{-\gamma}$, a maximum likelihood
   analysis gives $r_0 = \vro$ and $\gamma = \vgmm$.
   These values are remarkably consistent with measurements from much
   larger (i.e.\ non-linear) scales \citep{vikas+13}.  This
   implies that the majority of \ion{C}{4} absorption 
   may be associated to the massive halos ($M > 10^{12} \msun$)
   of galaxies that cluster with the quasar host. 

 \item Integrating column density estimates from the saturated \ciit\
   profiles, we set a strict lower limit to the metal mass within
   \rvir\ of $M_{\rm cool metal}^{\rm QPQ} > 10^7 \msun$.  Adopting
   conservative saturation and ionization corrections, we
   conservatively estimate $M_{\rm cool metal}^{\rm QPQ} > 10^8
   \msun$.  These values exceed estimates for the metal-mass of the
   cool CGM around present-day $L^*$ galaxies \citep{werk+14}. 
   These metals likely represent the early enrichment of halo gas
   predicted by chemical evolution models that study the formation and
   enrichment of the intragroup and intracluster medium. 

 \item We study the integrated incidence of strong, low-ion absorption arising
   from massive, $z \sim 2$ halos.  Under the assumption that all
   halos with mass $M_{\rm halo} > 10^{12} \msun$ exhibit properties
   similar to the QPQ sample and that strong \ion{C}{2} absorption
   indicates strong \ion{Mg}{2} absorption (as observed), we estimate
   the incidence of strong \ion{Mg}{2} lines to be
   $\mlxmg^{\rm QPQ}(>1{\rm \AA}) = 0.11$ for gas within 200\,kpc of
   these massive halos.  This integrated incidence is consistent with
   surveys of strong \ion{Mg}{2} along random sightlines
   \citep{seyffert+13}, implying that the majority of such absorbers
   are physically associated to massive halos.

 \item We compare the incidence and absorption strength of low-ions
   (\ion{C}{2}, \ion{Mg}{2}) between the QPQ-CGM and that for LBGs 
   at $z \sim 2$, quasar hosts at $z\sim 1$, LRGs at $z \sim 0.5$ and
   $L^*$ and sub-$L^*$ galaxies at $z\sim 0$.  
   On a physical scale,
   $\mrphys < 200$\,kpc, we conclude that the massive $z \sim 2$ halos
   hosting quasars represents the pinnacle of the cool CGM.  This
   conclusion is supported by our previous analysis of
   \ion{H}{1} \lya\ absorption (QPQ6).

 \item We examine the characteristics of the cool CGM for galaxies
   ranging from $z \sim 0$ dwarfs to massive,  $z \sim 0.5$ LRGs, and  
   the halos hosting quasars at $z \sim 1$ and $z \sim 2$.  We argue
   that the cool CGM is not dominated by astrophysical
   processes related to active star-formation or AGN feedback, nor
   cold gas accretion.  Instead, we propose that the cool CGM
   properties track halo mass with more massive halos exhibiting a
   more substantial cool CGM.  Of course, SFR and gas accretion are
   expected to scale with halo mass but we assert
   that the cool CGM primarily tracks the cumulative growth of
   galactic halos and their integrated enrichment history.

\end{itemize}

The results presented in this manuscript and previous publications of
our QPQ survey have defined the incidence and absorption strength of
cool gas surrounding the massive galaxies hosting luminous $z \sim 2$
quasars.  To our surprise, these galaxies exhibit a cool CGM with
properties exceeding any previous set of galaxies, including those
generated within cosmological simulations.  These results, therefore,
inspire the following lines of future inquiry:

\begin{enumerate}
\item A study of the CGM at high spectral resolution to resolve
  properties of the gas:  \ion{H}{1} surface density, kinematics,
  ionization state, metallicity, etc.  The majority of data analysis
  to date has been on low-resolution spectra with limited diagnostic
  power.  A manuscript presenting 11~sightlines with data similar to
  QPQ3 is forthcoming (Lau et al.\ in prep.).
\item A survey of the CGM surrounding halos with comparable mass to
  the hosts of $z \sim 2$ quasars ($\sim 10^{12.5} \msun$) but without
  a luminous AGN.  This would provide a test of our working hypothesis
  that the presence of a luminous quasar has minimal effect on the gas
  at CGM scales (hundreds kpc).  One viable approach is to survey
  the CGM of sub-mm galaxies, whose two-point clustering measurements
  are comparable to luminous quasars \citep{white12}.  Ongoing efforts
  to identify such sources across a wide area of the sky may yield
  several tens of projected SMG-quasar pairs for study.  
\item Study the CGM and surrounding IGM with narrow-band and IFU
  imaging  \citep{cantalupo14,martin14a}. 
\item New theoretical studies on the complex astrophysics of the CGM
  at $z > 2$.  It is now evident that even state-of-the-art zoom-in
  simulations have insufficient resolution to properly capture the
  astrophysics of this cool gas \citep[e.g.,
  QPQ6][]{crighton+14,fhp+14}. 
 Researchers may need to input
  sub-grid models, akin to those for star-formation and associated
  feedback, to examine global trends and establish the implications
  for galaxy formation. 
  
\end{enumerate}

\acknowledgements

JXP dedicates this manuscript to the memory of Arthur M. Wolfe, who
inspired his careers in the field.
JXP would also like to thank the Aspen Institute for Physics where
discussions at the Winter Conference stimulated several aspects of this work.
JXP and ML acknowledge support from the National
Science Foundation (NSF) grant AST-1010004 and AST-1412981. 
JXP thanks the Alexander
von Humboldt foundation for a visitor fellowship to the MPIA where
part of this work was performed, as well as the staff at MPIA for
their hospitality during his visits.
JFH acknowledges generous support from the Alexander von Humboldt
foundation in the context of the Sofja Kovalevskaja Award. The
Humboldt foundation is funded by the German Federal Ministry for
Education and Research.  
We thank D. Bowen for providing the \fcs\ values presented in the
Appendix and R. Bordoloi for providing measurements from the
COS-Dwarfs survey in advance of publication.  We further acknowledge 
R. Yates for providing outputs from his chemical evolution models in
advance of publication. 
We acknowledge the contributions of Sara Ellison, George Djorgovski,
Crystal Martin, Rob Simcoe, and Kate Rubin in obtaining some of the spectra analyzed in
this manuscript.

Much of the data presented herein were obtained at the W.M. Keck
Observatory, which is operated as a scientific partnership among the
California Institute of Technology, the University of California, and
the National Aeronautics and Space Administration. The Observatory was
made possible by the generous financial support of the W.M. Keck
Foundation.  Some of the Keck data were obtained through the NSF
Telescope System Instrumentation Program (TSIP), supported by AURA
through the NSF under AURA Cooperative Agreement AST 01-32798 as
amended.

Some of the data herein were obtained at the Gemini Observatory, which
is operated by the Association of Universities for Research in
Astronomy, Inc., under a cooperative agreement with the NSF on behalf
of the Gemini partnership: the NSF (United
States), the Science and Technology Facilities Council (United
Kingdom), the National Research Council (Canada), CONICYT (Chile), the
Australian Research Council (Australia), Minist\'{e}rio da
Ci\^{e}ncia, Tecnologia e Inova\c{c}\~{a}o (Brazil) and Ministerio de
Ciencia, Tecnolog\'{i}a e Innovaci\'{o}n Productiva (Argentina). 

The authors wish to recognize and acknowledge the very
significant cultural role and reverence that the summit of Mauna Kea
has always had within the indigenous Hawaiian community. We are most
fortunate to have the opportunity to conduct observations from this
mountain.


\begin{thebibliography}{164}
\expandafter\ifx\csname natexlab\endcsname\relax\def\natexlab#1{#1}\fi

\bibitem[{{Abazajian} {et~al.}(2009){Abazajian}, {Adelman-McCarthy},
  {Ag{\"u}eros}, {Allam}, {Allende Prieto}, {An}, {Anderson}, {Anderson},
  {Annis}, {Bahcall}, {Bailer-Jones}, {Barentine}, {Bassett}, {Becker},
  {Beers}, {Bell}, {Belokurov}, {Berlind}, {Berman}, {Bernardi}, {Bickerton},
  {Bizyaev}, {Blakeslee}, {Blanton}, {Bochanski}, {Boroski}, {Brewington},
  {Brinchmann}, {Brinkmann}, {Brunner}, {Budav{\'a}ri}, {Carey}, {Carliles},
  {Carr}, {Castander}, {Cinabro}, {Connolly}, {Csabai}, {Cunha}, {Czarapata},
  {Davenport}, {de Haas}, {Dilday}, {Doi}, {Eisenstein}, {Evans}, {Evans},
  {Fan}, {Friedman}, {Frieman}, {Fukugita}, {G{\"a}nsicke}, {Gates},
  {Gillespie}, {Gilmore}, {Gonzalez}, {Gonzalez}, {Grebel}, {Gunn},
  {Gy{\"o}ry}, {Hall}, {Harding}, {Harris}, {Harvanek}, {Hawley}, {Hayes},
  {Heckman}, {Hendry}, {Hennessy}, {Hindsley}, {Hoblitt}, {Hogan}, {Hogg},
  {Holtzman}, {Hyde}, {Ichikawa}, {Ichikawa}, {Im}, {Ivezi{\'c}}, {Jester},
  {Jiang}, {Johnson}, {Jorgensen}, {Juri{\'c}}, {Kent}, {Kessler}, {Kleinman},
  {Knapp}, {Konishi}, {Kron}, {Krzesinski}, {Kuropatkin}, {Lampeitl},
  {Lebedeva}, {Lee}, {Lee}, {Leger}, {L{\'e}pine}, {Li}, {Lima}, {Lin}, {Long},
  {Loomis}, {Loveday}, {Lupton}, {Magnier}, {Malanushenko}, {Malanushenko},
  {Mandelbaum}, {Margon}, {Marriner}, {Mart{\'{\i}}nez-Delgado}, {Matsubara},
  {McGehee}, {McKay}, {Meiksin}, {Morrison}, {Mullally}, {Munn}, {Murphy},
  {Nash}, {Nebot}, {Neilsen}, {Newberg}, {Newman}, {Nichol}, {Nicinski},
  {Nieto-Santisteban}, {Nitta}, {Okamura}, {Oravetz}, {Ostriker}, {Owen},
  {Padmanabhan}, {Pan}, {Park}, {Pauls}, {Peoples}, {Percival}, {Pier}, {Pope},
  {Pourbaix}, {Price}, {Purger}, {Quinn}, {Raddick}, {Fiorentin}, {Richards},
  {Richmond}, {Riess}, {Rix}, {Rockosi}, {Sako}, {Schlegel}, {Schneider},
  {Scholz}, {Schreiber}, {Schwope}, {Seljak}, {Sesar}, {Sheldon}, {Shimasaku},
  {Sibley}, {Simmons}, {Sivarani}, {Smith}, {Smith}, {Smol{\v c}i{\'c}},
  {Snedden}, {Stebbins}, {Steinmetz}, {Stoughton}, {Strauss}, {Subba Rao},
  {Suto}, {Szalay}, {Szapudi}, {Szkody}, {Tanaka}, {Tegmark}, {Teodoro},
  {Thakar}, {Tremonti}, {Tucker}, {Uomoto}, {Vanden Berk}, {Vandenberg},
  {Vidrih}, {Vogeley}, {Voges}, {Vogt}, {Wadadekar}, {Watters}, {Weinberg},
  {West}, {White}, {Wilhite}, {Wonders}, {Yanny}, {Yocum}, {York}, {Zehavi},
  {Zibetti}, \& {Zucker}}]{sdssdr7}
{Abazajian}, K.~N., {et~al.} 2009, \apjs, 182, 543

\bibitem[{{Adelberger} {et~al.}(2005{\natexlab{a}}){Adelberger}, {Shapley},
  {Steidel}, {Pettini}, {Erb}, \& {Reddy}}]{ass+05}
{Adelberger}, K.~L., {Shapley}, A.~E., {Steidel}, C.~C., {Pettini}, M., {Erb},
  D.~K., \& {Reddy}, N.~A. 2005{\natexlab{a}}, \apj, 629, 636

\bibitem[{{Adelberger} {et~al.}(2005{\natexlab{b}}){Adelberger}, {Steidel},
  {Pettini}, {Shapley}, {Reddy}, \& {Erb}}]{adel05}
{Adelberger}, K.~L., {Steidel}, C.~C., {Pettini}, M., {Shapley}, A.~E.,
  {Reddy}, N.~A., \& {Erb}, D.~K. 2005{\natexlab{b}}, \apj, 619, 697

\bibitem[{{Adelberger} {et~al.}(2003){Adelberger}, {Steidel}, {Shapley}, \&
  {Pettini}}]{adel03}
{Adelberger}, K.~L., {Steidel}, C.~C., {Shapley}, A.~E., \& {Pettini}, M. 2003,
  \apj, 584, 45

\bibitem[{{Ahn} {et~al.}(2012){Ahn}, {Alexandroff}, {Allende Prieto},
  {Anderson}, {Anderton}, {Andrews}, {Aubourg}, {Bailey}, {Balbinot}, {Barnes},
  \& et~al.}]{boss_dr9}
{Ahn}, C.~P., {et~al.} 2012, \apjs, 203, 21

\bibitem[{{Akerman} {et~al.}(2004){Akerman}, {Carigi}, {Nissen}, {Pettini}, \&
  {Asplund}}]{acn+04}
{Akerman}, C.~J., {Carigi}, L., {Nissen}, P.~E., {Pettini}, M., \& {Asplund},
  M. 2004, \aap, 414, 931

\bibitem[{{Anderson} {et~al.}(2013){Anderson}, {Bregman}, \&
  {Dai}}]{anderson13}
{Anderson}, M.~E., {Bregman}, J.~N., \& {Dai}, X. 2013, \apj, 762, 106

\bibitem[{{Andreon}(2012)}]{andreon12}
{Andreon}, S. 2012, \aap, 546, A6

\bibitem[{{Andrews} {et~al.}(2013){Andrews}, {Barrientos}, {L{\'o}pez}, {Lira},
  {Padilla}, {Gilbank}, {Lacerna}, {Maureira}, {Ellingson}, {Gladders}, \&
  {Yee}}]{andrews13}
{Andrews}, H., {et~al.} 2013, \apj, 774, 40

\bibitem[{{Antonucci}(1993)}]{Anton93}
{Antonucci}, R. 1993, \araa, 31, 473

\bibitem[{{Arrigoni} {et~al.}(2010){Arrigoni}, {Trager}, \&
  {Somerville}}]{arrigoni10b}
{Arrigoni}, M., {Trager}, S.~C., \& {Somerville}, R.~S. 2010, ArXiv e-prints

\bibitem[{{Arrigoni-Battaia} {et~al.}(2014){Arrigoni-Battaia}, {Hennawi}, \&
  {Prochaska}}]{fab14}
{Arrigoni-Battaia}, F., {Hennawi}, J.~F., \& {Prochaska}, J.~X. 2014, MNRAS,
  submitted

\bibitem[{{Bahcall} \& {Peebles}(1969)}]{bp69}
{Bahcall}, J.~N., \& {Peebles}, P.~J.~E. 1969, \apjl, 156, L7+

\bibitem[{{Bahcall} \& {Spitzer}(1969)}]{bs69}
{Bahcall}, J.~N., \& {Spitzer}, L.~J. 1969, \apj, 156, L63

\bibitem[{{Bajtlik} {et~al.}(1988){Bajtlik}, {Duncan}, \& {Ostriker}}]{bdo88}
{Bajtlik}, S., {Duncan}, R.~C., \& {Ostriker}, J.~P. 1988, \apj, 327, 570

\bibitem[{{Baldi} {et~al.}(2012){Baldi}, {Ettori}, {Molendi}, {Balestra},
  {Gastaldello}, \& {Tozzi}}]{baldi+12}
{Baldi}, A., {Ettori}, S., {Molendi}, S., {Balestra}, I., {Gastaldello}, F., \&
  {Tozzi}, P. 2012, \aap, 537, A142

\bibitem[{{Becker} {et~al.}(2013){Becker}, {Hewett}, {Worseck}, \&
  {Prochaska}}]{bhw+13}
{Becker}, G.~D., {Hewett}, P.~C., {Worseck}, G., \& {Prochaska}, J.~X. 2013,
  \mnras, 430, 2067

\bibitem[{{Bergeron}(1986)}]{bergeron86}
{Bergeron}, J. 1986, \aap, 155, L8

\bibitem[{{Bielby} {et~al.}(2013){Bielby}, {Hill}, {Shanks}, {Crighton},
  {Infante}, {Bornancini}, {Francke}, {H{\'e}raudeau}, {Lambas}, {Metcalfe},
  {Minniti}, {Padilla}, {Theuns}, {Tummuangpak}, \& {Weilbacher}}]{bielby+13}
{Bielby}, R., {et~al.} 2013, \mnras, 430, 425

\bibitem[{{Boksenberg} {et~al.}(2003){Boksenberg}, {Sargent}, \&
  {Rauch}}]{boks03}
{Boksenberg}, A., {Sargent}, W.~L.~W., \& {Rauch}, M. 2003, ArXiv Astrophysics
  e-prints

\bibitem[{{Bonaparte} {et~al.}(2013){Bonaparte}, {Matteucci}, {Recchi},
  {Spitoni}, {Pipino}, \& {Grieco}}]{bonaparte+13}
{Bonaparte}, I., {Matteucci}, F., {Recchi}, S., {Spitoni}, E., {Pipino}, A., \&
  {Grieco}, V. 2013, \mnras, 435, 2460

\bibitem[{{Bordoloi} {et~al.}(2014){Bordoloi}, {Tumlinson}, {Werk},
  {Oppenheimer}, {Peeples}, {Prochaska}, {Tripp}, {Katz}, {Dav{\'e}}, {Fox},
  {Thom}, {Ford}, {Weinberg}, {Burchett}, \& {Kollmeier}}]{bordoloi14}
{Bordoloi}, R., {et~al.} 2014, ArXiv e-prints

\bibitem[{{Borthakur} {et~al.}(2013){Borthakur}, {Heckman}, {Strickland},
  {Wild}, \& {Schiminovich}}]{borthakur+13}
{Borthakur}, S., {Heckman}, T., {Strickland}, D., {Wild}, V., \&
  {Schiminovich}, D. 2013, \apj, 768, 18

\bibitem[{{Bouch{\'e}} {et~al.}(2012){Bouch{\'e}}, {Murphy}, {P{\'e}roux},
  {Contini}, {Martin}, {Forster Schreiber}, {Genzel}, {Lutz}, {Gillessen},
  {Tacconi}, {Davies}, \& {Eisenhauer}}]{bouche12}
{Bouch{\'e}}, N., {et~al.} 2012, \mnras, 419, 2

\bibitem[{{Bovy} {et~al.}(2011){Bovy}, {Hennawi}, {Hogg}, {Myers},
  {Kirkpatrick}, {Schlegel}, {Ross}, {Sheldon}, {McGreer}, {Schneider}, \&
  {Weaver}}]{bovy11}
{Bovy}, J., {et~al.} 2011, \apj, 729, 141

\bibitem[{{Bovy} {et~al.}(2012){Bovy}, {Myers}, {Hennawi}, {Hogg}, {McMahon},
  {Schiminovich}, {Sheldon}, {Brinkmann}, {Schneider}, \& {Weaver}}]{bovy12}
---. 2012, \apj, 749, 41

\bibitem[{{Bowen} \& {Chelouche}(2011)}]{bc11}
{Bowen}, D.~V., \& {Chelouche}, D. 2011, \apj, 727, 47

\bibitem[{{Bowen} {et~al.}(2006){Bowen}, {Hennawi}, {M{\'e}nard}, {Chelouche},
  {Inada}, {Oguri}, {Richards}, {Strauss}, {Vanden Berk}, \& {York}}]{bhm+06}
{Bowen}, D.~V., {et~al.} 2006, \apjl, 645, L105

\bibitem[{{Cantalupo} {et~al.}(2014){Cantalupo}, {Arrigoni-Battaia},
  {Prochaska}, {Hennawi}, \& {Madau}}]{cantalupo14}
{Cantalupo}, S., {Arrigoni-Battaia}, F., {Prochaska}, J.~X., {Hennawi}, J.~F.,
  \& {Madau}, P. 2014, \nat, 506, 63

\bibitem[{{Chelouche} {et~al.}(2008){Chelouche}, {M{\'e}nard}, {Bowen}, \&
  {Gnat}}]{cmb+08}
{Chelouche}, D., {M{\'e}nard}, B., {Bowen}, D.~V., \& {Gnat}, O. 2008, \apj,
  683, 55

\bibitem[{{Chen} {et~al.}(2010{\natexlab{a}}){Chen}, {Helsby}, {Gauthier},
  {Shectman}, {Thompson}, \& {Tinker}}]{chg+10}
{Chen}, H., {Helsby}, J.~E., {Gauthier}, J., {Shectman}, S.~A., {Thompson},
  I.~B., \& {Tinker}, J.~L. 2010{\natexlab{a}}, \apj, 714, 1521

\bibitem[{{Chen}(2012)}]{chen12}
{Chen}, H.-W. 2012, \mnras, 427, 1238

\bibitem[{{Chen} {et~al.}(2001){Chen}, {Lanzetta}, \& {Webb}}]{clw01}
{Chen}, H.-W., {Lanzetta}, K.~M., \& {Webb}, J.~K. 2001, \apj, 556, 158

\bibitem[{{Chen} {et~al.}(2010{\natexlab{b}}){Chen}, {Wild}, {Tinker},
  {Gauthier}, {Helsby}, {Shectman}, \& {Thompson}}]{cwt+10}
{Chen}, H.-W., {Wild}, V., {Tinker}, J.~L., {Gauthier}, J.-R., {Helsby}, J.~E.,
  {Shectman}, S.~A., \& {Thompson}, I.~B. 2010{\natexlab{b}}, \apjl, 724, L176

\bibitem[{{Choi} {et~al.}(2014){Choi}, {Ostriker}, {Naab}, {Oser}, \&
  {Moster}}]{choi+14}
{Choi}, E., {Ostriker}, J.~P., {Naab}, T., {Oser}, L., \& {Moster}, B.~P. 2014,
  ArXiv e-prints

\bibitem[{{Churchill} {et~al.}(2013){Churchill}, {Trujillo-Gomez}, {Nielsen},
  \& {Kacprzak}}]{churchill13}
{Churchill}, C.~W., {Trujillo-Gomez}, S., {Nielsen}, N.~M., \& {Kacprzak},
  G.~G. 2013, ArXiv e-prints

\bibitem[{{Conroy} {et~al.}(2008){Conroy}, {Shapley}, {Tinker}, {Santos}, \&
  {Lemson}}]{conroy+08}
{Conroy}, C., {Shapley}, A.~E., {Tinker}, J.~L., {Santos}, M.~R., \& {Lemson},
  G. 2008, \apj, 679, 1192

\bibitem[{{Cooke} {et~al.}(2006){Cooke}, {Wolfe}, {Gawiser}, \&
  {Prochaska}}]{cwg+06}
{Cooke}, J., {Wolfe}, A.~M., {Gawiser}, E., \& {Prochaska}, J.~X. 2006, \apjl,
  636, L9

\bibitem[{{Cooksey} {et~al.}(2013){Cooksey}, {Kao}, {Simcoe}, {O'Meara}, \&
  {Prochaska}}]{cooksey+13}
{Cooksey}, K.~L., {Kao}, M.~M., {Simcoe}, R.~A., {O'Meara}, J.~M., \&
  {Prochaska}, J.~X. 2013, \apj, 763, 37

\bibitem[{{Coppin} {et~al.}(2008){Coppin}, {Swinbank}, {Neri}, {Cox},
  {Alexander}, {Smail}, {Page}, {Stevens}, {Knudsen}, {Ivison}, {Beelen},
  {Bertoldi}, \& {Omont}}]{coppin08}
{Coppin}, K.~E.~K., {et~al.} 2008, \mnras, 389, 45

\bibitem[{{Creasey} {et~al.}(2013){Creasey}, {Theuns}, \& {Bower}}]{creasey13}
{Creasey}, P., {Theuns}, T., \& {Bower}, R.~G. 2013, \mnras, 429, 1922

\bibitem[{{Crighton} {et~al.}(2011){Crighton}, {Bielby}, {Shanks}, {Infante},
  {Bornancini}, {Bouch{\'e}}, {Lambas}, {Lowenthal}, {Minniti}, {Morris},
  {Padilla}, {P{\'e}roux}, {Petitjean}, {Theuns}, {Tummuangpak}, {Weilbacher},
  {Wisotzki}, \& {Worseck}}]{crighton+11}
{Crighton}, N.~H.~M., {et~al.} 2011, \mnras, 414, 28

\bibitem[{{Crighton} {et~al.}(2013){Crighton}, {Hennawi}, \&
  {Prochaska}}]{crighton13}
{Crighton}, N.~H.~M., {Hennawi}, J.~F., \& {Prochaska}, J.~X. 2013, \apjl, 776,
  L18

\bibitem[{{Crighton} {et~al.}(2014){Crighton}, {Hennawi}, {Simcoe}, {Cooksey},
  {Murphy}, {Fumagalli}, {Prochaska}, \& {Shanks}}]{crighton+14}
{Crighton}, N.~H.~M., {Hennawi}, J.~F., {Simcoe}, R.~A., {Cooksey}, K.~L.,
  {Murphy}, M.~T., {Fumagalli}, M., {Prochaska}, J.~X., \& {Shanks}, T. 2014,
  ArXiv e-prints

\bibitem[{{Croft}(2004)}]{Croft04}
{Croft}, R.~A.~C. 2004, \apj, 610, 642

\bibitem[{{Crotts}(1989)}]{Crotts89}
{Crotts}, A.~P.~S. 1989, \apj, 336, 550

\bibitem[{{Davis} \& {Peebles}(1977)}]{dp77}
{Davis}, M., \& {Peebles}, P.~J.~E. 1977, \apjs, 34, 425

\bibitem[{{Diemand} {et~al.}(2007){Diemand}, {Kuhlen}, \& {Madau}}]{dkm07}
{Diemand}, J., {Kuhlen}, M., \& {Madau}, P. 2007, \apj, 667, 859

\bibitem[{{Diemer} {et~al.}(2013){Diemer}, {More}, \& {Kravtsov}}]{dmk13}
{Diemer}, B., {More}, S., \& {Kravtsov}, A.~V. 2013, \apj, 766, 25

\bibitem[{{D'Odorico} {et~al.}(2010){D'Odorico}, {Calura}, {Cristiani}, \&
  {Viel}}]{dcc+10}
{D'Odorico}, V., {Calura}, F., {Cristiani}, S., \& {Viel}, M. 2010, \mnras,
  401, 2715

\bibitem[{{Dunlop} {et~al.}(2003){Dunlop}, {McLure}, {Kukula}, {Baum}, {O'Dea},
  \& {Hughes}}]{dunlop03}
{Dunlop}, J.~S., {McLure}, R.~J., {Kukula}, M.~J., {Baum}, S.~A., {O'Dea},
  C.~P., \& {Hughes}, D.~H. 2003, \mnras, 340, 1095

\bibitem[{{Ebeling} {et~al.}(2014){Ebeling}, {Stephenson}, \&
  {Edge}}]{ebeling14}
{Ebeling}, H., {Stephenson}, L.~N., \& {Edge}, A.~C. 2014, \apjl, 781, L40

\bibitem[{{Faber} {et~al.}(2007){Faber}, {Willmer}, {Wolf}, {Koo}, {Weiner},
  {Newman}, {Im}, {Coil}, {Conroy}, {Cooper}, {Davis}, {Finkbeiner}, {Gerke},
  {Gebhardt}, {Groth}, {Guhathakurta}, {Harker}, {Kaiser}, {Kassin},
  {Kleinheinrich}, {Konidaris}, {Kron}, {Lin}, {Luppino}, {Madgwick},
  {Meisenheimer}, {Noeske}, {Phillips}, {Sarajedini}, {Schiavon}, {Simard},
  {Szalay}, {Vogt}, \& {Yan}}]{faber07}
{Faber}, S.~M., {et~al.} 2007, \apj, 665, 265

\bibitem[{{Fanidakis} {et~al.}(2013){Fanidakis}, {Macci{\`o}}, {Baugh},
  {Lacey}, \& {Frenk}}]{fanidakis13}
{Fanidakis}, N., {Macci{\`o}}, A.~V., {Baugh}, C.~M., {Lacey}, C.~G., \&
  {Frenk}, C.~S. 2013, \mnras, 436, 315

\bibitem[{{Farina} {et~al.}(2013){Farina}, {Falomo}, {Decarli}, {Treves}, \&
  {Kotilainen}}]{farina13}
{Farina}, E.~P., {Falomo}, R., {Decarli}, R., {Treves}, A., \& {Kotilainen},
  J.~K. 2013, \mnras, 429, 1267

\bibitem[{{Farina} {et~al.}(2014){Farina}, {Falomo}, {Scarpa}, {Decarli},
  {Treves}, \& {Kotilainen}}]{farina+14}
{Farina}, E.~P., {Falomo}, R., {Scarpa}, R., {Decarli}, R., {Treves}, A., \&
  {Kotilainen}, J.~K. 2014, ArXiv e-prints

\bibitem[{{Ferland} {et~al.}(2013){Ferland}, {Porter}, {van Hoof}, {Williams},
  {Abel}, {Lykins}, {Shaw}, {Henney}, \& {Stancil}}]{ferland13}
{Ferland}, G.~J., {et~al.} 2013, RMXAA, 49, 137

\bibitem[{{Font-Ribera} {et~al.}(2013){Font-Ribera}, {Arnau},
  {Miralda-Escud{\'e}}, {Rollinde}, {Brinkmann}, {Brownstein}, {Lee}, {Myers},
  {Palanque-Delabrouille}, {P{\^a}ris}, {Petitjean}, {Rich}, {Ross},
  {Schneider}, \& {White}}]{font13}
{Font-Ribera}, A., {et~al.} 2013, \jcap, 5, 18

\bibitem[{{Fumagalli} {et~al.}(2014){Fumagalli}, {Hennawi}, {Prochaska},
  {Kasen}, {Dekel}, {Ceverino}, \& {Primack}}]{fhp+14}
{Fumagalli}, M., {Hennawi}, J.~F., {Prochaska}, J.~X., {Kasen}, D., {Dekel},
  A., {Ceverino}, D., \& {Primack}, J. 2014, \apj, 780, 74

\bibitem[{{Fumagalli} {et~al.}(2011{\natexlab{a}}){Fumagalli}, {O'Meara}, \&
  {Prochaska}}]{fop11}
{Fumagalli}, M., {O'Meara}, J.~M., \& {Prochaska}, J.~X. 2011{\natexlab{a}},
  Science, 334, 1245

\bibitem[{{Fumagalli} {et~al.}(2013){Fumagalli}, {O'Meara}, {Prochaska}, \&
  {Worseck}}]{fop+13}
{Fumagalli}, M., {O'Meara}, J.~M., {Prochaska}, J.~X., \& {Worseck}, G. 2013,
  \apj, 775, 78

\bibitem[{{Fumagalli} {et~al.}(2011{\natexlab{b}}){Fumagalli}, {Prochaska},
  {Kasen}, {Dekel}, {Ceverino}, \& {Primack}}]{fpk+11}
{Fumagalli}, M., {Prochaska}, J.~X., {Kasen}, D., {Dekel}, A., {Ceverino}, D.,
  \& {Primack}, J.~R. 2011{\natexlab{b}}, \mnras, 418, 1796

\bibitem[{{Gauthier} \& {Chen}(2011)}]{gc11}
{Gauthier}, J.-R., \& {Chen}, H.-W. 2011, \mnras, 418, 2730

\bibitem[{{Haardt} \& {Madau}(2012)}]{hm12}
{Haardt}, F., \& {Madau}, P. 2012, \apj, 746, 125

\bibitem[{{Hennawi}(2004)}]{thesis}
{Hennawi}, J.~F. 2004, Ph.D.~Thesis

\bibitem[{{Hennawi} {et~al.}(2010){Hennawi}, {Myers}, {Shen}, {Strauss},
  {Djorgovski}, {Fan}, {Glikman}, {Mahabal}, {Martin}, {Richards}, {Schneider},
  \& {Shankar}}]{hennawi10}
{Hennawi}, J.~F., {et~al.} 2010, \apj, 719, 1672

\bibitem[{{Hennawi} \& {Prochaska}(2007)}]{QPQ2}
{Hennawi}, J.~F., \& {Prochaska}, J.~X. 2007, \apj, 655, 735

\bibitem[{{Hennawi} \& {Prochaska}(2013)}]{qpq4}
---. 2013, \apj, 766, 58 (QPQ4)

\bibitem[{{Hennawi} {et~al.}(2006{\natexlab{a}}){Hennawi}, {Prochaska},
  {Burles}, {Strauss}, {Richards}, {Schlegel}, {Fan}, {Schneider}, {Zakamska},
  {Oguri}, {Gunn}, {Lupton}, \& {Brinkmann}}]{qpq1}
{Hennawi}, J.~F., {et~al.} 2006{\natexlab{a}}, \apj, 651, 61

\bibitem[{{Hennawi} {et~al.}(2006{\natexlab{b}}){Hennawi}, {Strauss}, {Oguri},
  {Inada}, {Richards}, {Pindor}, {Schneider}, {Becker}, {Gregg}, {Hall},
  {Johnston}, {Fan}, {Burles}, {Schlegel}, {Gunn}, {Lupton}, {Bahcall},
  {Brunner}, \& {Brinkmann}}]{hso+06}
---. 2006{\natexlab{b}}, \aj, 131, 1

\bibitem[{{Hopkins} {et~al.}(2007){Hopkins}, {Richards}, \&
  {Hernquist}}]{hrh07}
{Hopkins}, P.~F., {Richards}, G.~T., \& {Hernquist}, L. 2007, \apj, 654, 731

\bibitem[{{Ivezi{\'c}} {et~al.}(2002){Ivezi{\'c}}, {Menou}, {Knapp}, {Strauss},
  {Lupton}, {Vanden Berk}, {Richards}, {Tremonti}, {Weinstein}, {Anderson},
  {Bahcall}, {Becker}, {Bernardi}, {Blanton}, {Eisenstein}, {Fan},
  {Finkbeiner}, {Finlator}, {Frieman}, {Gunn}, {Hall}, {Kim}, {Kinkhabwala},
  {Narayanan}, {Rockosi}, {Schlegel}, {Schneider}, {Strateva}, {SubbaRao},
  {Thakar}, {Voges}, {White}, {Yanny}, {Brinkmann}, {Doi}, {Fukugita},
  {Hennessy}, {Munn}, {Nichol}, \& {York}}]{ivezic02}
{Ivezi{\'c}}, {\v Z}., {et~al.} 2002, \aj, 124, 2364

\bibitem[{{Jenkins}(2009)}]{jenkins09}
{Jenkins}, E.~B. 2009, \apj, 700, 1299

\bibitem[{{Kwak} {et~al.}(2011){Kwak}, {Henley}, \& {Shelton}}]{kwak11}
{Kwak}, K., {Henley}, D.~B., \& {Shelton}, R.~L. 2011, \apj, 739, 30

\bibitem[{{Lanzetta}(1993)}]{lzt93}
{Lanzetta}, K.~M. 1993, in ASSL Vol. 188: The Environment and Evolution of
  Galaxies, ed. J.~M. {Shull} \& H.~A. {Thronson}, 237--+

\bibitem[{{Lee} {et~al.}(2013){Lee}, {Bailey}, {Bartsch}, {Carithers},
  {Dawson}, {Kirkby}, {Lundgren}, {Margala}, {Palanque-Delabrouille}, {Pieri},
  {Schlegel}, {Weinberg}, {Y{\`e}che}, {Aubourg}, {Bautista}, {Bizyaev},
  {Blomqvist}, {Bolton}, {Borde}, {Brewington}, {Busca}, {Croft}, {Delubac},
  {Ebelke}, {Eisenstein}, {Font-Ribera}, {Ge}, {Hamilton}, {Hennawi}, {Ho},
  {Honscheid}, {Le Goff}, {Malanushenko}, {Malanushenko}, {Miralda-Escud{\'e}},
  {Myers}, {Noterdaeme}, {Oravetz}, {Pan}, {P{\^a}ris}, {Petitjean}, {Rich},
  {Rollinde}, {Ross}, {Rossi}, {Schneider}, {Simmons}, {Snedden}, {Slosar},
  {Spergel}, {Suzuki}, {Viel}, \& {Weaver}}]{lee+13}
{Lee}, K.-G., {et~al.} 2013, \aj, 145, 69

\bibitem[{{Lee} {et~al.}(2012){Lee}, {Suzuki}, \& {Spergel}}]{lee+12}
{Lee}, K.-G., {Suzuki}, N., \& {Spergel}, D.~N. 2012, \aj, 143, 51

\bibitem[{{Loewenstein}(2013)}]{loewenstein13}
{Loewenstein}, M. 2013, \apj, 773, 52

\bibitem[{{Lopez} {et~al.}(2008){Lopez}, {Barrientos}, {Lira}, {Padilla},
  {Gilbank}, {Gladders}, {Maza}, {Tejos}, {Vidal}, \& {Yee}}]{lopez08}
{Lopez}, S., {et~al.} 2008, \apj, 679, 1144

\bibitem[{{Madau} {et~al.}(2001){Madau}, {Ferrara}, \& {Rees}}]{mfr01}
{Madau}, P., {Ferrara}, A., \& {Rees}, M.~J. 2001, \apj, 555, 92

\bibitem[{{Mainieri} {et~al.}(2011){Mainieri}, {Bongiorno}, {Merloni}, {Aller},
  {Carollo}, {Iwasawa}, {Koekemoer}, {Mignoli}, {Silverman}, {Bolzonella},
  {Brusa}, {Comastri}, {Gilli}, {Halliday}, {Ilbert}, {Lusso}, {Salvato},
  {Vignali}, {Zamorani}, {Contini}, {Kneib}, {Le F{\`e}vre}, {Lilly},
  {Renzini}, {Scodeggio}, {Balestra}, {Bardelli}, {Caputi}, {Coppa},
  {Cucciati}, {de la Torre}, {de Ravel}, {Franzetti}, {Garilli}, {Iovino},
  {Kampczyk}, {Knobel}, {Kova{\v c}}, {Lamareille}, {Le Borgne}, {Le Brun},
  {Maier}, {Nair}, {Pello}, {Peng}, {Perez Montero}, {Pozzetti},
  {Ricciardelli}, {Tanaka}, {Tasca}, {Tresse}, {Vergani}, {Zucca}, {Aussel},
  {Capak}, {Cappelluti}, {Elvis}, {Fiore}, {Hasinger}, {Impey}, {Le Floc'h},
  {Scoville}, {Taniguchi}, \& {Trump}}]{mainieri11}
{Mainieri}, V., {et~al.} 2011, \aap, 535, A80

\bibitem[{{Maller} \& {Bullock}(2004)}]{mb04}
{Maller}, A.~H., \& {Bullock}, J.~S. 2004, \mnras, 355, 694

\bibitem[{{Mandelbaum} {et~al.}(2005){Mandelbaum}, {Tasitsiomi}, {Seljak},
  {Kravtsov}, \& {Wechsler}}]{mandelbaum05}
{Mandelbaum}, R., {Tasitsiomi}, A., {Seljak}, U., {Kravtsov}, A.~V., \&
  {Wechsler}, R.~H. 2005, \mnras, 362, 1451

\bibitem[{{Martin} {et~al.}(2010{\natexlab{a}}){Martin}, {Papastergis},
  {Giovanelli}, {Haynes}, {Springob}, \& {Stierwalt}}]{martin10}
{Martin}, A.~M., {Papastergis}, E., {Giovanelli}, R., {Haynes}, M.~P.,
  {Springob}, C.~M., \& {Stierwalt}, S. 2010{\natexlab{a}}, \apj, 723, 1359

\bibitem[{{Martin}(2005)}]{martin05}
{Martin}, C.~L. 2005, \apj, 621, 227

\bibitem[{{Martin} {et~al.}(2010{\natexlab{b}}){Martin}, {Scannapieco},
  {Ellison}, {Hennawi}, {Djorgovski}, \& {Fournier}}]{martin+10}
{Martin}, C.~L., {Scannapieco}, E., {Ellison}, S.~L., {Hennawi}, J.~F.,
  {Djorgovski}, S.~G., \& {Fournier}, A.~P. 2010{\natexlab{b}}, \apj, 721, 174

\bibitem[{{Martin} {et~al.}(2014){Martin}, {Chang}, {Matuszewski}, {Morrissey},
  {Rahman}, {Moore}, \& {Steidel}}]{martin14a}
{Martin}, D.~C., {Chang}, D., {Matuszewski}, M., {Morrissey}, P., {Rahman}, S.,
  {Moore}, A., \& {Steidel}, C.~C. 2014, \apj, 786, 106

\bibitem[{{Matejek} \& {Simcoe}(2012)}]{ms12}
{Matejek}, M.~S., \& {Simcoe}, R.~A. 2012, \apj, 761, 112

\bibitem[{{Matsuda} {et~al.}(2004){Matsuda}, {Yamada}, {Hayashino}, {Tamura},
  {Yamauchi}, {Ajiki}, {Fujita}, {Murayama}, {Nagao}, {Ohta}, {Okamura},
  {Ouchi}, {Shimasaku}, {Shioya}, \& {Taniguchi}}]{matsuda04}
{Matsuda}, Y., {et~al.} 2004, \aj, 128, 569

\bibitem[{{Matteucci} \& {Gibson}(1995)}]{mg95}
{Matteucci}, F., \& {Gibson}, B.~K. 1995, \aap, 304, 11

\bibitem[{{McDonald} \& {Miralda-Escud{\'e}}(1999)}]{mcd99}
{McDonald}, P., \& {Miralda-Escud{\'e}}, J. 1999, \apj, 519, 486

\bibitem[{{M{\'e}nard} {et~al.}(2011){M{\'e}nard}, {Wild}, {Nestor}, {Quider},
  {Zibetti}, {Rao}, \& {Turnshek}}]{menard+11}
{M{\'e}nard}, B., {Wild}, V., {Nestor}, D., {Quider}, A., {Zibetti}, S., {Rao},
  S., \& {Turnshek}, D. 2011, \mnras, 417, 801

\bibitem[{{Mitchell} {et~al.}(1976){Mitchell}, {Culhane}, {Davison}, \&
  {Ives}}]{mitchell76}
{Mitchell}, R.~J., {Culhane}, J.~L., {Davison}, P.~J.~N., \& {Ives}, J.~C.
  1976, \mnras, 175, 29P

\bibitem[{{Mo} \& {White}(1996)}]{mw96}
{Mo}, H.~J., \& {White}, S.~D.~M. 1996, \mnras, 282, 347

\bibitem[{{Moller} \& {Kjaergaard}(1992)}]{moller92}
{Moller}, P., \& {Kjaergaard}, P. 1992, \aap, 258, 234

\bibitem[{{Moster} {et~al.}(2010){Moster}, {Somerville}, {Maulbetsch}, {van den
  Bosch}, {Macci{\`o}}, {Naab}, \& {Oser}}]{moster+10}
{Moster}, B.~P., {Somerville}, R.~S., {Maulbetsch}, C., {van den Bosch}, F.~C.,
  {Macci{\`o}}, A.~V., {Naab}, T., \& {Oser}, L. 2010, \apj, 710, 903

\bibitem[{{Narayanan} {et~al.}(2008){Narayanan}, {Charlton}, {Misawa}, {Green},
  \& {Kim}}]{narayanan08}
{Narayanan}, A., {Charlton}, J.~C., {Misawa}, T., {Green}, R.~E., \& {Kim},
  T.-S. 2008, \apj, 689, 782

\bibitem[{{Nestor} {et~al.}(2005){Nestor}, {Turnshek}, \& {Rao}}]{ntr05}
{Nestor}, D.~B., {Turnshek}, D.~A., \& {Rao}, S.~M. 2005, \apj, 628, 637

\bibitem[{{Nielsen} {et~al.}(2013){Nielsen}, {Churchill}, {Kacprzak}, \&
  {Murphy}}]{nck+13}
{Nielsen}, N.~M., {Churchill}, C.~W., {Kacprzak}, G.~G., \& {Murphy}, M.~T.
  2013, \apj, 776, 114

\bibitem[{{O'Meara} {et~al.}(2011){O'Meara}, {Prochaska}, {Chen}, \&
  {Madau}}]{omeara11}
{O'Meara}, J.~M., {Prochaska}, J.~X., {Chen}, H.-W., \& {Madau}, P. 2011,
  \apjs, 195, 16

\bibitem[{{Oppenheimer} \& {Dav{\'e}}(2006)}]{od06}
{Oppenheimer}, B.~D., \& {Dav{\'e}}, R. 2006, \mnras, 373, 1265

\bibitem[{{Ostriker} \& {Heisler}(1984)}]{oh84}
{Ostriker}, J.~P., \& {Heisler}, J. 1984, \apj, 278, 1

\bibitem[{{Peeples} {et~al.}(2013){Peeples}, {Werk}, {Tumlinson},
  {Oppenheimer}, {Prochaska}, {Katz}, \& {Weinberg}}]{peeples+14}
{Peeples}, M.~S., {Werk}, J.~K., {Tumlinson}, J., {Oppenheimer}, B.~D.,
  {Prochaska}, J.~X., {Katz}, N., \& {Weinberg}, D.~H. 2013, ArXiv e-prints

\bibitem[{{Pettini} {et~al.}(1998){Pettini}, {Kellogg}, {Steidel}, {Dickinson},
  {Adelberger}, \& {Giavalisco}}]{pks+98}
{Pettini}, M., {Kellogg}, M., {Steidel}, C.~C., {Dickinson}, M., {Adelberger},
  K.~L., \& {Giavalisco}, M. 1998, \apj, 508, 539

\bibitem[{{Portinari} {et~al.}(2004){Portinari}, {Moretti}, {Chiosi}, \&
  {Sommer-Larsen}}]{portinari04}
{Portinari}, L., {Moretti}, A., {Chiosi}, C., \& {Sommer-Larsen}, J. 2004,
  \apj, 604, 579

\bibitem[{{Prochaska} \& {Hennawi}(2009)}]{qpq3}
{Prochaska}, J.~X., \& {Hennawi}, J.~F. 2009, \apj, 690, 1558

\bibitem[{{Prochaska} {et~al.}(2013{\natexlab{a}}){Prochaska}, {Hennawi},
  {Lee}, {Cantalupo}, {Bovy}, {Djorgovski}, {Ellison}, {Wingyee Lau}, {Martin},
  {Myers}, {Rubin}, \& {Simcoe}}]{qpq6}
{Prochaska}, J.~X., {et~al.} 2013{\natexlab{a}}, \apj, 776, 136

\bibitem[{{Prochaska} {et~al.}(2013{\natexlab{b}}){Prochaska}, {Hennawi}, \&
  {Simcoe}}]{qpq5}
{Prochaska}, J.~X., {Hennawi}, J.~F., \& {Simcoe}, R.~A. 2013{\natexlab{b}},
  \apjl, 762, L19 (QPQ5)

\bibitem[{{Prochaska} {et~al.}(2011){Prochaska}, {Weiner}, {Chen}, {Mulchaey},
  \& {Cooksey}}]{pwc+11}
{Prochaska}, J.~X., {Weiner}, B., {Chen}, H.-W., {Mulchaey}, J., \& {Cooksey},
  K. 2011, \apj, 740, 91

\bibitem[{{Prochaska} \& {Wolfe}(2009)}]{pw09}
{Prochaska}, J.~X., \& {Wolfe}, A.~M. 2009, \apj, 696, 1543

\bibitem[{{Prochter} {et~al.}(2006){Prochter}, {Prochaska}, \&
  {Burles}}]{ppb06}
{Prochter}, G.~E., {Prochaska}, J.~X., \& {Burles}, S.~M. 2006, \apj, 639, 766

\bibitem[{{Quashnock} \& {vanden Berk}(1998)}]{qb98}
{Quashnock}, J.~M., \& {vanden Berk}, D.~E. 1998, \apj, 500, 28

\bibitem[{{Rakic} {et~al.}(2013){Rakic}, {Schaye}, {Steidel}, {Booth}, {Dalla
  Vecchia}, \& {Rudie}}]{rakic+13}
{Rakic}, O., {Schaye}, J., {Steidel}, C.~C., {Booth}, C.~M., {Dalla Vecchia},
  C., \& {Rudie}, G.~C. 2013, \mnras, 433, 3103

\bibitem[{{Rakic} {et~al.}(2012){Rakic}, {Schaye}, {Steidel}, \&
  {Rudie}}]{rakic12}
{Rakic}, O., {Schaye}, J., {Steidel}, C.~C., \& {Rudie}, G.~C. 2012, \apj, 751,
  94

\bibitem[{{Rao} {et~al.}(2006){Rao}, {Turnshek}, \& {Nestor}}]{rtn06}
{Rao}, S.~M., {Turnshek}, D.~A., \& {Nestor}, D.~B. 2006, \apj, 636, 610

\bibitem[{{Renzini} {et~al.}(1993){Renzini}, {Ciotti}, {D'Ercole}, \&
  {Pellegrini}}]{renzini93}
{Renzini}, A., {Ciotti}, L., {D'Ercole}, A., \& {Pellegrini}, S. 1993, \apj,
  419, 52

\bibitem[{{Ribaudo} {et~al.}(2011){Ribaudo}, {Lehner}, \& {Howk}}]{ribaudo11}
{Ribaudo}, J., {Lehner}, N., \& {Howk}, J.~C. 2011, \apj, 736, 42

\bibitem[{{Richardson} {et~al.}(2012){Richardson}, {Zheng}, {Chatterjee},
  {Nagai}, \& {Shen}}]{richardson12}
{Richardson}, J., {Zheng}, Z., {Chatterjee}, S., {Nagai}, D., \& {Shen}, Y.
  2012, \apj, 755, 30

\bibitem[{{Rosario} {et~al.}(2013){Rosario}, {Trakhtenbrot}, {Lutz}, {Netzer},
  {Trump}, {Silverman}, {Schramm}, {Lusso}, {Berta}, {Bongiorno}, {Brusa},
  {F{\"o}rster-Schreiber}, {Genzel}, {Lilly}, {Magnelli}, {Mainieri},
  {Maiolino}, {Merloni}, {Mignoli}, {Nordon}, {Popesso}, {Salvato}, {Santini},
  {Tacconi}, \& {Zamorani}}]{rtl+13}
{Rosario}, D.~J., {et~al.} 2013, \aap, 560, A72

\bibitem[{{Rubin} {et~al.}(2013){Rubin}, {Prochaska}, {Koo}, {Phillips},
  {Martin}, \& {Winstrom}}]{rubin+14}
{Rubin}, K.~H.~R., {Prochaska}, J.~X., {Koo}, D.~C., {Phillips}, A.~C.,
  {Martin}, C.~L., \& {Winstrom}, L.~O. 2013, ArXiv e-prints

\bibitem[{{Rudie} {et~al.}(2013){Rudie}, {Steidel}, {Shapley}, \&
  {Pettini}}]{rudie13}
{Rudie}, G.~C., {Steidel}, C.~C., {Shapley}, A.~E., \& {Pettini}, M. 2013,
  \apj, 769, 146

\bibitem[{{Rudie} {et~al.}(2012){Rudie}, {Steidel}, {Trainor}, {Rakic},
  {Bogosavljevi{\'c}}, {Pettini}, {Reddy}, {Shapley}, {Erb}, \&
  {Law}}]{rudie12}
{Rudie}, G.~C., {et~al.} 2012, \apj, 750, 67

\bibitem[{{Sasaki} {et~al.}(2014){Sasaki}, {Matsushita}, \& {Sato}}]{sms14}
{Sasaki}, T., {Matsushita}, K., \& {Sato}, K. 2014, \apj, 781, 36

\bibitem[{{Scherrer} \& {Weinberg}(1998)}]{sw98}
{Scherrer}, R.~J., \& {Weinberg}, D.~H. 1998, \apj, 504, 607

\bibitem[{{Scott} {et~al.}(2000){Scott}, {Bechtold}, {Dobrzycki}, \&
  {Kulkarni}}]{sbd+00}
{Scott}, J., {Bechtold}, J., {Dobrzycki}, A., \& {Kulkarni}, V.~P. 2000, \apjs,
  130, 67

\bibitem[{{Seyffert} {et~al.}(2013){Seyffert}, {Cooksey}, {Simcoe}, {O'Meara},
  {Kao}, \& {Prochaska}}]{seyffert+13}
{Seyffert}, E.~N., {Cooksey}, K.~L., {Simcoe}, R.~A., {O'Meara}, J.~M., {Kao},
  M.~M., \& {Prochaska}, J.~X. 2013, \apj, 779, 161

\bibitem[{{Shen} {et~al.}(2011){Shen}, {Madau}, {Aguirre}, {Guedes}, {Mayer},
  \& {Wadsley}}]{shen+11}
{Shen}, S., {Madau}, P., {Aguirre}, A., {Guedes}, J., {Mayer}, L., \&
  {Wadsley}, J. 2011, ArXiv e-prints

\bibitem[{{Shen} {et~al.}(2013{\natexlab{a}}){Shen}, {Madau}, {Conroy},
  {Governato}, \& {Mayer}}]{shen+14}
{Shen}, S., {Madau}, P., {Conroy}, C., {Governato}, F., \& {Mayer}, L.
  2013{\natexlab{a}}, ArXiv e-prints

\bibitem[{{Shen} {et~al.}(2013{\natexlab{b}}){Shen}, {Madau}, {Guedes},
  {Mayer}, {Prochaska}, \& {Wadsley}}]{smg+13}
{Shen}, S., {Madau}, P., {Guedes}, J., {Mayer}, L., {Prochaska}, J.~X., \&
  {Wadsley}, J. 2013{\natexlab{b}}, \apj, 765, 89

\bibitem[{{Shen} {et~al.}(2013{\natexlab{c}}){Shen}, {McBride}, {White},
  {Zheng}, {Myers}, {Guo}, {Kirkpatrick}, {Padmanabhan}, {Parejko}, {Ross},
  {Schlegel}, {Schneider}, {Streblyanska}, {Swanson}, {Zehavi}, {Pan},
  {Bizyaev}, {Brewington}, {Ebelke}, {Malanushenko}, {Malanushenko}, {Oravetz},
  {Simmons}, \& {Snedden}}]{yshen13}
{Shen}, Y., {et~al.} 2013{\natexlab{c}}, \apj, 778, 98

\bibitem[{{Shen} {et~al.}(2007){Shen}, {Strauss}, {Oguri}, {Hennawi}, {Fan},
  {Richards}, {Hall}, {Gunn}, {Schneider}, {Szalay}, {Thakar}, {Vanden Berk},
  {Anderson}, {Bahcall}, {Connolly}, \& {Knapp}}]{shen07}
---. 2007, \aj, 133, 2222

\bibitem[{{Sijacki} {et~al.}(2007){Sijacki}, {Springel}, {di Matteo}, \&
  {Hernquist}}]{Sijacki07}
{Sijacki}, D., {Springel}, V., {di Matteo}, T., \& {Hernquist}, L. 2007,
  \mnras, 380, 877

\bibitem[{{Silk} \& {Rees}(1998)}]{SilkRees98}
{Silk}, J., \& {Rees}, M.~J. 1998, \aap, 331, L1

\bibitem[{{Simcoe} {et~al.}(2006){Simcoe}, {Sargent}, {Rauch}, \&
  {Becker}}]{simcoe06}
{Simcoe}, R.~A., {Sargent}, W.~L.~W., {Rauch}, M., \& {Becker}, G. 2006, \apj,
  637, 648

\bibitem[{{Simpson} {et~al.}(2012){Simpson}, {Smail}, {Swinbank}, {Alexander},
  {Auld}, {Baes}, {Bonfield}, {Clements}, {Cooray}, {Coppin}, {Danielson},
  {Dariush}, {Dunne}, {de Zotti}, {Harrison}, {Hopwood}, {Hoyos}, {Ibar},
  {Ivison}, {Jarvis}, {Lapi}, {Maddox}, {Page}, {Riechers}, {Valiante}, \& {van
  der Werf}}]{simpson12}
{Simpson}, J.~M., {et~al.} 2012, \mnras, 426, 3201

\bibitem[{{Sivanandam} {et~al.}(2009){Sivanandam}, {Zabludoff}, {Zaritsky},
  {Gonzalez}, \& {Kelson}}]{siv09}
{Sivanandam}, S., {Zabludoff}, A.~I., {Zaritsky}, D., {Gonzalez}, A.~H., \&
  {Kelson}, D.~D. 2009, \apj, 691, 1787

\bibitem[{{Spitzer}(1978)}]{spitzer78}
{Spitzer}, L. 1978, {Physical processes in the interstellar medium} (New York
  Wiley-Interscience, 1978.~333 p.)

\bibitem[{{Springel} {et~al.}(2005){Springel}, {Di Matteo}, \&
  {Hernquist}}]{Springel05}
{Springel}, V., {Di Matteo}, T., \& {Hernquist}, L. 2005, \apjl, 620, L79

\bibitem[{{Steidel}(1993)}]{steidel93}
{Steidel}, C.~C. 1993, in ASSL Vol. 188: The Environment and Evolution of
  Galaxies, ed. J.~M. {Shull} \& H.~A. {Thronson}, 263--+

\bibitem[{{Steidel} {et~al.}(2010){Steidel}, {Erb}, {Shapley}, {Pettini},
  {Reddy}, {Bogosavljevi{\'c}}, {Rudie}, \& {Rakic}}]{steidel+10}
{Steidel}, C.~C., {Erb}, D.~K., {Shapley}, A.~E., {Pettini}, M., {Reddy}, N.,
  {Bogosavljevi{\'c}}, M., {Rudie}, G.~C., \& {Rakic}, O. 2010, \apj, 717, 289

\bibitem[{{Stinson} {et~al.}(2012){Stinson}, {Brook}, {Prochaska}, {Hennawi},
  {Shen}, {Wadsley}, {Pontzen}, {Couchman}, {Quinn}, {Macci{\`o}}, \&
  {Gibson}}]{sbp+12}
{Stinson}, G.~S., {et~al.} 2012, \mnras, 425, 1270

\bibitem[{{Stocke} {et~al.}(2013){Stocke}, {Keeney}, {Danforth}, {Shull},
  {Froning}, {Green}, {Penton}, \& {Savage}}]{stocke13}
{Stocke}, J.~T., {Keeney}, B.~A., {Danforth}, C.~W., {Shull}, J.~M., {Froning},
  C.~S., {Green}, J.~C., {Penton}, S.~V., \& {Savage}, B.~D. 2013, \apj, 763,
  148

\bibitem[{{Tejos} {et~al.}(2014){Tejos}, {Morris}, {Finn}, {Crighton},
  {Bechtold}, {Jannuzi}, {Schaye}, {Theuns}, {Altay}, {Le F{\`e}vre},
  {Ryan-Weber}, \& {Dav{\'e}}}]{tejos+14}
{Tejos}, N., {et~al.} 2014, \mnras, 437, 2017

\bibitem[{{Thom} {et~al.}(2012){Thom}, {Tumlinson}, {Werk}, {Prochaska},
  {Oppenheimer}, {Peeples}, {Tripp}, {Katz}, {O'Meara}, {Brady Ford},
  {Dav{\'e}}, {Sembach}, \& {Weinberg}}]{thom12}
{Thom}, C., {et~al.} 2012, \apjl, 758, L41

\bibitem[{{Tumlinson} {et~al.}(2013){Tumlinson}, {Thom}, {Werk}, {Prochaska},
  {Tripp}, {Katz}, {Dav{\'e}}, {Oppenheimer}, {Meiring}, {Ford}, {O'Meara},
  {Peeples}, {Sembach}, \& {Weinberg}}]{tumlinson+13}
{Tumlinson}, J., {et~al.} 2013, \apj, 777, 59

\bibitem[{{Tumlinson} {et~al.}(2011){Tumlinson}, {Thom}, {Werk}, {Prochaska},
  {Tripp}, {Weinberg}, {Peeples}, {O'Meara}, {Oppenheimer}, {Meiring}, {Katz},
  {Dav{\'e}}, {Ford}, \& {Sembach}}]{ttw+11}
---. 2011, Science, 334, 948

\bibitem[{{Turner} {et~al.}(2014){Turner}, {Schaye}, {Steidel}, {Rudie}, \&
  {Strom}}]{turner14}
{Turner}, M.~L., {Schaye}, J., {Steidel}, C.~C., {Rudie}, G.~C., \& {Strom},
  A.~L. 2014, ArXiv e-prints

\bibitem[{{van de Voort} \& {Schaye}(2012)}]{freeke12}
{van de Voort}, F., \& {Schaye}, J. 2012, \mnras, 423, 2991

\bibitem[{{Veilleux} {et~al.}(2009){Veilleux}, {Kim}, {Rupke}, {Peng},
  {Tacconi}, {Genzel}, {Lutz}, {Sturm}, {Contursi}, {Schweitzer}, {Dasyra},
  {Ho}, {Sanders}, \& {Burkert}}]{veilleux09}
{Veilleux}, S., {et~al.} 2009, \apj, 701, 587

\bibitem[{{Veilleux} {et~al.}(2013){Veilleux}, {Mel{\'e}ndez}, {Sturm},
  {Gracia-Carpio}, {Fischer}, {Gonz{\'a}lez-Alfonso}, {Contursi}, {Lutz},
  {Poglitsch}, {Davies}, {Genzel}, {Tacconi}, {de Jong}, {Sternberg}, {Netzer},
  {Hailey-Dunsheath}, {Verma}, {Rupke}, {Maiolino}, {Teng}, \&
  {Polisensky}}]{veilleux13}
---. 2013, \apj, 776, 27

\bibitem[{{Vikas} {et~al.}(2013){Vikas}, {Wood-Vasey}, {Lundgren}, {Ross},
  {Myers}, {AlSayyad}, {York}, {Schneider}, {Brinkmann}, {Bizyaev},
  {Brewington}, {Ge}, {Malanushenko}, {Malanushenko}, {Muna}, {Oravetz}, {Pan},
  {P{\^a}ris}, {Petitjean}, {Snedden}, {Shelden}, {Simmons}, \&
  {Weaver}}]{vikas+13}
{Vikas}, S., {et~al.} 2013, \apj, 768, 38

\bibitem[{{Wang} {et~al.}(2008){Wang}, {Carilli}, {Wagg}, {Bertoldi}, {Walter},
  {Menten}, {Omont}, {Cox}, {Strauss}, {Fan}, {Jiang}, \& {Schneider}}]{wang08}
{Wang}, R., {et~al.} 2008, \apj, 687, 848

\bibitem[{{Weiner} {et~al.}(2009){Weiner}, {Coil}, {Prochaska}, {Newman},
  {Cooper}, {Bundy}, {Conselice}, {Dutton}, {Faber}, {Koo}, {Lotz}, {Rieke}, \&
  {Rubin}}]{wcp+09}
{Weiner}, B.~J., {et~al.} 2009, \apj, 692, 187

\bibitem[{{Werk} {et~al.}(2013){Werk}, {Prochaska}, {Thom}, {Tumlinson},
  {Tripp}, {O'Meara}, \& {Peeples}}]{werk+13}
{Werk}, J.~K., {Prochaska}, J.~X., {Thom}, C., {Tumlinson}, J., {Tripp}, T.~M.,
  {O'Meara}, J.~M., \& {Peeples}, M.~S. 2013, \apjs, 204, 17

\bibitem[{{Werk} {et~al.}(2014){Werk}, {Prochaska}, {Tumlinson}, {Peeples},
  {Tripp}, {Fox}, {Lehner}, {Thom}, {O'Meara}, {Ford}, {Bordoloi}, {Katz},
  {Tejos}, {Oppenheimer}, {Dav{\'e}}, \& {Weinberg}}]{werk+14}
{Werk}, J.~K., {et~al.} 2014, ArXiv e-prints

\bibitem[{{Whitaker} {et~al.}(2012){Whitaker}, {van Dokkum}, {Brammer}, \&
  {Franx}}]{whitaker12}
{Whitaker}, K.~E., {van Dokkum}, P.~G., {Brammer}, G., \& {Franx}, M. 2012,
  \apjl, 754, L29

\bibitem[{{White} {et~al.}(2012){White}, {Myers}, {Ross}, {Schlegel},
  {Hennawi}, {Shen}, {McGreer}, {Strauss}, {Bolton}, {Bovy}, {Fan},
  {Miralda-Escude}, {Palanque-Delabrouille}, {Paris}, {Petitjean}, {Schneider},
  {Viel}, {Weinberg}, {Yeche}, {Zehavi}, {Pan}, {Snedden}, {Bizyaev},
  {Brewington}, {Brinkmann}, {Malanushenko}, {Malanushenko}, {Oravetz},
  {Simmons}, {Sheldon}, \& {Weaver}}]{white12}
{White}, M., {et~al.} 2012, \mnras, 424, 933

\bibitem[{{Wild} {et~al.}(2008){Wild}, {Kauffmann}, {White}, {York}, {Lehnert},
  {Heckman}, {Hall}, {Khare}, {Lundgren}, {Schneider}, \& {vanden
  Berk}}]{wkw+08}
{Wild}, V., {et~al.} 2008, \mnras, 388, 227

\bibitem[{{Yates}(2014)}]{yates14}
{Yates}, R. 2014, MNRAS, submitted

\bibitem[{{Yoon} \& {Putman}(2013)}]{yp13}
{Yoon}, J.~H., \& {Putman}, M.~E. 2013, \apjl, 772, L29

\bibitem[{{Zaritsky} {et~al.}(2004){Zaritsky}, {Gonzalez}, \&
  {Zabludoff}}]{zaritsky04}
{Zaritsky}, D., {Gonzalez}, A.~H., \& {Zabludoff}, A.~I. 2004, \apjl, 613, L93

\bibitem[{{Zhu} {et~al.}(2014){Zhu}, {M{\'e}nard}, {Bizyaev}, {Brewington},
  {Ebelke}, {Ho}, {Kinemuchi}, {Malanushenko}, {Malanushenko}, {Marchante},
  {More}, {Oravetz}, {Pan}, {Petitjean}, \& {Simmons}}]{zhu+14}
{Zhu}, G., {et~al.} 2014, \mnras

\bibitem[{{Zwaan} {et~al.}(2005{\natexlab{a}}){Zwaan}, {Meyer},
  {Staveley-Smith}, \& {Webster}}]{zms+05}
{Zwaan}, M.~A., {Meyer}, M.~J., {Staveley-Smith}, L., \& {Webster}, R.~L.
  2005{\natexlab{a}}, \mnras, 359, L30

\bibitem[{{Zwaan} {et~al.}(2005{\natexlab{b}}){Zwaan}, {van der Hulst},
  {Briggs}, {Verheijen}, \& {Ryan-Weber}}]{zvb+05}
{Zwaan}, M.~A., {van der Hulst}, J.~M., {Briggs}, F.~H., {Verheijen}, M.~A.~W.,
  \& {Ryan-Weber}, E.~V. 2005{\natexlab{b}}, \mnras, 364, 1467

\end{thebibliography}

\appendix

{\bf APPENDIX A: \ion{C}{4} Search}

To enable the evaluation of the cross-correlation of strong
\ion{C}{4} absorption to quasars \xic\ from our dataset
($\S$~\ref{sec:CIV}), we have first
performed a survey for strong \ion{C}{4} absorption ($\mwciv \ge
0.3$\AA) along random samples of quasar sightlines.  Although previous
surveys for \ion{C}{4} absorption have been performed
\citep[e.g.][]{boks03,dcc+10,cooksey+13}, none of these were fully
satisfactory for our planned analysis.
Therefore, we chose to survey a modest sample ($\sim 80$) of quasar
spectra taken from the SDSS survey.  Although we have access to
spectra with much higher resolution, these SDSS spectra have
comparable data quality to the majority of our quasar pair sample  
and therefore suffers from similar systematic uncertainties.
Specifically, we analyzed the SDSS spectra of \nsdss\,quasars drawn
randomly from the cohort defined by \cite{omeara11} to survey $z<2$
Lyman limit absorption using UV spectrometers on the Hubble Space
Telescope ({\it HST}).
These have magnitudes $g < 18.3$, yielding relatively high quality
SDSS data, and emission redshifts $\mzem \approx 2.5$.  From these, 
we excluded any quasar exhibiting very strong associated absorption at
$z \approx \mzem$.

A majority of these data had been previously surveyed for \ion{C}{4}
absorption by \cite{cooksey+13}.  We began our search with their
\ion{C}{4} candidate
list\footnote{http://ucolick.org/$\sim$xavier/SDSSCIV/index.htmlV}, vetted these candidates
and identified new systems by-eye using custom software.  
We adopted and then modified, as necessary, the quasar continua implemented by
those authors using a similar algorithm as for
the quasar pair analysis.  
Table~\ref{tab:sdss_civ} lists the quasars surveyed
and all of the \ion{C}{4} absorption systems discovered.  We measured
equivalent widths \wciv\ for these systems using the same algorithms
for the quasar pair spectra ($\S$~\ref{sec:ew}).

\begin{figure}
\includegraphics[width=3.7in,angle=90]{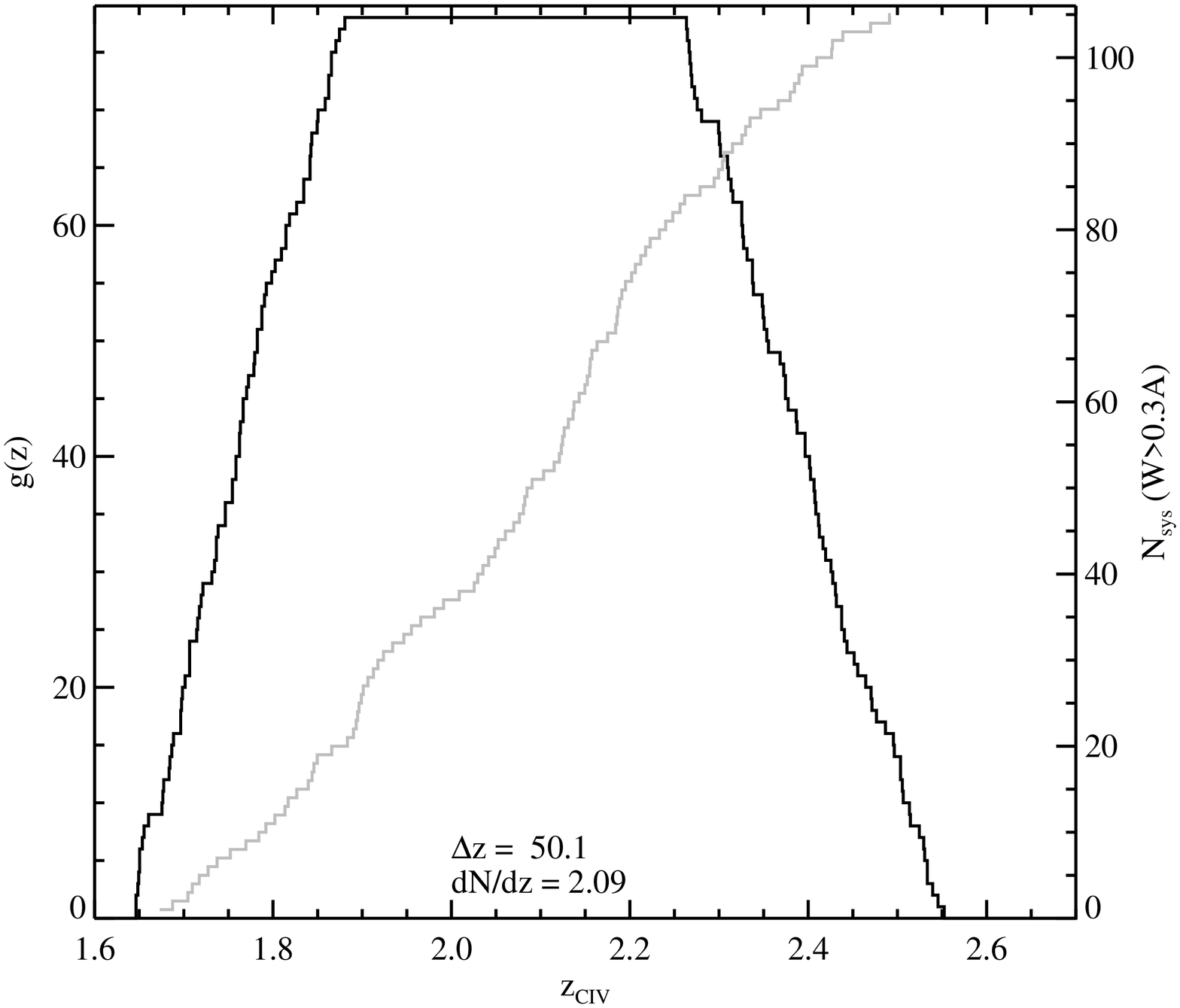}
\caption{The black curve depicts the sensitivity function $g(z)$ of
  the survey we have performed for strong \civt\ absorption ($\mwciv
  \ge 0.3$\AA) in a set of SDSS quasar spectra defined by
  \cite{omeara11}.  The total search path $\Delta z = \int g(z) dz =
  50.1$.  The gray curve shows the cumulative incidence of intervening
  \ion{C}{4} systems satisfying the \wciv\ limit discovered in the
  survey.  This gives an estimate for the incidence of $\ell(z) =
  2.09$ with an approximately 10\%\ Poisson error.  Previous
  estimations are in good agreement with our more precise measurement
  \citep[e.g.][]{boks03,dcc+10,cooksey+13}.
}
\label{fig:dndz}
\end{figure}

Figure~\ref{fig:dndz} presents a summary of the survey path and the
set of \ion{C}{4} absorbers discovered.  For completeness, we have
listed all of the systems, not only those that enter into the
statistical sample.   
Note that for our equivalent width limit of 0.3\AA, such systems
represent $>5\sigma$ detections.
Regarding the statistical survey, we limit the
search of each quasar to the wavelengths $\lambda_r \ge 1238.8$\AA\ in
the quasar rest-frame and to the spectral region $\delta v \le
-4000\mkms$ of the emission redshift.  These criteria define the
$z_{\rm start}$ and $z_{\rm end}$ values listed in
Table~\ref{tab:sdss_civ}.  For the \nsdss\ quasars surveyed, we
constructed a total redshift path $\Delta z = 50.1$ and have
discovered \nciv\ systems satisfying $\mwciv > 0.3$\AA\ within the
search regions.   This gives an incidence of strong \civt\ absorption
at $<z>=2.1$ of $\ell_{0.3 \rm \AA}^{\rm CIV}(z) = \dndz$.  The error reported
reflects Poisson uncertainty from limited sample size.  We estimate
an additional $\approx 10\%$ systematic uncertainty owing to continuum
placement, line-blending, and mis-identifications (e.g.\ a pair of
coincident absorption lines masquerading as a \ion{C}{4} doublet).

We may compare our results to previous estimates of $\ell^{\rm CIV}(z)$ for
strong \ion{C}{4} systems.  \cite{cooksey+13} reported 
$\ell^{\rm CIV}_{0.6 \rm \AA}(z) = 1.04$
for $\mwciv \ge 0.6$\,\AA. 
For $\mwciv \ge 0.3$\AA, they estimate a completeness of less than
50\%\ but correct for this incompleteness to estimate 
$\ell^{\rm CIV}_{0.3 \rm \AA}(z) = 2.3$.
This value is fully consistent with our results given the
uncertainties associated with incompleteness and sample variance.
\cite{dcc+10} performed a \ion{C}{4} survey in high quality,
echelle spectra at $z \sim 2$ achieving a sensitivity to $\mwciv \ll
0.3$\AA.  They discovered 28~\ion{C}{4} absorbers with $\mwciv >
0.3$\AA\ within their $\Delta z=17.9$ survey path giving 
$\ell^{\rm CIV}_{0.3 \rm \AA}(z) = 1.6$, but they defined systems
within windows of only 50\,\kms\ which is below the SDSS resolution
and much lower than our analysis windows.  We consider their results
to be fully consistent, especially given their large sample variance.

\begin{deluxetable*}{lccccccccccccccccc}
\tablewidth{0pc}
\tablecaption{SDSS \ion{C}{4} SURVEY \label{tab:sdss_civ}}
\tabletypesize{\scriptsize}
\tablehead{\colhead{Quasar} &
\colhead{$z_{\rm q}$} &
\colhead{$z_{\rm start}^a$} &
\colhead{$z_{\rm end}^b$} 
& \colhead{$z_{\rm CIV}$}
& \colhead{$W_{\rm CIV}$}
& \colhead{$\sigma(W_{\rm CIV})$}
\\
&&&& (\AA) & (\AA) }
\startdata
J023924.48$-09$0138.6&2.472&1.778&2.426&2.393&0.44&0.04\\
&&&2.495&1.17&0.03\\
J075158.65$+42$4522.9&2.453&1.763&2.407&2.137&0.53&0.03\\
&&&2.305&0.65&0.03\\
J075547.83$+22$0450.1&2.319&1.656&2.275&1.794&0.44&0.06\\
&&&2.284&0.17&0.04\\
J080132.02$+19$2317.5&2.542&1.834&2.495&2.117&0.24&0.04\\
J080620.47$+50$4124.4&2.432&1.746&2.387&1.794&0.08&0.04\\
&&&1.810&0.70&0.04\\
J081014.62$+20$4021.4&2.484&1.788&2.438&1.780&0.73&0.03\\
&&&1.817&0.34&0.03\\
&&&2.084&0.42&0.03\\
&&&2.190&0.45&0.03\\
&&&2.361&0.19&0.03\\
J081114.66$+17$2057.4&2.307&1.646&2.263&\\
J083326.82$+08$1552.0&2.572&1.858&2.525&2.071&0.19&0.05\\
&&&2.187&1.14&0.04\\
J085417.60$+53$2735.2&2.418&1.735&2.373&\\
J091301.01$+42$2344.7&2.311&1.649&2.267&1.711&0.41&0.04\\
&&&1.897&1.62&0.05\\
&&&2.046&0.43&0.04\\
&&&2.068&0.37&0.04\\
&&&2.207&1.01&0.05\\
&&&2.316&0.24&0.03\\
\enddata
\tablenotetext{a}{Starting redshift for the \ion{C}{4} search.  Defined as $(1+z_q)*1238.8/1548.195 - 1$ with $z_{\rm q}$ the emission redshift of the quasar.}
\tablenotetext{b}{Ending redshift for the \ion{C}{4} search.  This is defined as 4000\kms\ blueward of the quasar redshift.}
\tablecomments{[The complete version of this table is in the electronic edition of the Journal.  The printed edition contains only a sample.]}
\end{deluxetable*}

\noindent
{\bf APPENDIX B: Luminous Red Galaxy Analysis}

To generate the CGM maps of \ion{Mg}{2} for the LRG sample
(Figure~\ref{fig:CII_map}), we made the following assumptions and
approximations.  First, we adopted the covering fractions \fcs\
measured by \cite{bc11} for strong \ion{Mg}{2} absorbers ($\mwmgii >
0.3$\AA). Specifically, they measured
$\mfc =[0.245, 0.193, 0.032, 0.01]$ for $\mrphys = [25., 75., 125.,
175]$\,kpc.  Although these results were derived from a
photometrically-selected sample of LRGs, they are in good agreement with
the (smaller) spectroscopic study of \cite{gc11}.

Second, we assume that the absorbers associated with LRGs and having
$\mwmgii > 0.3$\AA\ exhibit a \wmgii\ distribution similar to the
intervening population along `random' quasar sightlines.  This \wmgii\
distribution is well-described by an exponential, 
$f(\mwmgii) \propto \exp[-\mwmgii/\mwmgii^*]$, 
with \wmgii$^* = 0.7$\AA\ \citep{ntr05,ppb06}. 
Integrating this distribution from 0.3\AA\ to $\infty$, we
calculate an average equivalent width $\mavmgs = 1$\AA\ for
these strong systems.  This means that an ensemble of sightlines with
a covering fraction $\mfc^S$ to strong \ion{Mg}{2} absorption has an
average equivalent width $\mfc^S \, \mavmgs$.
We further assume that sightlines with $\mwmgii < 0.3$\AA\
have an equivalent width distribution that follows a power-law
to 0.01\AA\ with
exponent $\gamma = -3$.  This gives an average equivalent width for
such weak absorbers, \avmgw=0.02\AA.
Altogether, an ensemble of sightlines has

\begin{equation}
\mtavmg = (1-\mfc^S)\mavmgw + \, \mfc^S \, \mavmgs
\label{eqn:avmg}
\end{equation}
Finally, to generate the CGM maps for LRGs in
Figure~\ref{fig:CII_map}, we combine the \fcs\ values of
\cite{bc11} and the \wmgii\ distributions for weak/strong systems. 
Together these provide the \wmgii\ distribution shown in
Figure~\ref{fig:CII_map} as a function of \rphys.

\begin{figure}
\includegraphics[width=5in,angle=90]{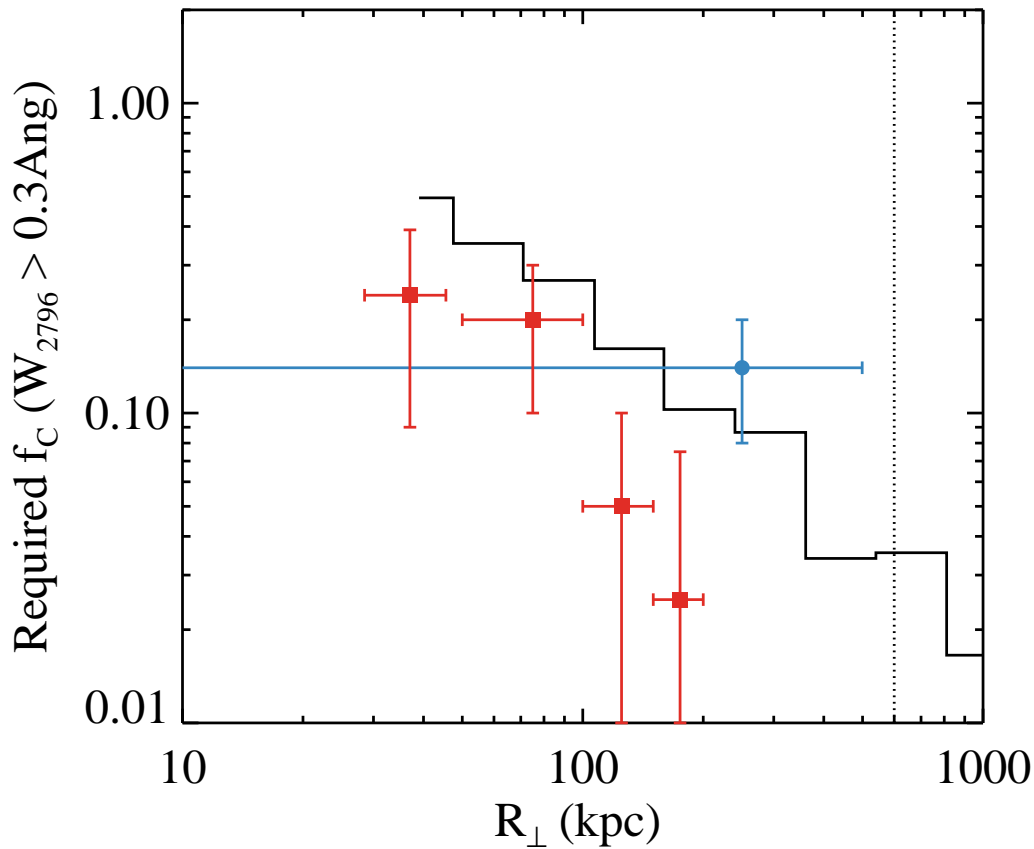}
\caption{Estimated covering fraction \fc\ (black curve) required for strong
  \ion{Mg}{2} absorption systems ($\mwmgii > 0.3$\AA) to reproduce the
  average equivalent width observed in spectral stacks probing LRGs
  \citep{zhu+14}.  
  See the Appendix for a discussion of the methodology.
  The colored points show estimations of the covering
  fraction from \cite[red][]{bc11} and \cite[blue][]{gc11}.
}
\label{fig:lrg_fc}
\end{figure}

Figure~\ref{fig:lrg_fc} presents the \fcs\ values from the literature
for a limiting equivalent width of 0.3\AA\ versus impact parameter
\citep{bc11,gc11}.
Overplotted on these data is the curve showing the \fcs\ values required
to reproduce the measured \tavmg\ values from Z14, ignoring
the contribution from weak absorbers (i.e.\ the first term in
Equation~\ref{eqn:avmg}). 
At $\mrphys < 100$\,kpc it is evident that a
large \fcs\ value is required to reproduce the observations.  We
conclude that the central, red-and-dead galaxy of these massive halos
exhibits a cool and enriched CGM.

At larger radii ($\mrphys \approx \mrvir$), however, the
measurements may be reproduced with a very small cover fraction
($\mfcs < 5\%$), consistent with the low \fcs\ values inferred from
direct analysis.  We contend that the \tavmg\ measurements of
Z14 are dominated by such rare sightlines. 
We further
posit that these represent sightlines penetrating the gas associated to
satellite galaxies of the dark matter halos hosting LRGs.  Such gas may
be the ISM of these galaxies and/or their own `local' CGM. 
One need not invoke an extremely diffuse medium with
$\mwmgii \ll 0.3$\AA\ throughout the LRG halo to explain the
Z14 measurements.  And, we also emphasize
that this interpretation follows from the results on \ion{Mg}{2}
observed for the CGM of lower mass populations
\citep[e.g.][]{werk+13}.

Taking the above conclusion one step further, we may test whether all of
the mass in the Z14 measurements at $\mrphys > \mrvir$ could be
associated to the ISM of individual galaxies.  Integrating the mass
profile of their 2-halo term, they represent an average ratio of Mg$^+$ ions to
dark matter: $\rho(\rm Mg^+)/\rho_{\rm DM} \approx  10^{-8}$.
Now consider the mass density of Mg$^+$ in present-day galaxies.
At $z \sim 0$, surveys of 21cm emission\footnote{
  Survey of the damped \lya\ systems (DLAs) give even larger values at
  $z \approx 0.5$ \citep{rtn06,pw09}.}
give a cosmological \ion{H}{1} mass density
$\Omega_{\rm HI} = 4.3 \sci{-4} h^{-1}_{70}$ \citep{zms+05,martin10}.  
It is reasonable to assume that
all of the Mg associated with this atomic gas\footnote{There could be
  additional Mg$^+$ in molecular gas.} is Mg$^+$.  Therefore,
to estimate the mass density of Mg$^+$ in the gas-phase, one must
adopt a metallicity [Mg/H] and a dust depletion $D_{\rm Mg}$.
For the former, \cite{zvb+05} estimate the ISM of the population of
galaxies dominating $\rho_{\rm HI}$ to be $\approx 1/2$ solar.  For the latter, 
the Galactic ISM shows $D_{\rm Mg} \approx 0.3$ for all but the densest
gas \citep{jenkins09}.   Combining these factors, we find

\begin{equation}
\rho({\rm Mg^+})_{\rm ISM} = \rho_c \Omega_{\rm HI} \ltp \frac{Mg}{H} \rtp \, 
 \ltp \frac{m_{\rm Mg}}{m_p} \rtp \, D_{\rm Mg}  = 5.2 \sci{-8} \,
 \rho_c
\end{equation}
We can compare this quantity to the dark matter density today, 
$\rho_{\rm DM} = \Omega_{\rm DM} \rho_c$ with $\Omega_{\rm DM} = 0.22$.  This gives the ratio
of Mg$^+$ to dark matter of  $\rho({\rm Mg^+})_{\rm ISM}/\rho_{\rm DM}
= 2 \sci{-7}$.  This exceeds the Z14 estimates by over an order of
magnitude.  Even if we allow for a greater depletion factor \citep[dense gas
in the Galaxy exhibits $D_{\rm Mg} \lesssim 0.1$][]{jenkins09}, 
the gas in the ISM easily accounts for the Z14 results.  In
fact, we are be forced to conclude that the Z14 measurements are
biased low, perhaps due to dust obscuration \citep{oh84} and/or a
systematic underestimate of the column densities from their equivalent
width measurements (i.e.\ line-saturation).

\end{document}